\documentclass[twoside,12pt]{article}
\usepackage{epsfig}
\usepackage{graphicx}
\usepackage{amssymb}

\begin{document}
\newcommand{\p}{\partial}
\newcommand{\ls}{\left(}
\newcommand{\rs}{\right)}
\newcommand{\beq}{\begin{equation}}
\newcommand{\eeq}{\end{equation}}
\newcommand{\beqa}{\begin{eqnarray}}
\newcommand{\eeqa}{\end{eqnarray}}
\newcommand{\bdm}{\begin{displaymath}}
\newcommand{\edm}{\end{displaymath}}
\newcommand{\fps}{f_{\pi}^2 }
\newcommand{\mks}{m_{{\mathrm K}}^2 }
\newcommand{\ms}{m_{{\mathrm K}}^{*} }
\newcommand{\mk}{m_{{\mathrm K}} }
\newcommand{\msq}{m_{{\mathrm K}}^{*2} }
\newcommand{\rhos}{\rho_{\mathrm s} }
\newcommand{\rhob}{\rho_{\mathrm B} }
\title{Kaon production in heavy ion reactions at intermediate energies 
\footnote{To appear in Prog. Part. Nucl. Phys., [nucl-th/0507017]}}
\author{Christian Fuchs\\
Institut f\"ur Theoretische Physik der Universit\"at T\"ubingen,\\
Auf der Morgenstelle 14, D-72076 T\"ubingen, Germany
}
\maketitle  
\begin{abstract}
The article reviews the physics related to kaon and antikaon production 
in heavy ion reactions at intermediate energies. Chiral dynamics predicts 
substantial modifications of the kaon properties in a dense nuclear 
environment. The status of the theoretical predictions as well as 
experimental evidences for medium effects such as repulsive/attractive 
mass shifts for $K^+/K^-$ are reviewed. 
In the vicinity of the thresholds, and even more pronounced 
below threshold, the production of strangeness is a highly collective 
process. Starting from elementary reaction channels the phenomenology 
of $K^+$ and $K^-$ production, i.e. freeze-out densities, time scales etc. 
as derived from experiment and theoretical 
transport calculations is presented. Below threshold kaon production 
shows a high sensitivity on the nuclear compression reached in heavy 
ion reactions. 
This allows to put constraints on the nuclear equation-of-state which 
are finally discussed.  
\end{abstract}
\tableofcontents
\section{Introduction}
For more than two decades strangeness production has been one of the 
major research topics of heavy ion physics at intermediate energies. 
First experiments on strangeness production were already performed in the 
early 80ties at the LBL in Berkeley \cite{lbl}. These were, however, 
restricted to collisions of light nuclei. 
In the middle of the 90ties the whole 
field obtained a significant push 
when the KaoS \cite{kaos94} and the FOPI \cite{fopi95} Collaborations at 
the Gesellschaft f\"ur Schwerionenforschung (GSI) in Darmstadt/Germany 
started to deliver high precision data for kaon and antikaon production 
in heavy systems. The kaon spectrometer KaoS has stopped 
data acquisition in 2004 and will be dismantled at the GSI. Hence it is 
a proper time to draw some resume on what has been achieved during the 
last twenty years and which problems still have to be settled. 
The  present article tries to summarize the status 
of the field and to draw conclusions from the various experimental and
theoretical efforts. 

Strangeness production in heavy ion reactions at intermediate energies 
is of high interest since it opens the possibility to attack several 
fundamental questions of nuclear and hadron physics which are not 
only interesting by themselves but have also astrophysical 
implications and/or are related to 
fundamental aspects of Quantum Chromo Dynamics (QCD). 

What makes the production of strange hadrons special is 
the fact that strangeness is exactly conserved in hadronic reactions. 
Open strangeness can be produced by the creation of kaon 
($K^+ ({\bar u}s),K^0 ({\bar d}s)$) - antikaon 
($K^- (u{\bar s}),{\bar K}^0 (d {\bar s})$) pairs or by kaon-hyperon pairs. 
The hyperons carrying one strange quark are $\Lambda (uds)$ and 
$\Sigma$ ($\Sigma^-(dds),~\Sigma^0(uds),~\Sigma^+(uus)$) 
 hyperons. The production of hidden strangeness through 
$\phi (s{\bar s})$ mesons is possible but 
suppressed according to the Okubo-Zweig-Iizuka 
selection rule \cite{OZI}. Since data on $\phi$ production 
are extremely scarce at intermediate energies, the present article 
will mainly focus on kaon and antikaon production. 

A first consequence of strangeness conservation is the fact that $K^+$ 
mesons, once produced, cannot be absorbed by 
the surrounding nucleons. This results in a rather long mean free 
path of about 7 fm of  $K^+$ mesons in nuclear matter and makes them 
 a suitable 'penetrating' probe for the dense fireball produced 
in heavy ion reactions. Final state interactions such as elastic 
kaon-nucleon scattering or the propagation in potentials 
influence the dynamics but do not change the total yields. 
A second consequence of strangeness conservation 
are the high production thresholds. The cheapest way to 
produce an $s{\bar s}$ 
pair is the reaction $NN\longrightarrow N\Lambda K^+$  which has a 
threshold of $E_{\rm lab} =1.58$ GeV for the incident 
nucleon. When the 
incident energy per nucleon in a heavy ion reaction is below this 
value one speaks about {\it subthreshold} kaon production. Subthreshold 
kaon production is in particular interesting since it ensures that the 
kaons originate from  the high density phase of the reaction. The missing 
energy has to be provided either by the Fermi motion of the nucleons or 
by energy accumulating multi-step reactions. Both processes exclude significant
distortions from surface effects if one goes sufficiently far below 
threshold. In combination with the long mean free path subthreshold $K^+$ 
production is an ideal tool to probe compressed nuclear matter in relativistic 
heavy ion reactions. Indeed, one of the
major motivations to start the kaon project at the GSI was to explore 
the nuclear equation-of-state (EOS) at supra-normal densities, i.e. 
significantly above nuclear saturation density 
$\rho_0 \simeq 0.16~{\rm fm}^{-3}$. A better knowledge of the high 
density behavior of the nuclear EOS is relevant for astrophysical 
purposes, e.g. for the understanding of neutron stars and supernovae 
explosions. Heavy ion reactions provide the only possibility to attack  
this question experimentally and to constrain theoretical models 
above saturation density. After a more than thirty years quest for 
the nuclear EOS it seems that by studying subthreshold $K^+$ 
production substantial progress could be achieved in the recent years.

A second push of more theoretical nature was given to the field when 
chiral models became more and more popular in the late 80ties and 
early 90ties. 
Chiral symmetry is an exact symmetry of QCD 
in the limit of massless quarks.  
Since the up and down current quark masses 
are small, i.e. of the order of 5-10 MeV, 
this symmetry is still approximately fulfilled. In nature it is, 
however, spontaneously broken by the 
non-vanishing -- and large -- expectation value of the scalar chiral quark 
condensate  $\langle {\bar q}q \rangle$ of the QCD vacuum (for a 
pedagogical review see e.g. \cite{koch}). 
The spontaneous symmetry breaking, similar to the 
spontaneous magnetization of a ferromagnet which breaks the symmetry of the 
underlying Hamiltonian, implies the existence of massless Goldstone 
bosons which are the pions. The small pion mass of 140 MeV ensures that 
the concept of chiral symmetry is a fundamental feature of low 
energy hadron physics. By the extension to the full SU(3) sector the 
pseudoscalar meson octet of $\pi,\eta,K$ and ${\bar K}$ mesons plays now the 
role of the Goldstone bosons. With a strange quark mass of about 150 MeV 
the explicit symmetry breaking is much larger in the strange sector than 
in the SU(2) sector but 
the general concepts of chiral symmetry and its spontaneous breaking 
are believed to be still valid. They do not only allow to 
understand the origin of the 
pseudoscalar meson masses. Chiral perturbation theory (ChPT) is 
also considered as the  {\it exact}, QCD based, theory of 
the pion-nucleon interaction at low energies \cite{gasser82}. 

Hence it was a natural step to extend ChPT to the SU(3) flavor sector 
\cite{kaplan86}. However, the situation turned out to be more
complicated than in pure SU(2). 
While the $\pi N$ and the $K N$ interactions can be treated 
perturbatively, the  ${\bar K} N$ interaction is already around threshold 
dominated by the presence of resonances. This makes a perturbative treatment 
of the antikaon-nucleon interaction impossible. Instead, non-perturbative 
approaches are required, which can be achieved 
within chiral dynamics, however, by the price of loosing of a well defined
expansion scheme. This essential difference between kaons and antikaons 
will be reflected at many points in the following discussions. It is 
the main reason why conclusions on in-medium modifications of kaon 
properties are much firmer than those for antikaons. 

Such medium modifications are closely connected to the conjecture of a partial 
restoration of chiral symmetry in dense matter. The chiral 
condensate  $\langle {\bar q}q \rangle$ is expected to be reduced at 
finite density and/or temperature \cite{brown96,weise,tuebingenC} which should 
be reflected, e.g., in shifts of the corresponding meson masses. 
Heavy ion reactions open thus the possibility to test
fundamental concepts of hadron physics and QCD. A strong reduction of the 
$K^-$ mass, as also supported by the investigation of kaonic atoms 
\cite{gal94}, will have severe astrophysical consequences. It can lead 
to $K^-$ condensation in neutron stars \cite{kaplan86,brown94,prakash94,li97b} 
which, due to additional negative charge, increases the proton 
fraction and softens thus the equation-of-state for the neutron star. 
The onset of $K^-$ condensation is reached when the electron 
chemical potential starts to exceed that of the kaons $\mu_e \ge m^*_{K^-}$ 
which might happen at around 4-5 times nuclear saturation density.  
This lowers the maximal neutron star mass to about 1.5 solar masses 
and the core of a supernova, if heavier than this value, will collapse 
into a black hole. This, on the other hand will lead to a large 
number of low mass black holes in the universe \cite{bethe94} 
(for a recent review see \cite{weber05}). 

In the late 80ties/ early 90ties first mean field 
calculations \cite{kaplan86,lutz94} predicted already a 
moderately repulsive $K^+$ potential ($V \sim + (20\div 30)$ MeV) and a 
strongly attractive $K^-$ potential ($V \sim - (100\div 200)$ MeV) at 
nuclear saturation density. Such a value of the $K^+$ potential is 
in accordance with the $K^+N$ scattering length and theoretical 
estimates for the potential remained fairly stable over the years. 
For antikaons the 
situation is more complex. The resonant character of the $K^-N$ 
interaction makes the mean field picture highly questionable 
and the size of the in-medium $K^-$ potential is even not yet  
completely settled from the theoretical side. 

A major challenge studying strangeness production in 
heavy ion reactions is to verify (or falsify) 
the conjecture of the existence 
of these two potentials, to determine their size from experiment and 
to put constraints on theoretical models. 
However, heavy ion reactions are highly dynamical processes. At intermediate 
energies the phase space distributions of the colliding nuclei are far 
from global and even local equilibrium over most of the duration of 
the reaction \cite{fuchs03}. Since observables are not snapshots but the 
results of space-time integrals over the entire reaction dynamics 
it is unavoidable to account for this dynamical evolution. 
This means that the link between experiment and the underlying 
physics has to be provided by dynamical transport models. Semi-classical 
transport equations of a Boltzmann type can be derived from 
non-equilibrium quantum field theory \cite{dani84,btm90}. Corresponding 
Boltzmann-Uehling-Uhlenbeck (BUU) \cite{buu,rbuu} 
models or, alternatively, the 
Quantum-Molecular-Dynamics (QMD) approach \cite{Ai91} are well 
established transport models and explain successfully a 
large variety of hadronic 
observables in heavy ion reactions, such as e.g. collective flow 
pattern or spectra and abundances of newly produced particles 
\cite{bass98,cassing99}. Most of the results presented in this 
review are based on the QMD approach. Details of the T\"ubingen 
QMD model can be found in \cite{shekhter03,uma97}. 

Such hadronic transport models propagate one-body phase space 
distributions in self-consistent potentials and account for 
elastic and inelastic two-body scattering processes. They contain 
basic quantum features such as final state Pauli-blocking for fermions. 
As soon as particle production comes into play the transport approach 
becomes a coupled-channel problem. The various hadron species are coupled 
via production and absorption processes and by their mean fields. 
Transport models are generally 
mean field approaches which rely on the quasi-particle approximation (QPA) 
for stable particles and narrow resonances. $K^+$ mesons can safely be 
treated within such a framework since they retain good quasi-particle 
properties also in a dense environment. Antikaons, in contrast, seem to 
develop complex structures in their spectral functions which would require 
to account for off-shell effects beyond the quasi-particle approximation. 
There exist first attempts to go in this direction \cite{cassing00} 
but most available calculations on $K^-$ production are based on the QPA.

The review is now organized as follows: Chapter 2 gives an overview 
of the theoretical predictions for the in-medium modifications 
which the kaons should experience in a dense nuclear environment. Since 
chiral perturbation theory is considered as the most suitable tool to study 
the interactions of pseudoscalar mesons with nucleons 
at low energies the chapter starts with a short outline of the 
derivation of effective kaon-nucleon models based on chiral dynamics. 
Mean field models as well as more elaborated coupled channel dynamics 
which are required by the resonant structure of the antikaon-nucleon 
interactions, are briefly discussed. For completeness also the expected 
scenario in the complementary case, i.e. in a high temperature, baryon 
dilute but pion dominated, environment is mentioned. 
Chapter 3 summarizes the phenomenology of strangeness 
production in  heavy ion collisions around threshold energies. 
Starting from elementary processes, i.e. strangeness 
production and exchange reactions, the presently accepted 
scenarios for kaon and antikaon production are developed 
as they can be deduced from 
experiments and corresponding dynamical simulations. This concerns 
the questions of system size dependences, freeze-out 
densities, time scales etc.. An important question is in this context 
the degree of equilibration which the $K^\pm$ mesons reach or do not 
reach in the expanding system. This question is intimately connected 
with the possibility to probe the early high density phase of a 
heavy ion reaction. If the kaons would have time to 
equilibrate they would loose their memory on the early reaction stages.  
Chapter 4 turns then to the treatment of strangeness production 
within dynamical transport models. Here we focus on the $K^+$ 
production and discuss exemplarily the realization 
within the T\"ubingen QMD approach. The treatments within transport 
models used by other groups are similar in principle, but can differ 
in details. Finally 
a comparison of the predictions from various transport models 
for selected pion and kaon observables is given. The search for signatures 
of in-medium mass shifts, or more generally, in-medium potentials is 
discussed in Chapter 5. Mass shifts lead first of all to shifts of the 
production threshold which are reflected in the total particle yields. 
Since conclusions are only possible relative to a free mass scenario they 
have to be based on transport simulations. The data situation strongly supports 
the in-medium mass scenario, at least concerning the $K^+$ mesons. These observations 
are complemented by the study of dynamical observables such as the collective 
in-plane and out-of-plane pattern.  Finally Chapter 6 turns to the original 
issue which motivated the kaon program at the GSI, namely to extract 
information on the nuclear equation-of-state. The measurement of $K^+$ excitation 
functions down to energies far below threshold can 
be considered as a breakthrough. Before coming to the interpretation 
of the data a brief summary on the present status of the theoretical 
prediction for the nuclear EOS is given.  The dependence of the 
kaon production on the compression achieved in heavy ion reactions 
puts constraints on the EOS which are discussed in detail. Finally 
the consistency with information from other sources such as 
nucleon and neutron stars is outlined. 

The review closes with a summary of the major results.

\section{Kaons in dense matter}
\subsection{Chiral SU(3) Lagrangian}
The  natural framework to study the interaction between pseudoscalar 
mesons and baryons at low energies is chiral perturbation theory (ChPT). 
Kaplan and Nelson were the first to apply the chiral Lagrangian to 
the properties of kaons in nuclear matter \cite{kaplan86}.  Later on  
this framework has  been used by many other authors 
\cite{brown96,brown94,brown92,kaiser95,brown96b,waas96,waas96b,waas97,lutz98,mao99,lutz02,lutz02b}.   
The corresponding chiral SU(3)$_L\times$SU(3)$_R$ Lagrangian used
by Kaplan and Nelson reads
\begin{eqnarray}
{\cal L}&=&{1\over 4}f^2{\rm Tr}\partial^\mu\Sigma\partial_\mu\Sigma^+
+{1\over 2}f^2\Lambda[{\rm Tr}M_q(\Sigma-1)+{\rm h.c.}]
+{\rm Tr}{\bar B}(i\gamma^\mu\partial_\mu-m_B)B\nonumber\\
&+&i{\rm Tr}{\bar B}\gamma^\mu[V_\mu, B]
+D{\rm Tr}{\bar B}\gamma^\mu\gamma^5\{A_\mu, B\}
+F{\rm Tr}{\bar B}\gamma^\mu\gamma^5[A_\mu, B]\nonumber\\
&+&a_1{\rm Tr}{\bar B}(\xi M_q\xi+{\rm h.c.})B
+a_2{\rm Tr}{\bar B}B(\xi M_q\xi+{\rm h.c.})\nonumber\\
&+&a_3[{\rm Tr}M_q\Sigma+{\rm h.c.}]{\rm Tr}{\bar B}B.
\label{lag1}
\end{eqnarray}
The degrees of freedom in the Lagrangian (\ref{lag1}) are 
the baryon octet $B$ 
\begin{equation}
   B = \left( \begin{array}{ccc} {\Lambda \over \sqrt{6}} + {\Sigma^0 \over
   \sqrt{2}} &   \Sigma^+ &  p  \\
  \Sigma^- &  {\Lambda \over \sqrt{6}} - {\Sigma^0 \over \sqrt{2}} & n \\
  \Xi^- & \Xi^0  & - {2 \over \sqrt{6}} \Lambda  \end{array} \right)
\label{octet1}
\end{equation}
with a degenerate mass $m_B$, and the pseudoscalar meson octet $\phi$
\begin{equation}
  \phi = \sqrt{2} \left( \begin{array}{ccc} {{\eta_8} \over \sqrt{6}} +
{{\pi^0} \over \sqrt{2}} & \pi^+ & K^+ \\
 \pi^- & {{\eta_8} \over \sqrt{6}} - {{\pi^0} \over \sqrt{2}} & K^0 \\
 K^-  &  \overline{K^0} & - {2 \over \sqrt{6}} \eta_8 \end{array} \right)
\label{octet2}
\end{equation}
entering into the chiral pseudoscalar meson fields 
\begin{equation}
\Sigma=\exp (2i\phi/f_\pi)~~{\rm and}~~
\xi=\sqrt\Sigma=\exp(i\phi/f_\pi)~~.
\label{field1}
\end{equation}
The pseudoscalar meson decay constants are equal in the SU(3)$_V$ 
limit and given by the weak pion decay constant $f_\pi \simeq 93$ MeV.  
 The current quark mass matrix which is responsible for  
explicit chiral symmetry breaking is given by 
\begin{equation}
 M_q = \left( \begin{array}{ccc} m_q & 0 & 0 \\
                                0 & m_q & 0 \\
                                0 & 0 & m_s  \end{array} \right)
\label{mquark}
\end{equation}
if one neglects the small difference 
between the up and down quark masses 
($m_u \simeq m_d \equiv m_q \simeq  5.5$ MeV). 
The constants $F$ and $D$ are the  SU(3) axial vector 
couplings with $F+D = g_{\rm A}$ which determine the pseudo-vector 
meson-baryon coupling strengths through corresponding 
Goldberger-Treiman relations. 

Chiral symmetry of QCD is explicitely broken by the finite, but small 
quark masses (\ref{mquark}). However, compared to the chiral symmetry 
breaking scale $\Lambda_\chi \simeq 4\pi f_\pi \sim 1$ GeV the up 
and down quark masses and also the strange quark mass 
($m_s \simeq 150$ MeV) are small and the QCD Lagrangian is still 
approximately chirally invariant. The same holds for the mesonic 
sector of the chiral Lagrangian (\ref{lag1}). This allows a systematic 
expansion in powers of hadron momenta and light quark masses over  
$\Lambda_\chi$, i.e. chiral perturbation theory \cite{gasser82}. In 
the baryonic sector chiral symmetry is broken due to the 
baryon mass $m_B$. In the  Lagrangian  $m_B$ is degenerate for the 
baryon octet and the mass spectrum has to be fixed through the 
expansion coefficients.

The mesonic vector $V_\mu$ and axial vector 
$A_\mu$ currents are defined as
\begin{equation}
V_\mu={1\over 2}(\xi^+\partial_\mu\xi+\xi\partial_\mu\xi^+)~~{\rm and}~~
A_\mu={i\over 2}(\xi^+\partial_\mu\xi-\xi\partial_\mu\xi^+),
\end{equation}
respectively. 

The treatment of the full Lagrangian (\ref{lag1}) is complicated since 
it leads automatically to a coupled channel problem. This approach 
has been pursued by several authors 
\cite{kaiser95,waas96,waas96b,waas97,lutz98,mao99,lutz02,lutz02b}. However, 
for many applications, in particular studying kaon properties 
at the mean field level, an effective chiral Lagrangian based 
on kaon and nucleon degrees of freedom can be used. 
We discuss the effective $KN$ Lagrangian approach first and 
turn then to the more involved coupled channel problem. 

\subsection{Effective chiral Lagrangian}
The Lagrangian (\ref{lag1}) can be reduced to an effective Lagrangian 
by the steps outlined below \cite{kaplan86,ko95}. One should, however, 
be aware that the following expansion is not chiral perturbation theory since 
it mixes contributions of different order in ChPT. 

First the 
pseudoscalar meson field $\Sigma$ is expanded up to order  $1/\fps$ and 
only the kaon field $K$ is kept. The terms  
involving the axial vector current do not contribute to the kaon mass 
and can be ignored. The first two terms in Eq. (\ref{lag1}) are the 
kinetic and the mass term 
\begin{equation}
\partial^\mu\bar K\partial_\mu K-\Lambda(m_q+m_s)\bar KK~~,
\end{equation}
from where one can identify the kaon mass 
\begin{equation}
m_{\rm K}^2=\Lambda(m_q+m_s) ~~.
\end{equation}
The kaon field is given by 
\begin{equation}
K=\left(\matrix{K^+ \cr
                K^0 \cr}\right)~~
{\rm and} ~~\bar K=(K^- ~~\bar {K^0})~~~.
\end{equation}
Keeping explicitly only nucleon and kaon degrees of freedom, 
the third and fourth terms in 
Eq. (\ref{lag1}) lead to the Dirac equation for the nucleon field 
\begin{equation}
N=\left(\matrix{p \cr
                n \cr}\right)~~{\rm and}~~\bar N=(\bar p~~ \bar n),
\end{equation}
and the Weinberg-Tomozawa $KN$ interaction term \cite{weinberg}
\begin{equation}
{\bar N}(i\gamma^\mu\partial_\mu-m_B)N
-{3i\over 8\fps}{\bar N}\gamma^\mu N 
\bar K \buildrel \leftrightarrow\over \partial_\mu ~~.
\label{wt1}
\end{equation}
 
The last three terms in Eq.  (\ref{lag1}) can be similarly worked out 
and lead to a scalar  $KN$ interaction, the so-called Kaplan-Nelson 
term \cite{kaplan86}
\begin{eqnarray}
{\rm Tr}{\bar B}(\xi M_q\xi+{\rm h.c.})B&=&2m_q{\bar N}N-{{\bar N}N\over 2\fps}
(m_q+m_s){\bar K}K \nonumber\\
{\rm Tr}{\bar B}B(\xi M_q\xi+{\rm h.c.})&=&2m_s{\bar N}N-{{\bar N}N\over \fps}
(m_q+m_s){\bar K}K \nonumber\\
\,[{\rm Tr}M_q\Sigma+{\rm h.c.}]{\rm Tr}{\bar B}B&=&2(2m_q+m_s){\bar N}N
-{2{\bar N}N\over \fps}(m_q+m_s){\bar K}K ~~.
\end{eqnarray}
Combining these expressions, the full Lagrangian reads
\begin{eqnarray}
{\cal L}&=&{\bar N}(i\gamma^\mu\partial_\mu-m_B)N
+\partial^\mu{\bar K}\partial_\mu K-\Lambda(m_q+m_s){\bar K}K\nonumber\\
&-&{3i\over 8\fps}{\bar N}\gamma^\mu N
\bar K \buildrel \leftrightarrow\over \partial_\mu K
+[2m_qa_1+2m_sa_2+2(2m_q+m_s)a_3]{\bar N}N\nonumber\\
&-&{{\bar N}N{\bar K}K\over 2\fps}(m_q+m_s)(a_1+2a_2+4a_3)
~~~.
\label{lag2}
\end{eqnarray}
Now one can fix the remaining free parameters $a_1,~a_2,~{\rm and}~a_3$ 
from the nucleon mass   
\begin{eqnarray}
m_N=m_B-2[a_1m_q+a_2m_s+a_3(2m_q+m_s)] 
\label{mb}
\end{eqnarray}
and the kaon-nucleon sigma term $\Sigma_{\rm KN}$ can be identified 
\begin{equation}
\Sigma_{\rm KN}= -{1\over 2}(m_q+m_s)(a_1+2a_2+4a_3)~~.
\label{knsig1}
\end{equation}
Eq. (\ref{knsig1}) is obtained using eq. (\ref{mb}) for the 
nucleon mass and the definition of the sigma term 
\begin{equation}
\Sigma_{\rm KN}=
{1\over 2}(m_q+m_s)\Big[{1\over 2}\frac{\partial m_N}{\partial m_q}+
\frac{\partial m_N}{\partial m_s}\Big]
\label{knsig2}
\end{equation}
The effective chiral 
kaon-nucleon Lagrangian reads now up to order $(1/\fps)$
\begin{eqnarray}
{\cal L}&=&{\bar N}(i\gamma^\mu\partial_\mu-m_N)N
+\partial^\mu{\bar K}\partial_\mu K
-(\mks -{\Sigma_{KN}\over \fps}{\bar N}N){\bar K}K\nonumber\\
&-&{3i\over 8\fps}{\bar N}\gamma^\mu N
\bar K \buildrel \leftrightarrow\over \partial_\mu K~~~.
\label{lag3}
\end{eqnarray}
It contains a vector interaction, the Weinberg-Tomozawa term, 
which is repulsive for kaons and 
attractive for antikaons due to $g$-parity. The attractive scalar 
interaction, the Kaplan-Nelson term, is equal for kaons and antikaons. 
The strength of the 
Kaplan-Nelson term depends thereby on the magnitude of the kaon-nucleon 
sigma term $\Sigma_{\mathrm{KN}}$.
The Weinberg-Tomozawa term is current algebra while the 
Kaplan-Nelson interaction is next to leading order in ChPT.

In contrast to the pion-nucleon-sigma term 
which is experimentally well determined from pion-nucleon scattering 
($\Sigma_{\pi N} \simeq 45$ MeV), the kaon-nucleon-sigma term is 
a relatively uncertain quantity since it is related to the 
strangeness content of the nucleon. In addition to the explicit 
breaking from the quark masses (\ref{mquark}) chiral symmetry is 
spontaneously broken by the large expectation values of the scalar 
quark condensate $\langle {\bar q}q \rangle 
\simeq ((230\pm 25)~{\rm MeV})^3$ of the QCD vacuum. The pion, 
respectively the complete pseudoscalar meson octet assuming SU(3)$_V$ 
symmetry, plays the role of the Goldstone boson of chiral symmetry 
breaking. Thus, like for the pion, the kaon mass $m_{\rm K}$ 
can be related to the vacuum quark condensates 
$ \langle {\bar q}q \rangle $ by the 
Gell-Mann-Oakes-Renner (GOR) relation \cite{gor} 
\begin{equation}
m_{\mathrm K}^2 = \frac{1}{2}( m_u + m_s) 
\langle {\bar u}u + {\bar s}s \rangle 
\quad .
\label{renner}
\end{equation}
$ m_u$ in eq. (\ref{renner}) 
is the up-quark current mass and $ m_s$ is the 
strange quark mass. In the  nucleon the right hand side of eq. 
(\ref{renner}) defines the kaon-nucleon sigma term 
\begin{equation}
\Sigma_{\rm KN} = \frac{1}{2}( m_u + m_s) 
\langle N | {\bar u}u + {\bar s}s | N \rangle 
\quad .
\label{sigkn} 
\end{equation}
Kaon-nucleon scattering yields values for the isospin averaged 
sigma term of about 
$\Sigma_{\mathrm KN} \simeq 400$ MeV whereas lattice QCD predicts 
values between 300-450 MeV \cite{brown96,dong95,Borasoy02}, 
heavy baryon ChPT~\cite{Borasoy98} 
$\Sigma_{\rm KN} = 380\pm 40$ MeV (I=1) and 
chiral quark model calculations \cite{inoue04} 
$\Sigma_{\rm KN} = 386$ MeV. Thus $\Sigma_{\mathrm KN}$ 
can range from $2m_\pi$ up to 450 MeV, however, with the tendency 
of a value around 400 MeV to establish. 
 
\subsubsection{Mean field dynamics}
For estimates of kaon mass shifts in nuclear matter and kaon 
dynamics in heavy ion reactions the above Lagrangian (\ref{lag3}) is 
usually applied in mean field approximation. Already in the early 
90ties mean field calculations were carried out in the Nambu-Jona-Lasinio 
(NJL) model \cite{lutz94}. In the context of chiral SU(3) dynamics 
the mean field approximation means to treat 
the $KN$ interaction at the tree level.  
The in-medium Klein-Gordon equation for the kaons follows from (\ref{lag3}) 
via the Euler-Lagrange equations 
\beq
\left[ \partial_\mu \partial^\mu \pm \frac{3i}{4\fps} j_\mu \partial^\mu 
+ \left( \mks - \frac{\Sigma_{\mathrm{KN}}}{\fps} \rhos \right) 
\right] \phi_{\mathrm{K^\pm}} (x) = 0
\quad .
\label{kg1}
\eeq
Here $j_\mu = \langle {\bar N}\gamma_\mu N  \rangle $ is the 
nucleon four-vector current and 
$\rhos= \langle {\bar N} N  \rangle  $ the scalar baryon density. 
With the vector potential 
\beq
V_\mu = \frac{3}{8\fps} j_\mu 
\label{v1}
\eeq
and an effective kaon mass $\ms$ defined as \cite{fuchs98}
\beq
\ms = \sqrt{ \mks - \frac{\Sigma_{\mathrm{KN}}}{\fps} \rhos 
     + V_\mu V^\mu }
\label{mstar1}
\eeq
the Klein-Gordon Eq. (\ref{kg1}) can be written as
\beq
\left[ \left( \partial_\mu \pm i V_\mu \right)^2  + \msq \right] 
\phi_{{\mathrm K}^\pm} (x) = 0 
\quad . 
\label{kg2}
\eeq
Thus the vector field is introduced by minimal coupling 
into the Klein-Gordon with opposite signs for $K^+$ and 
$K^-$ while the effective mass $\ms$ is equal for both.
Introducing  effective 
momenta ($k_{\mu}^* = (E^*, {\bf k}^*)$) as well  
\beq
k^{*}_{\mu} = k_{\mu} \mp V_\mu
\label{keff}
\eeq
the Klein-Gordon equation (\ref{kg1},\ref{kg2}) reads in 
momentum space 
\beq
\left[ k^{*2} - \msq \right] \phi_{\mathrm{K} } (k) = 0~~~.
\label{kg3}
\eeq
Eq. (\ref{kg3}) is just the mass-shell constraint for quasi-particles 
inside the nuclear medium. There exists now a complete analogy to the 
quasi-particle picture for the nucleons in relativistic mean field 
theory, e.g. in the Walecka model of Quantum Hadron Dynamics (QHD) 
\cite{serot88} where the nucleon obeys  
an effective Dirac equation 
\bdm
\left[ {\slash \!\!\! k}^* - m^* \right] u (k) = 0
\edm
for the in-medium nucleon spinors $u$. 
From Eqs. (\ref{kg3}) the dispersion relation follows 
\beqa
E ({\bf k}) = k_0 = \sqrt{ {\bf k}^{*2} + \msq } \pm V_0
\quad .
\label{disp1}
\eeqa
In nuclear matter at rest where the space-like components of the 
vector potential vanish, i.e. ${\bf V} = 0$ and 
${\bf k}^* ={\bf k}$, Eq. (\ref{disp1}) 
reduces to 
\beq 
E ({\bf k}) = \sqrt{ {\bf k}^2 
+ \mks - \frac{\Sigma_{\mathrm{KN}}}{\fps}\rhos + V_{0}^2 } 
\pm V_0 
\quad .
\label{disp2}
\eeq
Eq. (\ref{disp1}) accounts for the full Lorentz structure, a fact 
which comes into play when  heavy ion collisions are considered 
where one has to transform between different 
reference frames, e.g. the center-of-mass frame of the colliding nuclei 
and the frame where a kaon is created. Like in electrodynamics the 
spatial components of the vector field give rise to a Lorentz force 
\cite{fuchs98} as discussed in detail in Chap. 4. 

Now one can also introduce the kaon optical potential through the 
in-medium dispersion relation   
\beqa
0= k_{\mu}^{*2} - \msq = k_{\mu}^{2} - \mks - 2\mk U_{\rm opt} 
\quad .
\label{mass2}
\eeqa
In mean field approximation the difference between the mass shell conditions 
for kinetic and canonical quantities is simply given by the optical or 
Schroedinger equivalent potential 
\beqa
U_{\rm opt}(\rho ,{\bf k}) 
=  -\Sigma_S \pm \frac{k_{\mu} V^{\mu}}{\mk} 
+ \frac{\Sigma_S^2 - V_{\mu}^2}{2\mk}  
=  \pm \frac{\ k_{\mu} V^{\mu}}{\mk} 
- \frac{\Sigma_{\mathrm{KN}}}{\fps 2\mk}\rhos ~.
\label{uopt}
\eeqa
Here we introduced the total scalar kaon self-energy 
\beqa  
\Sigma_S \equiv m_{{\mathrm K}} - \ms 
\approx \frac{1}{2 m_{\mathrm{K}} }\left( 
\frac{\Sigma_{\mathrm{KN}}}{\fps}\rhos - V_{\mu}^2 \right)
\quad .
\label{sigs}
\eeqa
From Eq. (\ref{mass2}) follows 
\beq
\frac{{\bf k}^2}{2\mk} + U_{\rm opt} = \frac{{\bf k}^{2}_{\infty}}{2\mk}
\eeq
where $|{\bf k}_{\infty}|$ is the 
asymptotic momentum ${\bf k}^{2}_{\infty} = k_{0}^2 - \mks $ 
of the incoming particle in the Schroedinger equation. 
Thus  $U_{\rm opt}$ of Eq. (\ref{uopt}) corresponds 
to the potential in the non-relativistic Schroedinger equation. 
Again the optical potential is of identical structure 
as the central part of the optical potential 
for nucleons \cite{serot88}. However, in the latter case 
an additional spin-dependent 
part of the interaction can be obtained from a Fouldy-Wouthousen 
transformation of the Dirac equation.

In this context it should be mentioned that in the literature 
it is often not distinguished  between in-medium 'mass' 
and in-medium energy shifts, i.e. the energy (\ref{disp2}) 
at zero momentum is identified 
as an in-medium mass $\ms \equiv E ({\bf k}=0)$. In order to strengthen 
the analogy with the relativistic mean field picture for baryons 
and to distinguish clearly between the different Lorentz properties 
 - $\ms$ given by Eq. (\ref{mstar1}) is by definition a Lorentz scalar - 
we will distinguish between these two quantities in the 
discussion of medium effects. 

To call $\ms$ the in-medium mass is also consistent with 
the general picture of a reduction of meson masses 
in dense and hot hadronic matter. From the Hellmann-Feynman theorem 
follows a reduction of the non-strange quark condensate 
$\langle {\bar u}u + {\bar d}d \rangle$ which is linear 
to first order in the nuclear density\footnote{In 
isospin symmetric nuclear matter the density is related to the 
Fermi momentum by  $\rho = 2\,k_F^3/(3\,\pi^2 )$} $\rho$ \cite{cohen92,brown96}
\begin{equation}
\frac{ \langle \rho \mid {\bar u}u +  {\bar d}d \mid \rho \rangle}
{\langle {\bar u}u + {\bar d}d \rangle} \simeq 1 - 
\frac{\Sigma_{\pi N}}{f_{\pi}^2 m_{\pi}^2}   \rho + \cdots
\label{cond1}
\quad .
\end{equation}
Model calculations \cite{lutz94,weise,tuebingenC}, 
e.g. within the NJL model, predicted a similar reduction of 
the chiral condensate. Assuming 
an analogous behavior of the strange condensate yields 
\begin{equation}
\frac{ \langle \rho \mid {\bar u}u +  {\bar s}s \mid \rho \rangle}
{\langle {\bar u}u + {\bar s}s \rangle} \simeq 1 - 
\frac{\Sigma_{\mathrm{KN}}}{f_{\pi}^2 m_{\mathrm K}^2}   \rho + \cdots
\quad .
\label{cond2}
\end{equation}
Combining now the GOR relation (\ref{renner}) with (\ref{cond2}), 
the effective kaon mass scales with density as 
\begin{equation}
\msq = \mks - \frac{\Sigma_{\mathrm{KN}}}{f_{\pi}^2} \rho 
+ {\cal O}\left( k_F^4 \right)
\label{kmass1} 
\end{equation}
which is exactly the form of eq. (\ref{mstar1}) and consistent with the 
picture of partial restoration of chiral symmetry. 
\subsubsection{Higher order corrections}
The importance of higher order contributions beyond tree level 
can be estimated from the empirical kaon-nucleon scattering lengths. 
Such a comparison gives also a feeling how far the mean field model 
complies with kaon-nucleon scattering.

To lowest order in density the mass or energy shift of a meson is 
generally given by the forward scattering length, in the case of kaons 
\begin{eqnarray}
\Delta \,E_{K}^2 ({\bf k}=0) = E_{K}^2 -\mk^2 = 
-\pi \left( 1+\frac{\mk}{m_N} \right)
\left( a^{(I=0)}_{KN}+3\,a^{(I=1)}_{KN} \right)
\,\rho +{\cal O}\left( k_F^4 \right)~~.
\label{lowdensity}
\end{eqnarray}
The empirical values of the isospin $I=0$ and $I=1$ 
$K^+$-nucleon scattering lengths are 
$a_{K^+N}^{(I=0)}\simeq 0.02 $ fm and 
$a_{K^+N}^{(I=1)}\simeq -0.32 $ fm \cite{martin81} which leads to a 
repulsive mass shift of about 28 MeV at 
nuclear saturation density ($k_F \simeq 265 $ MeV).
Higher order corrections in the density expansion of (\ref{lowdensity}) 
were found to be small for $K^+$. 
The $k_F^4 $ correction was found to increases the
repulsive $K^+$-mass shift only by about $20 \%$ 
\cite{lutz98} compared to expression (\ref{lowdensity}).  
This fact makes the density expansion useful in the $K^+$ sector. 
The empirical scattering lengths can now 
be compared to the tree level Weinberg-Tomozawa interaction which 
yields $a_{K^+N}^{(I=0)} = 0 $ fm and 
$a_{K^+N}^{(I=1)}\simeq -0.585 $ fm \cite{kaiser95}. Thus current algebra 
and the corresponding effective $KN$ Lagrangian (\ref{lag3}) is 
in rough qualitative agreement with the constraints from 
low energy $K^+$ nucleon scattering. 
In the effective model the too large vector 
repulsion (\ref{v1}) is compensated by the attractive scalar 
Kaplan-Nelson potential (\ref{kmass1}). Compared to the empirical 
values, the repulsion is now, however, overcompensated \cite{ko95}. 

One way to overcome this problem  at the mean field level 
has been suggested by Brown and Rho \cite{brown96b}. They 
subsumed higher correlations of chiral order $Q^3$ and above 
into a medium dependence of the pion decay constant which 
should scale in matter similar as the chiral condensate: 
Assuming the Gell-Mann-Oakes-Renner relation (\ref{renner}) 
to be still valid  in the medium, one obtains the following 
relation for the in-medium pion decay constant $f_{\pi}^{*}$ 
\begin{eqnarray}
{f_\pi^{*2}\over f_\pi^2} = {m_\pi^2\over m_\pi^{*2}}
{\langle \rho \mid {\bar q}q \mid  \rho \rangle \over \langle {\bar q}q\rangle }~~.
\label{gmor}
\end{eqnarray}
From ChPT \cite{weise97} and $\pi$-mesonic atoms \cite{kienle04} the 
(s-wave)  pion mass is known to change only slightly with density.  
Using the empirical values of $m_\pi^*(\rho _0)/m_\pi \approx 1.05 $, 
one obtains
\begin{eqnarray}\label{fpi}
f_\pi ^{*2}(\rho _0) / f_\pi^2 \approx 0.6
\end{eqnarray}
at nuclear saturation density $\rho_0 \simeq 0.16~{\rm fm}^{-3}$. 
A dropping pion decay constant 
enhances both, the vector repulsion and the scalar attraction. However, 
as found in one-loop ChPT the additional attraction is  counterbalanced 
by the range term which is of the same order 
as the Kaplan-Nelson term. Similar results 
have been found for pions \cite{ericson92}. Moreover, such a dropping 
of the pion decay constant seems to be supported by the potentials 
extracted from pionic atoms \cite{friedman04}. At the mean field level 
these results can be incorporated by replacing $\fps \mapsto f_{\pi}^{*2}$ 
only in the vector potential (\ref{v1}).  

\begin{figure}[h]
\unitlength1cm
\begin{picture}(13.,8.0)
\put(1.5,0){\makebox{\epsfig{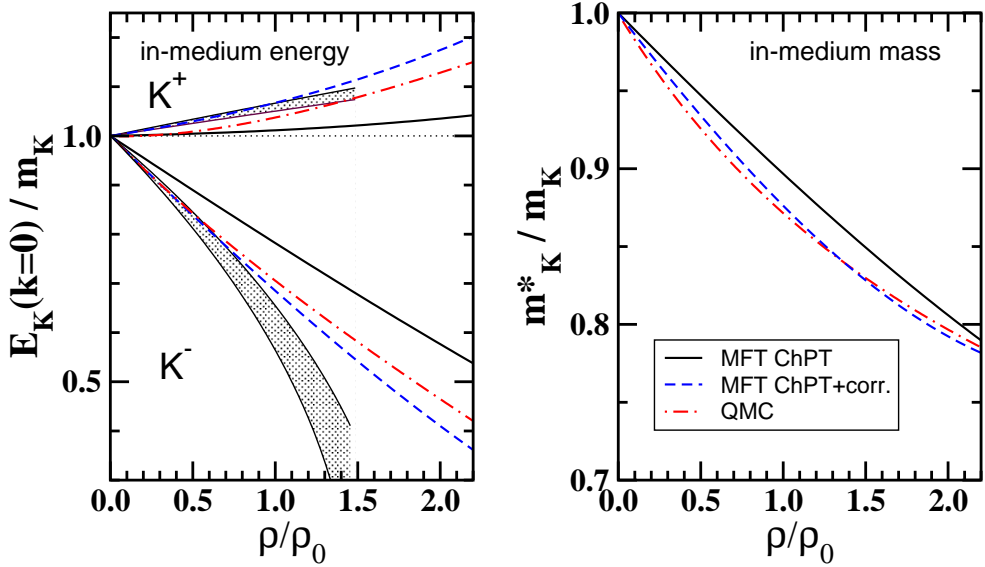}}}
\end{picture}
\caption{In-medium kaon energy (left) and quasi-particle mass (right) 
in the chiral mean field theory (MFT ChPT) and including 
higher order corrections (MFT ChPT+corr.) \protect\cite{brown96b}. 
Results from the mean field quark-meson-coupling (QMC) model \cite{qmc98} 
are shown as well. The bands represent the values extracted from 
empirical $K^{+}N$ scattering and $K^{-}$ atoms 
\protect\cite{batty97}.
}
\label{kmass1_fig}
\end{figure}

Fig. \ref{kmass1_fig} shows the in-medium energy shift of 
$K^+$ and $K^-$ and the in-medium mass defined by (\ref{mstar1}) in nuclear 
matter. MFT denotes thereby the Lagrangian (\ref{lag3}) with a value of 
$\Sigma_{\mathrm{KN}} = 350$ MeV which has originally been used 
by Li and Ko \cite{ko95}. MFT+corr. denotes the mean field model proposed by 
Brown et al. \cite{brown96b} including the above mentioned 
higher order corrections with  
a value of $\Sigma_{\mathrm{KN}} = 450$ MeV. The MFT and MFT+corr.
curves shwon in Fig.  \ref{kmass1_fig} are obtained by Eqs. 
(\ref{mstar1},\ref{disp2}) with the corresponding values 
for $\Sigma_{\mathrm{KN}}$, however in the MFT+corr. case 
with the additional replacement of  $\fps$  by $f_{\pi}^{*2}$ 
in the vector field (\ref{v1}). The empirical 
energy shifts are shown as well in Fig. \ref{kmass1_fig}. For $K^+$ 
the value is obtained by Eq. (\ref{lowdensity}) from the 
empirical $K^+ N$ scattering length. The $K^- $ band corresponds to the 
empirical iso-spin averaged $K^- N$ scattering length of 
${\bar a}_{K^{-}N}=0.62\pm 0.5$ fm extracted from kaonic atom 
data \cite{batty97}. The strength of the empirical 
$K^- $-nucleus potential is in the analysis of 
Batty et al. \cite{batty97} quite robust 
against a variation of the  nuclear wave functions and 
the inclusion of $p$-wave interactions. The $s$-wave 
potential shown in Fig. \ref{kmass1_fig} gives therefore a 
good impression of the values determined from kaonic atoms.

Similar results have been obtained with slightly modified versions 
of the effective chiral Lagrangian \cite{mao99}, in the 
quark-meson-coupling model \cite{qmc98} shown in Fig. \ref{kmass1_fig}, 
and in relativistic mean field 
calculations where kaons are coupled to static $\sigma , \omega, \rho$ and 
$\delta$ meson background fields \cite{schaffner97}.

In this context it is worth to mention that the philosophy behind the 
$KN$ mean field model is similar to that of 
effective relativistic nucleon-meson 
Lagrangiens of QHD \cite{serot88,rmf}. Both are designed to describe 
in-medium properties, in the latter case nuclear matter and finite nuclei. 
In both cases the structure of the interaction complies with the knowledge 
from free scattering, however, the models do not pretend to give a 
quantitative description of free scattering data. In the present case 
 this is true for the $K^+$ sector. The situation becomes, however, much 
less satisfying turning to the $K^-$ sector. 

Again one can use the low density theorem to estimate the medium 
effects to leading order in density. The  empirical scattering lengths
$a_{K^-N}^{(I=0)}\simeq (-1.70+i\,0.68 )$ fm and
$a_{K^-N}^{(I=1)}\simeq (0.37 +i\,0.60)$ fm \cite{martin81,Iwasaki}
imply according to (\ref{lowdensity}) a repulsive mass shift of $
23$ MeV and a width of $\Gamma_{K^{-}}
\simeq 147 $ MeV at saturation density. In contrast to $K^+$, the 
next order correction to the density expansion of 
Eq. (\ref{lowdensity}) is large, resulting in a total 
repulsive mass shift of $55 $ MeV and a width of $\Gamma_{K^-}
\simeq 195 $ MeV \cite{lutz98}. First of all, this questions the 
convergence of a density expansion for the $K^{-}$-mode. Moreover, 
the leading terms suggest a repulsive $K^-$ potential which stands in clear 
contradiction to the empirical knowledge from kaonic atom data 
(${\bar a}_{K^-N} = 0.62\pm0.5$ fm) \cite{batty97} 
suggesting sizable attraction at small density . Finally the
empirical $K^-N$ scattering lengths are in disagreement
with the Weinberg-Tomozawa term which predicts an attractive mass 
shift. These facts imply that perturbation theory is not applicable in the 
$K^-$ sector. The reason lies in the existence of a resonance, the 
$\Lambda (1405)$ close the $K^-p$ threshold which makes the  $K^-p$ 
interaction repulsive at threshold. The appearance of resonances 
requires generally a non-perturbative treatment of two-body scattering 
processes.  

\subsection{Non-perturbative coupled channel dynamics}
A non-perturbative calculation of the full two-body scattering 
amplitude $T$ requires to solve the Lippmann-Schwinger equation, 
respectively its relativistic counterpart, the Bethe-Salpeter 
equation. For kaon-nucleon scattering the Bethe-Salpeter equation reads 
schematically 
\beq
T_{KN\rightarrow KN} = V_{KN\rightarrow KN} -i \int V_{KN\rightarrow MB} 
G_B D_M  T_{MB\rightarrow KN}
\label{bsk1}
\eeq
where $G_B$ and $D_M$ are baryon and meson propagators, respectively
\beqa
G_B(p) = \frac{1}{\not{p} - m_B +i\epsilon}~~,~~
D_M(k) = \frac{1}{k^2 - m_M^2 +i\epsilon}~~.
\eeqa
The Bethe-Salpeter equation (\ref{bsk1}) iterates the $KN$ interaction 
kernel $V$  to infinite order. It is a coupled channel equation since 
it contains not only kaon and nucleon degrees of freedom but involves 
the complete baryon ($B=N,\Lambda ,\Sigma$) 
and pseudoscalar meson ($M=\pi,K$) octet (\ref{octet1}) and 
 (\ref{octet2}) of the chiral Lagrangian. The coupling to 
$\Xi$'s and $\eta$'s can be neglected. The $\Lambda ,~\Sigma$ and $\pi$ 
degrees of freedom are, however, essential for $KN$ scattering.  
\begin{figure}[h]
\unitlength1cm
\begin{picture}(13.,9.0)
\put(2.5,0){\makebox{\epsfig{file=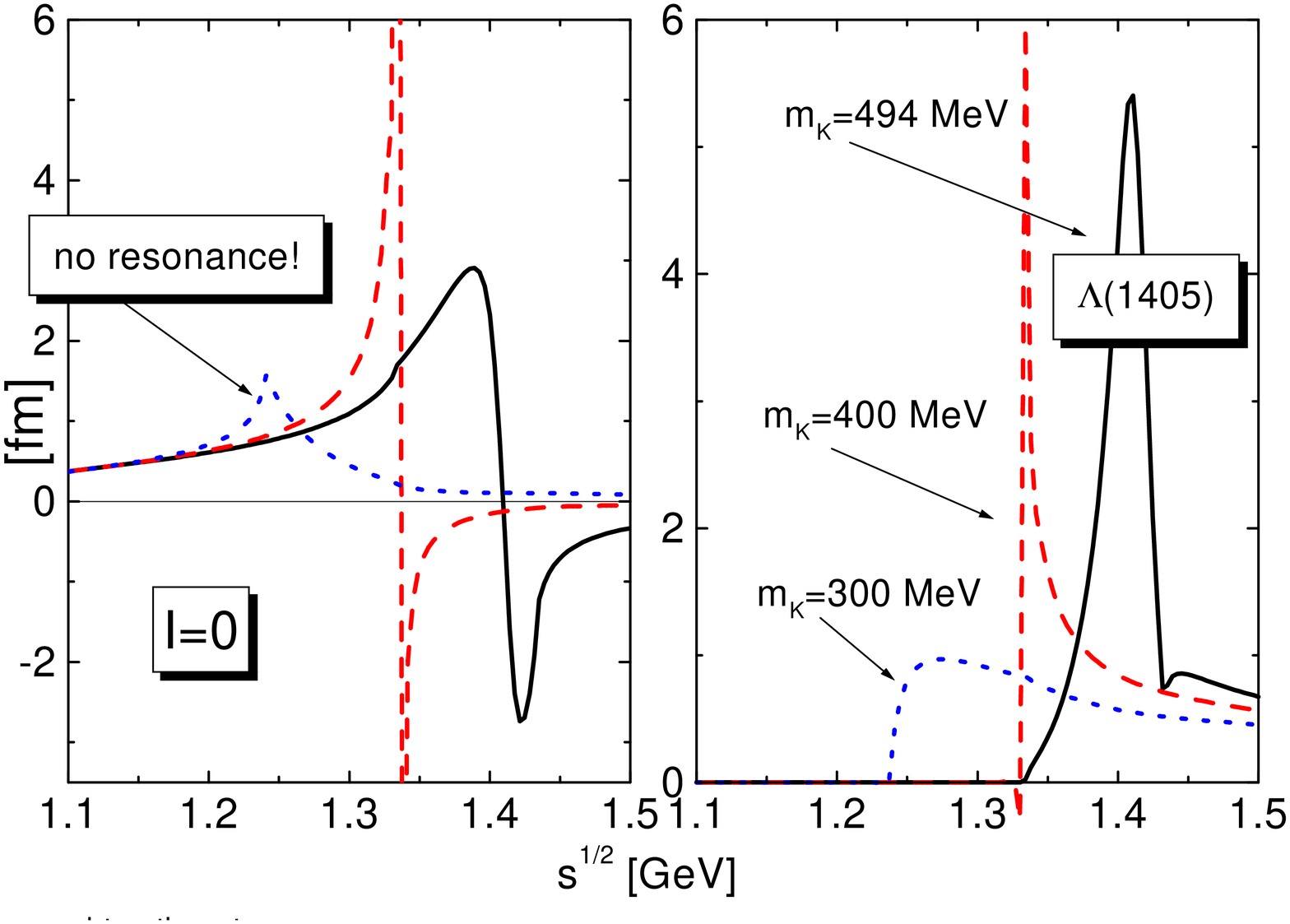,width=13cm}}}
\end{picture}
\caption{Real (left) and imaginary (right) 
part of the isospin zero s-wave $K^-$-nucleon 
scattering amplitude from the iterated Weinberg-Tomozawa interaction 
in a coupled channel calculation. The figure is taken from 
\protect\cite{lutz02b}. 
}
\label{kscatt_fig1}
\end{figure}

Hence the strategy is different to the $\pi N$ sector where 
a perturbative expansion of the $SU(2)$ chiral Lagrangian has been 
demonstrated to be very successful \cite{bernard97}. In the $SU(3)_V$ sector 
the interaction kernel $V$ rather than directly the scattering amplitude 
has to be expanded. This kernel is then iterated to all orders in the 
Bethe-Salpeter equation. The leading order in the expansion of the 
$KN$ interaction is current algebra, i.e. the Weinberg-Tomozawa 
term (\ref{wt1}) which is of chiral order $Q$. Fig. \ref{kscatt_fig1} 
shows the real and imaginary part of the isospin zero $s$-wave $K^-$-nucleon 
scattering amplitude from the iterated Weinberg-Tomozawa interaction 
in a coupled channel calculation from Ref. \cite{lutz02b}. It is nicely 
demonstrated that using the physical kaon mass the $\Lambda (1405)$  is 
dynamically generated as a pole in the $K^-$-proton scattering amplitude. 
A decrease of the $K^-$ mass leads to a disappearance of the $\Lambda (1405)$ 
which will be crucial for the discussion of in-medium effects. 

Expanding $V$ beyond current algebra the corresponding coefficients have 
to be fixed by $KN$ scattering data. Coupled channel calculations 
for $s$-wave scattering with the  interaction kernel truncated at 
chiral order $Q^2$ were first 
carried out by Kaiser et al. \cite{kaiser95}. 
$p$-wave contributions have been taken into account at the one-loop 
level by Kolomeitsev et al. \cite{kolom95} and in coupled channel 
calculations by Lutz et al. 
\cite{lutz02b}.  Higher partial waves have been taken into account in 
the G-Matrix calculations of Ramos and Oset \cite{oset01} and 
by Tolos et al. \cite{tolos01,tolos02}, in the latter case 
with an interaction kernel $V$ 
based on the J\"ulich meson-exchange potential \cite{juelich90}. 

After fixing the model parameters from free $NK$ scattering one is 
now able to systematically incorporate medium effects and to determine 
thus  in-medium scattering amplitudes, mass shifts and spectral 
functions. Medium modifications of the Bethe-Salpeter-equation (\ref{bsk1}) 
are the following ones:
\begin{itemize}
\item {\bf Pauli-blocking} of intermediate nucleon states: The Pauli 
principle is of course not active for hyperons and 
suppresses $NK$ excitations compared to $Y\pi$ excitations. 
\item {\bf Self-consistency:} This means a self-consistent dressing of the 
$K^-$ propagator 
\begin{equation}
 D_{K^-} (k) \mapsto  D_{K^-}^* (k)= \frac{1}{k^2 - \mks - \Pi_{K^-}}
\label{kmprop}
\end{equation}
by the in-medium kaon self-energy $\Pi_{K^-}$. Since $K^-$ mesons 
receive a substantial width in the medium $\Pi_{K^-}$ is generally complex 
\begin{eqnarray}
\Re \Pi_{K^-} (k) = 2E({\bf k})\Re U_{\rm opt}(E, {\bf k})~~ ,~~ 
\Im \Pi_{K^-} (k) = -2E({\bf k}) \Gamma_{K^-}(E, {\bf k})
\label{uopt3}
\end{eqnarray}
\item {\bf Dressing of the nucleon propagator:} At finite nuclear 
density the nucleon propagator is dressed by the nucleon self-energy 
$\Sigma_N$ due to the interaction with the surrounding nucleons 
\begin{equation}
 G_{N}^* (p) = \frac{1}{\not{p} - m_N + \Sigma_N + i\epsilon}~~.
\end{equation}
Nucleons are still good quasi-particles and thus 
$\Sigma_N = \Sigma_S  + \gamma_\mu \Sigma^{\mu}_V $ is real. Scalar 
and vector contributions of $\Sigma_N$ can e.g. be taken from the 
Walecka model of nuclear matter \cite{serot88}. The same holds for 
the other baryons of the baryon octet (\ref{octet1}) where self-energy 
contributions can e.g. be derived from simple counting of non-strange quarks, 
e.g. $\Sigma_\Lambda = 2/3 \Sigma_N$.
\item {\bf Dressing of the pion propagator:} Analogous to the kaons 
the intermediate pion propagator  $D_{\pi}  \mapsto  D_{\pi}^*$ is 
dressed by a pion self-energy $\Pi_{\pi}$ due to $\Delta$-hole or 
$N$-hole excitations in the nuclear medium. 
\end{itemize}
The effect of Pauli blocking was first pointed out by Koch \cite{koch94} 
and later on studied in detail by Waas, Kaiser, Rho 
and Weise \cite{waas96b,waas97}. 
Pauli blocking effects were found to play a dominant role since the 
attractive $K^-N$ interaction is reduced at finite densities. This   
acts effectively as a repulsive force which shifts the $\Lambda (1405)$ 
resonance above the $K^- p$ threshold and leads to a dissolution of this 
resonance at densities above $2-3 \rho_0$. Since the existence of the 
$\Lambda (1405)$ was, on the other hand, the origin of the repulsive 
$K^- N$ scattering length at threshold, a shift or a dissolution of this 
resonance causes an in-medium $K^-$ potential which is now close 
to the tree-level result predicted by the attractive Weinberg-Tomozawa 
term.  However, self-consistency, i.e. the dressing of the $K^-$ propagator 
by the attractive potential counteracts the Pauli effect. As pointed out by 
Lutz \cite{lutz98} a decreasing $K^-$ mass results in a negative shift of 
the $\Lambda (1405)$ , regarded as a bound $K^- p$ state, and compensates 
the positive Pauli shift to large extent. The position of the 
  $\Lambda (1405)$ pole stays now fairly constant but the resonance 
is still substantially broadened and dissolves at high densities (as can be 
seen from the schematic calculation shown in Fig. \ref{kscatt_fig1}.) 
The influence of dressing of nucleon and hyperon propagator 
due to short-range $NN$ and $NY$ correlations has been 
investigated in \cite{kolom03}. 
\subsubsection{In-medium potentials}
The dressing of nucleon and hyperon propagators is generally included 
in such type of calculations but the effects are of minor importance. 
A strong influence has, however, a dressing of the pion propagator by 
an attractive pion potential which arises due to nucleon- and $\Delta$-hole 
excitations in the medium.  As shown by Ramos and Oset \cite{oset01} 
and also found by Tolos et al. \cite{tolos01} the thresholds of the 
$\pi \Lambda$ and $\pi \Sigma$ channels are lowered resulting in less 
attraction for the in-medium $K^-$ potential. 

\begin{figure}[h]
\unitlength1cm
\begin{picture}(9.,7.5)
\put(3.5,0){\makebox{\epsfig{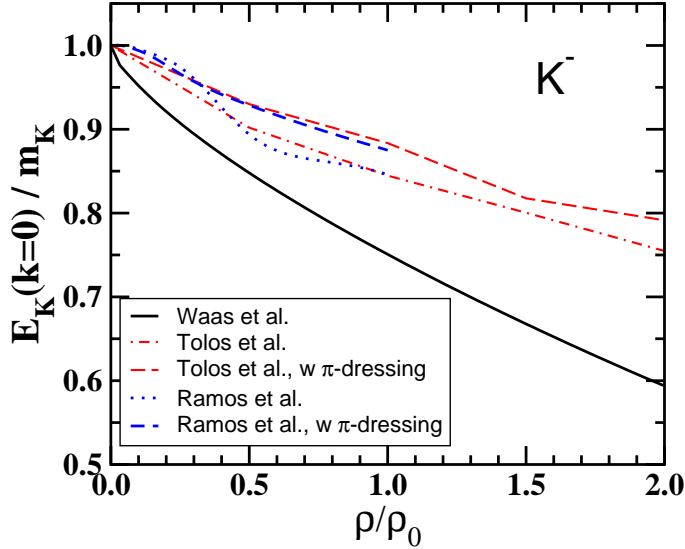}}}
\end{picture}
\caption{In-medium $K^-$ energy obtained in coupled channel 
calculations which includes Pauli-blocking 
(Waas et al., \protect\cite{waas96b}), kaon dressing 
(Ramos et al., \protect\cite{oset01}) and pion dressing 
(Ramos et al., \protect\cite{oset01}, Tolos et al. \protect\cite{tolos01}).  
}
\label{kpot_fig1}
\end{figure}

Fig. \ref{kpot_fig1} shows the single particle energy or 'in-medium mass
shift' $E({\bf k}=0)= \mk + \Re U_{\rm opt} (E,{\bf k}=0)$ 
for antikaons obtained in various coupled channel calculations. This 
quantity can be compared to the mean field picture although such a 
comparison has to be taken with care. At finite densities the antikaons 
aquire a substantial in-medium width and do no more behave like good 
quasi-particles, as assumed in a mean field picture. In 
particular at low momenta the spectral functions can be of complex structure 
without a well defined quasi-particle pole which makes the interpretation 
of the in-medium self-energy $\Pi_K$ in terms of on-shell potentials 
questionable. However, transport models are usually formulated in terms of 
quasi-particles. Hence, we do not want to refrain from this 
comparison. The microscopic coupled channel calculations 
deliver an attractive in-medium potential which is significantly smaller 
than in the mean field approaches, in particular when a self-consistent 
dressing of the kaon propagator 
is taken into account, and even smaller when pion dressing is included. 
\begin{figure}[h]
\unitlength1cm
\begin{picture}(9.,7.0)
\put(3.5,0){\makebox{\epsfig{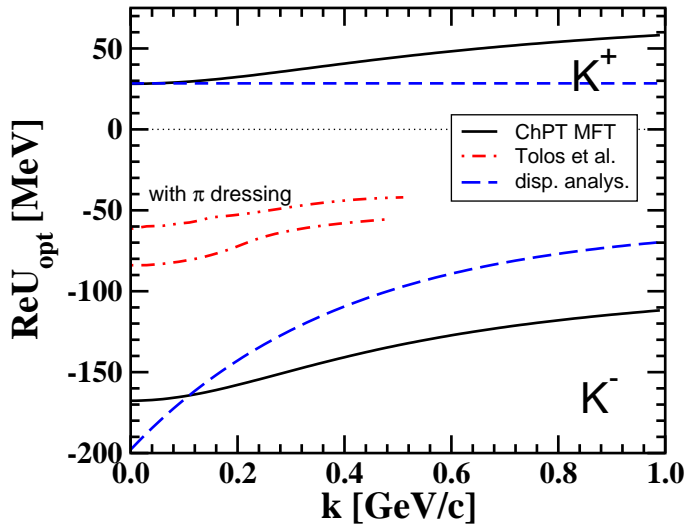}}}
\end{picture}
\caption{Optical kaon potentials at nuclear saturation density. 
Results from the chiral mean field approach (ChPT MFT), from 
coupled channel calculations (Tolos et al. \protect\cite{tolos01}) 
and from the dispersion analysis of \protect\cite{sibirtsev99}) 
are compared.
}
\label{kpot_fig2}
\end{figure}
This fact is also reflected in the optical potential (real part) 
shown in Fig. \ref{kpot_fig2}. We compare the momentum dependence of 
$\Re U_{\rm opt}$  at saturation density obtained in various 
approaches. Results are taken 
from the chiral mean field approach \cite{brown96b}, 
denoted in Fig. \ref{kmass1_fig} as MFT ChPT+corr., the coupled channel 
calculations of Tolos et al. \cite{tolos01}, with and without pion dressing, 
and a dispersion analysis of $K^+N$ and $K^-N$ scattering amplitudes  
by Sibirtsev and Cassing \cite{sibirtsev99}. 
Since the definition used to extract the optical potential from the 
self-energy $\Pi$ varies in the literature, this comparison is based 
on relation (\ref{uopt3}) which has been used in  \cite{tolos01} 
instead of Eq. (\ref{mass2}). For $K^+$ the magnitudes of the potential are 
consistent, i.e. the dispersion analysis agrees with the 
mean field approach at zero momentum. It predicts 
an almost momentum independent potential while $U_{\rm opt}$ 
is slightly rising as a function of momentum 
in mean field models. For  $K^-$ all models predict 
a considerably reduced attraction at high kaon momenta, however, 
the potential depths strongly deviate. The self-consistent coupled 
channel calculations from Schaffner et al. \cite{schaffner00} 
predict an even smaller potential which is of the size of -32 MeV 
at saturation density. 

The dispersion analysis of 
Ref. \cite{sibirtsev99} comes close to the mean field result which is, 
however, not astonishing since the authors disregarded the repulsive 
contributions from the $\Sigma (1385)$ and $\Lambda (1405)$ resonances 
according to the argument that these resonances should dissolve at finite 
density. They found their parameterization of the  $K^-$ potential 
consistent with data from $p+A$ reactions \cite{sibirtsev99}.

All the microscopic approaches predict $K^-$ potentials of only 
moderate attraction and are thus in stark contrast to the mean 
field picture and the standard analysis of kaonic atoms \cite{gal94}.  
The latter suggests a strongly attractive on-shell $K^-$ potential of 
about 200 MeV at $\rho_0$. It is not yet fully clear if the 
microscopic approaches which comply with
kaon-nucleon scattering data, can explain kaonic atoms as well. However, 
there are indications that kaonic atoms explore the antikaon 
potential only at the nuclear surface \cite{baca00,lutz00b} 
and weak $K^-$ potentials describe the available data as well 
\cite{ramos01}. A final answer 
would require to account for the full off-shell behavior of self-energy
and spectral properties of a bound $K^-$ state. However, such calculations 
have not yet been performed.

\begin{figure}[h]
\begin{minipage}[h]{185mm}
\unitlength1cm
\begin{picture}(18.5,7.3)
\put(0.,0.){\makebox{\epsfig{file=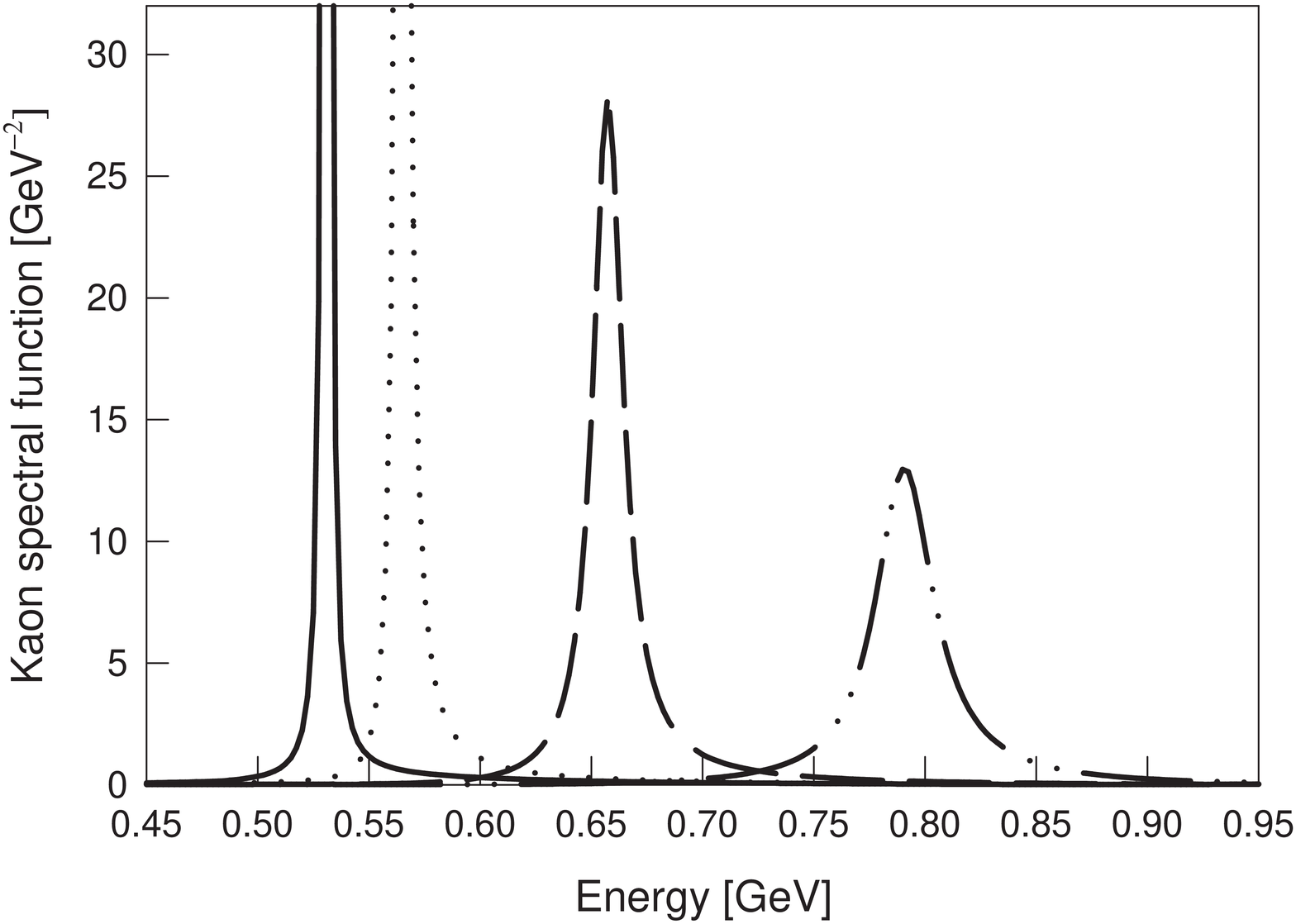,width=9.5cm}}}
\put(9.0,0.3){\makebox{\epsfig{file=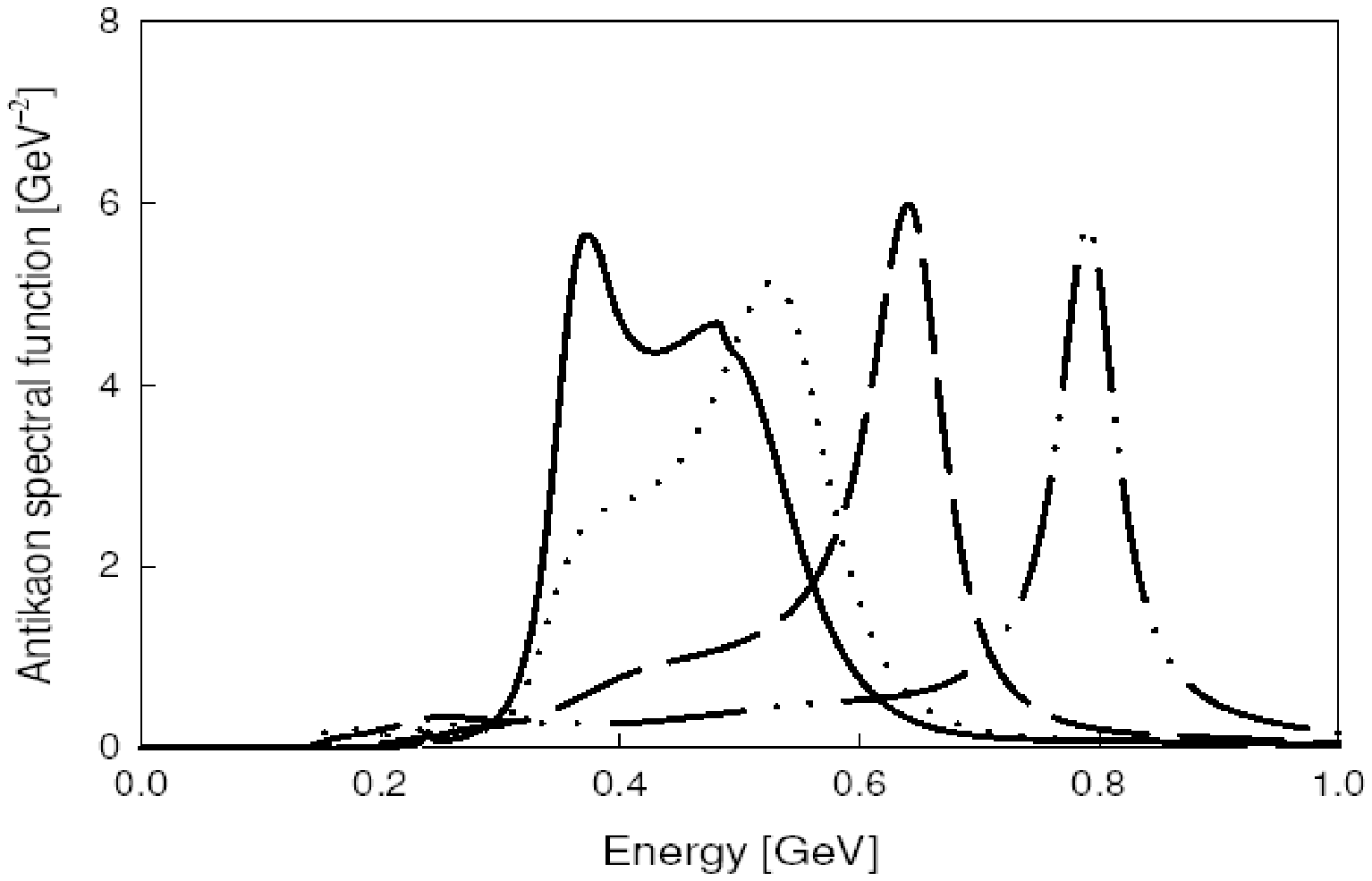,width=9.3cm}}}
\end{picture}
\end{minipage}
\caption{In-medium kaon (left) and antikaon (right) spectral 
functions from coupled channel calculations of \protect\cite{lutz04b} 
at saturation density. Results are shown for different kaon momenta:
$p_K=0$ (solid line), $200\,$MeV (dotted line), $400\,$MeV (dashed line),
and $600\,$MeV (dash-dot-dot line). 
}
\label{spectral_fig}
\end{figure}
 Figure \ref{spectral_fig} demonstrates finally the validity of the 
quasi-particle picture. It shows the kaon and antikaon 
spectral functions at saturation density obtained from coupled 
channel calculations from Ref. \cite{lutz04b} (including pion dressing) 
for different momenta. The kaons have still a clear quasi-particle peak 
which in the medium acquires a finite width. The latter is, however, 
quite small (less than $5\,$MeV) for small momenta and increases 
up to $15\,$MeV at a momentum above $400\,$MeV which is still 
moderate. Hence the  quasi-particle picture and the mean field 
approximation are well justified. As already stressed several times, the 
situation for $K^-$ mesons is quite different. In particular at low 
momenta the spectral functions are broad and of complex structure. 
At larger momenta a quasi-particle peak may still be visible but also 
here substantial strength is shifted to lower momenta. Thus the mean 
field approximation is questionable for the antikaons.

\subsubsection{In-medium cross sections}
Coupled channel calculations may predict sizeable in-medium 
modifications of the pion-induced $K^-$ production cross sections and 
the corresponding absorption cross sections $\pi Y \longleftrightarrow N K^-$. 
The fact that the s-wave $\Lambda(1405)$ resonance lies only 27 MeV 
below the $K^- p$ threshold implies a strong coupling to this state
and requires a non-perturbative treatment. The melting of the 
$\Lambda(1405)$ and $\Sigma (1385)$ bound states due to Pauli 
blocking of the intermediate states in the BS-equation (\ref{bsk1}) leads to a
dramatic increase in particular of the  $\pi \Sigma \longrightarrow N K^-$ 
cross section at threshold. In \cite{schaffner00} the enhancement 
factor was found to be more than one order of magnitude at $\rho_0$.  
However, self-consistency shifts the $K^-$ mass below threshold and 
decreases the available phase space which counteracts the enhancement due 
to a melting $\Lambda(1405)$. In the calculations of Schaffner et al. 
\cite{schaffner00} the  $\pi \Sigma \longrightarrow N K^-$ is then only 
enhanced by a factor of two and the $\pi \Lambda \longrightarrow N K^-$ 
is hardly affected at all. In the self-consistent calculations 
of Lutz and Korpa \cite{lutz02} the predicted in-medium modifications of 
these cross sections are practically opposite. They account in addition 
for the full in-medium modifications of the $ K^-$ spectral 
distributions and obtain a strong 
enhancement of the $\pi \Lambda \longrightarrow N K^-$ cross section 
due to the coupling to the  $\Sigma (1385)$ but almost no changes for the 
$\pi \Sigma \longrightarrow N K^-$ channel. The G-matrix calculations of 
Tolos et al. \cite{tolos01} came to opposite conclusions, namely an almost 
complete suppression of the pion induced reactions in the nuclear
environment. Such strong modifications of the $K^-$ production cross 
sections and the corresponding absorption cross sections would have severe 
consequences for the  $K^-$ dynamics in heavy ion reactions.
\subsection{Kaons in pion matter}
In heavy ion reactions at intermediate energies the matter is baryon 
dominated, at ultra-relativistic energies at CERN-SPS or at RHIC the 
matter is, however, pion dominated. For completeness it is thus instructive 
to consider this case as well. The problem of medium modifications 
experienced by kaons in a hot pion gas has been dressed in the early 90ties 
in Refs. \cite{shuryak91} and \cite{blaizot91} where the authors came, 
however, to opposite conclusions concerning the kaon mass shifts. Recently 
this problem has been picked up by Martemyanov et al. \cite{boris04}. 
In Ref.  \cite{boris04} the kaon self-energy has been determined in a 
model independent way to leading order in pion density, 
based on ChPT at low temperatures and experimental 
phase shifts at high temperatures. 

Analogous to nuclear matter (\ref{lowdensity}) the kaon self-energy 
$\Pi_K (k^{2},E)$ can be  expressed in terms of the $\pi K$ forward 
scattering amplitudes for on-shell
pions and off-shell kaons. The necessary on-shell $\pi K$ amplitudes 
have been evaluated in ChPT to order $p^{4}$ 
by several authors (see e.g. \cite{pik} and references therein). 
Near the threshold, the isospin-even $(+)$ and odd $(-)$ 
$\pi K$ scattering amplitudes can be expressed in terms of scattering
lengths and effective ranges $a_{\ell }^{(\pm) }$ and $b_{\ell }^{(\pm) }$ 
\beq
A^{(\pm )}(s,t,k^2) = 8\pi \sqrt{s}\left[ a_{0}^{(\pm) }
+p^{*2}(b_{0}^{(\pm)}+3a_{1}^{(\pm) })
+\frac{3}{2}ta_{1}^{(\pm) }\right] 
+ c^{(\pm)}(k^{2}-m_{\rm K}^{2}) 
\eeq
with $p^{*}=p^{*}(\sqrt{s},m_{\pi },m_{\rm K})$ the
c.m. momentum of the $\pi K$ system \cite{boris04}. The $K$-meson self-energy is then obtained by 
integration over the corresponding pionic Bose distributions 
\beq
d\rho_{\pi }=\frac{d^{3}k_{\pi }}{(2\pi )^{3}}\left( {\rm exp}(\frac{E_{\pi
}-\mu _{\pi }}{T})-1\right) ^{-1} 
\eeq
with the scalar pion density $d\rho_{s\pi }=d\rho_{\pi }/(2E_{\pi })$  
\beq
-\Pi_K (k^{2},E) =\int A^{(+)}(s,0,k^2)
(d\rho_{s\pi ^{+}}+d\rho_{s\pi ^{0}}+d\rho_{s\pi
^{-}}) +\int A^{(-)}(s,0,k^2)(-d\rho_{s\pi ^{+}}+d\rho_{s\pi ^{-}})~~.  
\label{SE}
\eeq
To lowest order ChPT isospin symmetric pion matter does not change the
kaon dispersion law. The leading order effect appears at the 
one loop level. The mean field, i.e. the scalar mass shift 
$\delta m_{\rm K}$ and 
the vector potential $V_K$ follow from the self-energy at the on-shell 
point which can be expressed in terms of $s$- and $p$-wave 
scattering lengths and $s$-wave effective ranges:
\begin{equation}
\Pi_K (\mks,m_{\rm K})= 
-4\pi \rho_{\pi}\frac{m_{\pi }+m_{\rm K}}{m_{\pi }}a_{0}^{(+)}
~~~.
\label{sigm}
\end{equation}
With $\delta m_{\rm K}+V_{K}=\Pi_K (\mks,m_{\rm K})/2m_{\rm K}$. From 
(\ref{sigm}) follows also the vector potential 
\begin{equation}
V_{K}=-\frac{2\pi \rho_\pi}{m_{\pi }+m_{\rm K}}
\left[ a_{0}^{(+)}+2m_{\pi} m_{\rm K}(b_{0}^{(+)}+3a_{1}^{(+)})\right]~~ .  
\label{VD}
\end{equation}
Since ChPT is only valid at temperatures well below the pion mass 
in Ref. \cite{boris04} the high temperature behavior has been  based on 
a more phenomenological approach which parameterizes the experimental 
phase shifts and matches smoothly with the one-loop ChPT 
low-temperature limit. 
\begin{figure}[h]
\unitlength1cm
\begin{picture}(9.,7.5)
\put(3.5,0){\makebox{\epsfig{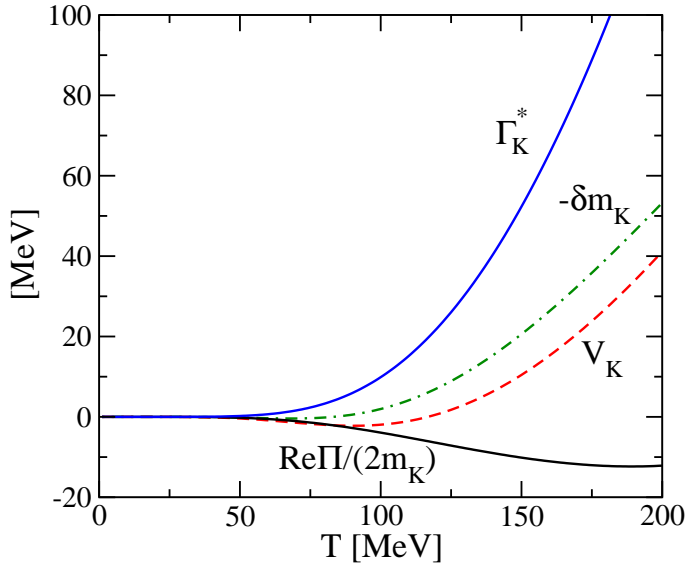}}}
\end{picture}
\caption{Medium modifications of kaons in an isospin-symmetric hot pion 
gas: self-energy $ \Re \Pi_K/(2m_{\rm K})$, 
mass shift $- \delta \mk $, vector potential $V_K$, and kaon collision width 
$\Gamma^{*}_K$ versus temperature $T$. 
Results are taken from \protect\cite{boris04}. 
}
\label{kpionmatter_fig}
\end{figure}
The corresponding kaon self-energy at threshold, the mass shift 
and the vector potential are shown in Fig. \ref{kpionmatter_fig} 
as a function of temperature. At $T=170$ MeV one obtains 
a negative mass shift $\delta M_{K}=-33$ MeV and a repulsive 
vector potential of $V_{K}=21$ MeV. There exists a remarkable 
analogy to the nuclear matter case: the kaon mass shift at 
high temperatures is large and negative, the vector potential 
is large and positive, 
their sum is relatively small and negative. Kaons are therefore 
bound in pion matter
similar to nucleons or antikaons in nuclear matter. The vector 
potential is, however, $C$-even, distinct from the case of 
nuclear matter. In addition both, kaons and antikaons aquire 
a substantial in medium width $\Gamma^{*}_K$ at finite temperature 
\cite{boris04}. 

In-medium modifications of the kaons have direct implications on the 
$\phi$ meson properties. At ultra-relativistic energies 
an inconsistency has been observed between the  $\phi $ 
yields measured through the dilepton and  $K\bar{K}$ channels. 
In Pb+Pb collisions at CERN/SPS energies ($E_{\rm lab} =158$ AGeV)
the $\mu ^{+}\mu ^{-}$ yield from $\phi $ decays was measured 
by the NA50 Collaboration \cite{NA50} and the $K\bar{K}$
channel by the NA49 \cite{NA49}. The leptonic channel was 
found to be enhanced by a factor of two to four compared to the mesonic 
channel. A similar observation has been made in Au+Au 
collisions at RHIC ($\sqrt{s_{NN}}=200$ GeV) by 
PHENIX \cite{phenix1,phenix2} which measured simultaneously the $\phi \rightarrow
e^{+}e^{-} $ and $K\bar{K}$ channels. 
At RHIC energies this apparent enhacement of the leptonic channel 
can qualitatively be understood by an enhanced in-medium $\phi$ width 
$\Gamma _{\phi }^{{\rm med}}\sim 2\div 3 \Gamma _{\phi }^{{\rm vac}}$ 
and the final state interaction of the corresponding kaons inside 
the fireball \cite{boris04,phiko02}. 

\subsection{Chiral symmetry restoration?}
At the end of this chapter I want to shortly address the question of 
chiral symmetry restoration. The spontaneously broken chiral symmetry 
of QCD manifests itself in the large expectation value of the scalar 
quark condensate $ \langle {\bar q}q \rangle$ of the QCD vacuum while 
the small but finite current quark masses are responsible for the explicit 
chiral symmetry breaking of QCD. In the chiral limit 
of vanishing current quark masses, the Goldstone bosons of spontaneous 
chiral symmetry breaking, i.e. the pion, respectively the full 
pseudo-scalar obtect, become massless. This fact is guaranteed by the 
Gell-Mann-Oakes-Renner relation. However, through the GOR relation   
the pseudoscalar meson masses are also directly proportional to 
the scalar condensates. The GOR relation is of leading order in the 
quark masses. By in-medium  chiral symmetry restoration one understands 
now the fact that dropping pseudoscalar meson masses are 
caused by a  reduction of the condensate at finite nuclear density 
and/or temperature. The meson masses approach the chiral limit 
although the explicit symmetry breaking is still valid.  
\begin{figure}[h]
\unitlength1cm
\begin{picture}(13.0,10.5)
\put(1.5,0){\makebox{\epsfig{file=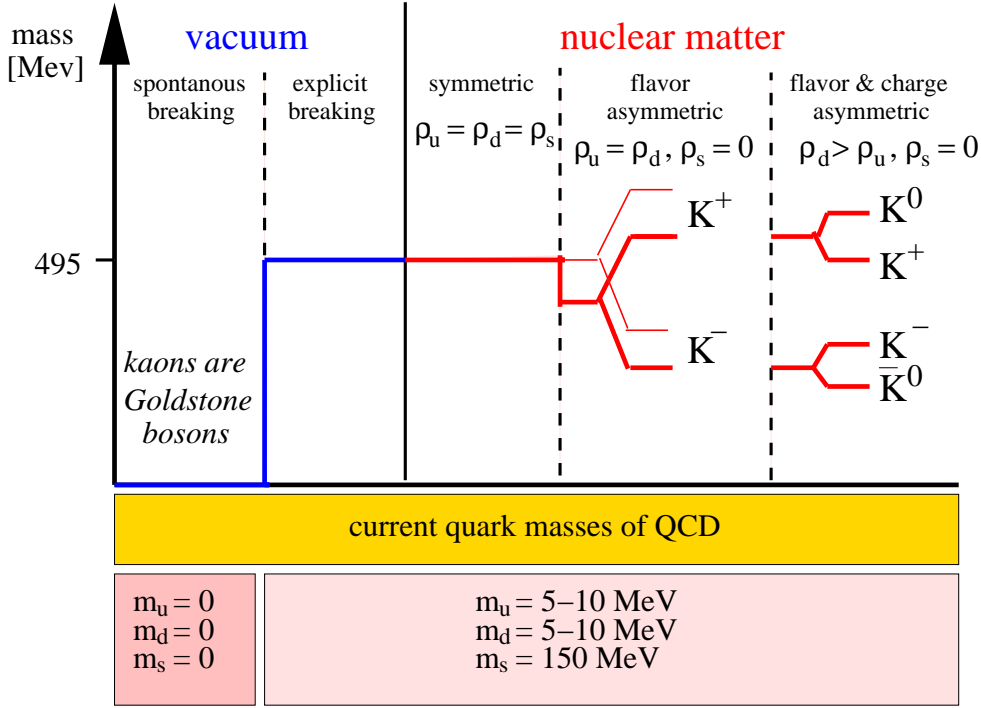,width=13cm}}}
\end{picture}
\caption{Schematic representation of kaon and antikaon energy shifts in 
dense nuclear matter due to SU(3) symmetry breaking 
($\rho_u = \rho_d, \rho_s=0$). The Weinberg-Tomozawa term leads to 
the $K^+/K^-$ mass splitting (thin line) while the Kaplan-Nelson 
term is responsible for the common mass shift (thick line).   
}
\label{symm_fig}
\end{figure}

How such a scenario is connected with the medium properties 
obtained from chiral dynamics is, however, a non-trivial question. 
The leading term 
in the chiral pseudoscalar meson-nucleon interaction is the 
Weinberg-Tomozawa term. At finite nuclear density it is responsible 
for the splitting of the energy levels between the 
degenerate flavor eigenstates 
$K^+ (u{\bar s})$ and $ K^- ({\bar u}s)$ due to SU(3) flavor 
symmetry breaking. As indicated in Fig. \ref{symm_fig}, 
SU(3) flavor symmetry is broken by the 
non-vanishing up and down quark densities $\rho_{u/d}$ while the 
strange quark density  $\rho_{s}$ is still zero. 
Isospin symmetric pion matter, in contrast, is flavor symmetric 
and there occurs no mass splitting between  kaons and antikaons.  
Charge symmetry breaking which occurs in isospin asymmetric nuclear 
matter leads to an additional mass splitting of the different 
isospin states  $K^+$ and $K^0 (d{\bar s})$, respectively 
 $K^-$ and ${\bar K}^0 ({\bar d}s)$. 

In chiral coupled channel dynamics the leading 
order Weinberg-Tomozawa kaon-nucleon interaction 
is iterated to infinite order. The kaons obtain medium modifications, 
i.e. mass shifts and changes of their spectral distributions. However, 
in this framework their origin can not easily be traced back to a 
restoration of chiral symmetry. Such an interpretation is to some extent 
possible 
when the Kaplan-Nelson term in the KN-interaction is taken into account. 
The    Kaplan-Nelson term is of order $ \langle {\bar q}q \rangle$ 
and therefore directly related to the in-medium condensates 
via the GOR relation. In 
the mean field picture the two effects, mass splitting 
(Weinberg-Tomozawa) and dropping mass (Kaplan-Nelson) are clearly 
separated and the latter one can be interpreted in terms of a 
partial chiral symmetry restoration. Fig. \ref{symm_fig} 
illustrates this scenario schematically. 

In coupled channel 
chiral dynamics it is no more straightforward to disentangle {\it ordinary} 
hadronic many-body effects from a change of the QCD vacuum.  
The principle connection of hadronic many-body effects with basic 
QCD quantities such as the chiral condensate is 
provided by the Hellman-Feynman 
theorem which relates $ \langle {\bar q}q \rangle$ with the derivative 
of the QCD-Hamiltonian with respect to the current quark masses. 
In nuclear matter the condensate is in the same way obtained from the 
total energy density ${\cal E}$
\beq
 \langle \rho| {\bar q}q |\rho \rangle = \langle {\bar q}q \rangle + 
\frac{1}{2} \frac{d {\cal E}}{dm_q}~~~.
\label{hell}
\eeq
Eq. (\ref{hell}) implies that many-body correlations, albeit based on 
hadronic degrees of freedom, provide contributions to the in-medium quark 
condensate. The practical use of this type of quark-hadron duality is, 
however, rather limited. While there exist sophisticated models for 
the treatment of hadronic many-body correlations, such as e.g. 
Brueckner theory, one would have in addition to know the dependence of 
these correlations on the current quark masses. For pionic correlations,  
lets call them ${\cal C}_\pi$,  such a procedure may in principle be 
possible making use of the GOR relation, i.e. 
$\partial  {\cal C}_\pi /\partial m_q = 
\partial  {\cal C}_\pi / \partial m_\pi m_\pi/2m_q$. The derivative 
of the nucleon mass is given by the pion-nucleon $\Sigma$ term 
$ d m_N/ d  m_q = \Sigma_{\pi N}/m_q$ and analogous relations exist 
for the strange quark. However, these derivatives are only valid to 
leading order in ChPT which restricts their applications to the 
mean field level. Attempts to proceed in such a  direction have been 
performed, e.g., by Cohen et al. \cite{cohen92} and more 
recently by Chanfray and Ericson \cite{chanfray04}.

\section{Phenomenology of strangeness production}
\subsection{Elementary reaction channels}
We start with a short discussion of the various reaction
channels which determine the kaon and antikaon production at
intermediate energies, the role these channels play in the
complex reaction dynamics and their experimental and theoretical
knowledge. The various processes are summarized in
Tab. \ref{tab_cross1}. 
\begin{figure}[h]
\unitlength1cm
\begin{picture}(6.,4.5)
\put(5.5,0){\makebox{\epsfig{file=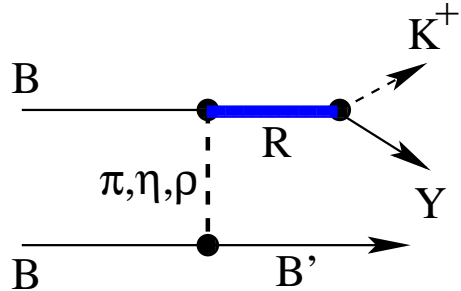,width=6.0cm}}}
\end{picture}
\caption{Diagrammatic representation of the reaction 
$BB \longrightarrow BYK^+$ within the one-boson-exchange + resonance 
model. 
}
\label{fig_resmodel}
\end{figure}
They can be classified in primary and secondary
strangeness production reactions, strangeness and charge exchange
reactions and strangeness absorption reactions. Strangeness production
can be subdivided into two classes, namely baryon induced 
($BB \longrightarrow BYK^+$, $BB \longrightarrow BBK^+K^-$) and pion 
induced ($\pi B \longrightarrow YK^+$,  $\pi B \longrightarrow BK^+K^-$) 
reactions. $B$ stands here for a baryon which can be either 
a nucleon or a nucleon resonance 
($N,\Delta , N^*$) and $Y$ for hyperons ($\Lambda ,
\Sigma$). The first class falls into two subclasses: primary reactions
with two nucleons in the entrance channel and secondary reactions with
at least one nucleon resonance in the entrance channel. Meson induced
reactions are by definition of secondary type. Both involve at least a
two step process with first the production of the intermediate 
resonance or meson. Primary
reactions play the dominant role for kaon and antikaon production at
higher energies and contribute dominantly to the high momentum part
of the spectra. However, at subthreshold energies the production rates
are dominated by the secondary type reaction mechanism. The same
feature appears already in proton-nucleus collisions
\cite{cassing94,paryev99,sibirtsev99,rudy02}. 
Other processes like the production of
multi-strange baryons ($\Xi, \Omega$) or hidden strangeness ($\phi$) 
are subdominant and will be discussed separately. 
\begin{table}
\begin{center}
\begin{tabular}{|c|c|c|}
\hline
 $ K^+$  & $K^-$  & type\\ 
\hline\hline
$BB \longrightarrow BYK^+$ & $BB \longrightarrow BBK^+K^-$ &
strangeness production\\
 &   & primary/secondary \\
\hline
$\pi B \longrightarrow YK^+$ & $\pi B \longrightarrow BK^+K^-$ &
strangeness production\\
 &   & secondary \\
\hline
 & $B Y \longrightarrow BB K^-$  & strangeness exchange\\
\hline
 & $\pi Y \longleftrightarrow B K^-$  & strangeness exchange\\
\hline
$YK^+ \longrightarrow \pi B$  &   & strangeness absorption\\
\hline
$BK^+ \longleftrightarrow B K^{+/0}$ & 
$BK^- \longleftrightarrow B K^- (\overline{K}^{0})$  & elastic/charge exchange\\
\hline
\end{tabular}
\end{center}
\caption{\label{tab_cross1}
Elementary hadronic reactions which are relevant for kaon dynamics 
at intermediate energies. $B$ stands for nucleons or nucleon 
resonances ($N,\Delta , N^*$) and $Y$ for hyperons ($\Lambda ,
\Sigma$). 
}
\end{table}

{\bf $BB \longrightarrow BYK^+$:}\\
The reaction $pp\longrightarrow p\Lambda K^+$ sets the threshold for 
$K^+$ production in free space ($T_p =1.58$ GeV). Historically this
cross section was overestimated for a long time \cite{ran80}, in
particular close to threshold. The situation improved when the COSY-11 
Collaboration delivered data for $pp\longrightarrow p\Lambda K^+$ close
to threshold \cite{cosy11}. An open question is still the isospin 
dependence of this channel. Isotopic relations predict 
$(pn\longrightarrow n\Lambda K^+) = 2(pp\longrightarrow p\Lambda K^+)$
while recent proton-deuteron data indicate that the $pn$ cross section
might be even larger \cite{anke04}. Already rather early it was
noticed that secondary reactions $(N\Delta \longrightarrow N Y K^+)$, 
where the $\Delta$ resonance plays the role of an energy storage,   
contribute essentially to the subthreshold kaon yield in HIC's 
\cite{xiong90,huang93,bali94,hartnack94,giessen94,brat97} while 
multi-step processes of higher order 
with two $\Delta$ resonances in the entrance channel are
suppressed \cite{bali94,ko96,hartnack01}. Reactions with a $\Delta$ in the exit
channel are energetically suppressed. Since no experimental data exist 
for cross sections involving nucleon resonances one has here to rely on
model calculations. Commonly used parameterizations are based on 
one-boson-exchange models \cite{sibirtsev95,li98} or on more
microscopic one-boson-exchange + resonance models 
\cite{wilkin97,tsushima97,sibirtsev98,tsushima99}. Both models are 
usually applied at the tree level. The resonance model assumes 
thereby that the heavy meson production process runs over the excitation 
of an intermediate nucleon resonance $R=N^*,\Delta$ as depicted in
Fig. \ref{fig_resmodel}. The model parameters, i.e. coupling constants 
can be fixed from data on $\pi N \longrightarrow YK^+$ reactions. As 
already shown in \cite{wilkin97} the resonant s-channel is dominant in 
$pp \longrightarrow p\Lambda K^+$, non-resonant t-channel
contributions can be parameterized through $K^*$ exchange. 
This type of resonance models are popular 
and generally quite successful in the description of 
heavy meson production ($\rho, \omega,
\phi$) close to threshold 
\cite{pirner97,post01,titov01,mosel03,omega03,phi03}.

{\bf $\pi B \longrightarrow YK^+$:}\\
The importance of this channel for HIC's was first pointed out by Fuchs
et al. \cite{fuchs97b} and confirmed by the Giessen group 
\cite{brat97}. Depending
on energy and system size the pion-induced reactions 
$\pi N \longrightarrow YK^+,\pi \Delta \longrightarrow YK^+$ can
contribute up to about 50\% of the total kaon yield. Naturally the
contribution is largest in heavy systems. The corresponding cross
sections are relatively well constrained by pion-proton scattering
data. The parameterizations
of Tsushima et al. \cite{tuebingen} obtained within an 
one-boson-exchange + resonance 
model are presently used in most transport calculations.

{\bf $YK^+ \longrightarrow \pi B $:}\\
At subthreshold energies strangeness absorption plays no significant 
role. That the probability for such an event is, however, not
complete vanishing has been demonstrated in \cite{pal01}. Since
strangeness absorption would be the only possibility to drive kaons towards
chemical equilibrium the chemical kaon freeze-out occurs already at a
non-equilibrium stage \cite{brat00}.  

{\bf $BB \longrightarrow BBK^+K^-$:}\\
The reaction $pp \longrightarrow pp K^+K^-$ sets the threshold for 
$K^-$ production in free space ($T_p = 2.50$ GeV). 
The history of this reaction is similar to that of 
$pp \longrightarrow p \Lambda K^+$. Also here early predictions 
\cite{zwermann} overestimated this cross section in the vicinity of 
the threshold by 
about two orders of magnitude, as it turned out when COSY-11 provided 
the first data point close to threshold \cite{cosy1101}. 
Tree level OBE calculations based on pion and kaon exchange are able 
to reproduce the near threshold behavior \cite{sibirtsev97}. In contrast to 
$pp \longrightarrow p \Lambda K^+$ this reaction seems to be dominated
by t-channel exchange. Hence the resonance model which is based on 
resonant s-channel exchange has not been applied. 
Reactions which involve nuclear resonances in the initial 
and/or final state have not been investigated theoretically and are 
neglected in corresponding transport calculations.

{\bf $\pi B \longrightarrow B K^+ K^-$:}\\
The $\pi N \longrightarrow N K^+ K^-$ and $\pi N \longrightarrow N \pi
K^+ K^-$ cross sections are experimentally well
constrained. Calculations based on tree level one-boson-exchange 
have been performed 
in \cite{sibirtsev97}. Transport calculations 
\cite{cassing97a,hartnack01} 
estimate this channel of about equal importance as baryon induced 
primary $K^-$ production.

{\bf $B Y \longleftrightarrow B B K^-$:}\\
As first pointed out by Ko \cite{ko83}, 
strangeness exchange is the dominant mechanism for $K^-$ production at 
subthreshold energies. The relative importance of the two possibilities, 
through nucleon-hyperon scattering or pion absorption are not yet completely
settled. More than 20 years ago Ko \cite{ko83} 
estimated the contribution of this channel 
to about 10\%. The $N Y \longrightarrow NN K^-$ 
can in principle easily be determined in a OBE picture since the cross section
factorizes in the pion exchange and the experimentally known 
$\pi Y \longleftrightarrow N K^-$ part. The importance of this channels 
is, however, still controversial. In \cite{cassing97a} 
it was found to contribute  only a few percent to the total antikaon
yield in heavy ion reactions whereas a recent reanalysis of this
channel by Barz et al. \cite{barz03} stresses its importance (about
25\%) for $p+A$ reactions.

{\bf $\pi Y \longleftrightarrow B K^-$:}\\
This channel is experimentally well known from $K^-$-proton
scattering (for a recent partial wave analysis see \cite{hyslop}). 
The cross sections are large, e.g. 40-70 mb for $N K^-\longrightarrow
\pi Y$ depending on relative momentum. 
Strangeness exchange reactions are essential for the $K^-$ 
dynamics in heavy ion collisions. The cross sections are generally
large, i.e. of the order of mbarn, whereas the strangeness production 
is of the order of $\mu$barn. In the standard scenarios strangeness 
exchange through pion absorption is one of the major sources for the $K^-$ yield 
at subthreshold energies \cite{brat97,aichelin03}. The reason lies in the fact that the
threshold for strangeness production through hyperons in association 
with $K^+$ mesons is about 1 GeV below the direct $K^+ K^-$ production 
threshold and these strange quarks can be transfered to antikaons by 
pion absorption. This drives the $Y-K^-$ system towards chemical
equilibrium whereas for anti-strangeness no comparable process
exists. Hence, antikaons freeze out later than kaons
\cite{pal01,aichelin03}. However, due to the strong coupling 
of the  $K^-$-nucleon system to the $\Lambda (1405)$ large medium 
effects have to be expected for the $\pi \Sigma \longleftrightarrow N K^-$ 
cross section and this reaction has 
theoretically been studied in great detail 
\cite{schaffner00,lutz02,oset01,tolos01,cassing03}. The medium 
dependence of the cross sections is, however, still controversial and 
ranges from moderate enhancement close to threshold to a 
strong suppression (see discussion in chapter 5). 
This, of course would change the complete scenario 
of   $K^-$ production and absorption. 

{\bf $BK  \longleftrightarrow B K, ~ 
B\overline{K}\longleftrightarrow B\overline{K}$   :}\\
Because of strangeness conservation the kaon final state interaction
is at low energies dominated by elastic kaon-nucleon scattering. This
does in principle not affect the total yields. However, 
elastic scattering changes the
shape of the momentum spectra significantly. Rescattering makes the
spectra harder and changes the angular distributions
\cite{fang94b,wang97a,wang98c}. In systems with large isospin asymmetry charge
exchange $pK^+  \longleftrightarrow n K^0$ may also slightly affect
the final yields. The elastic antikaon cross section is large 
($\sim 20-100$ mb), in particular at low momenta, which is the result 
of the strong coupling to the $\Lambda(1405)$ resonance. It competes,
however, with $K^-$ absorption. The corresponding charge 
exchange cross section 
 $pK^-  \longleftrightarrow n \overline{K}^0$ is comparatively small 
and does not significantly influence the $K^-$ dynamics.
\subsection{Hidden strangeness production}
In addition to open strangeness production through $K{\bar K}$ pairs 
or kaon-hyperon pairs hidden strangeness can be produced through 
$\phi (s{\bar s})$ mesons. The $\phi$ is a $0^- (1^{--})$ vector meson 
with the same quantum numbers as the $\omega$ and has a mass of 1020 MeV. 
However, compared to the $\omega$ 
the production of $\phi$ mesons is strongly suppressed 
due to the Okubo-Zweig-Iizuka (OZI) \cite{OZI} selection rule which 
forbids the appearance of disconnected quark line diagrams. 
According to the OZI estimate $\phi$ mesons can only be produced 
due to a small admixture of non-strange light quarks in their wave
function. The corresponding mixing angle $\theta_{mix}$
is equal to $\theta_{mix} \approx 3.7^o$\cite{PDG}. According to this 
mixing angle  the ratio of
$\phi$ and $\omega$ mesons cross sections should at comparable
energies naively be equal to
\begin{equation}
R_{\phi/\omega} = \tan^2 \theta_{mix}\cdot F \approx 4.2\times 10^{-3}\cdot F~,
\label{naive}
\end{equation}
where $F$ is a correction due to the different phase space factors 
for $\phi$ and $\omega$ mesons.
Experimentally, the ratio $R_{\phi/\omega}$ is in $pp\rightarrow
pp\phi(\omega)$  reactions, however, known to be one order of 
magnitude larger than the naive expectation \cite{DISTO}.
Elementary $\phi$ production has been calculated 
within several approaches, mostly within meson exchange models 
\cite{chung97,titov02,Tsushima03}. In Refs. \cite{omega03} and \cite{phi03} 
the cross section for $\omega$ and $\phi$ meson production in
$pp$ collision have been calculated within the resonance model. 
The experimental data are well reproduced and the 
large violation of the OZI prediction 
for the $\phi$ over $\omega$ production, observed experimentally, 
could be explained in \cite{phi03} without the introduction of 
additional parameters. 
The reactions of the type $NN\rightarrow NN\phi$ and 
  $\pi N\rightarrow N\phi$ should be OZI suppressed compared to 
open strangeness production as well, whereas the DISTO results  
\cite{DISTO} show that almost 50\% of the $pp\longrightarrow pp K^+ K^-$ 
cross section is resonant, i.e. running over an intermediate $\phi$. 
When parameterizations of the $NN\longrightarrow NN K^+ K^-$ are used 
in transport calculations the resonant parts of these cross sections 
are usually implicitly included. 

The FOPI 
Collaboration measured the $\phi$ production in Ni+Ni reactions at 
1.93 AGeV and came to the conclusion that at least 
about 20\% of the observed $K^-$ mesons originate from $\phi$ decays 
\cite{fopiphi}. Due to the large coupling to the $K{\bar K}$ channel 
(the partial $\Gamma (\phi \rightarrow K^+ K^-)$ decay width is about 50\%) 
one can expect that the $\phi$ reacts sensitive on medium modifications of 
the kaons which will mainly result in a broadening of the $\phi$ width 
in hadronic matter (see e.g. \cite{boris04} and refs. therein). The  $\phi$ 
production in heavy ion reactions has been 
studied in \cite{chung97,barz02}. However, also with the inclusion of medium 
effects the corresponding $\phi$ yields are underestimated by a factor 
two to three by these transport calculations which indicates that  $\phi$ 
production in heavy ion reactions is not yet fully understood. The situation 
will probably be improved when the HADES Collaboration will be able to 
measure the $\phi$ production through the $e^+ e^-$ decay channel. The 
dilepton spectra should provide a very clear access to a broadening 
of the $\phi$ and the underlying in-medium modifications of the kaons.

\subsection{Densities and time scales}
Since the conjecture that the early and dense phase of heavy ion reactions 
can be probed by $K^+$ production has been put forward by Aichelin and
Ko about two decades ago \cite{AiKo85} strong efforts were undertaken
to settle this question, both from the experimental and the theoretical
side. The original motivation was to study the nuclear
equation of state at supra-normal densities. This point will be
addressed in detail in Chap. 6. The prerequisite for such an
enterprise is  of course the fact that $K^+$ mesons originate indeed
from the early phase {\it and} from  supra-normal densities.  
Both features were predicted by transport calculations 
in the early 90ties \cite{AiKo85,bali94,huang93,hartnack94,li95b}. 
At that time experimental data on subthreshold kaon production were 
scarce and the same was true for the elementary cross sections which
serve as input for the transport calculations. Hence these calculations
were burdened with large uncertainties but the predictions on
densities and time scales turned out to be stable. A more recent 
analysis is e.g. given in \cite{pal01}. 
\subsubsection{Conditions for strangeness production}
\begin{figure}[h]
\unitlength1cm
\begin{picture}(9.,9.5)
\put(1.5,0){\makebox{\epsfig{file=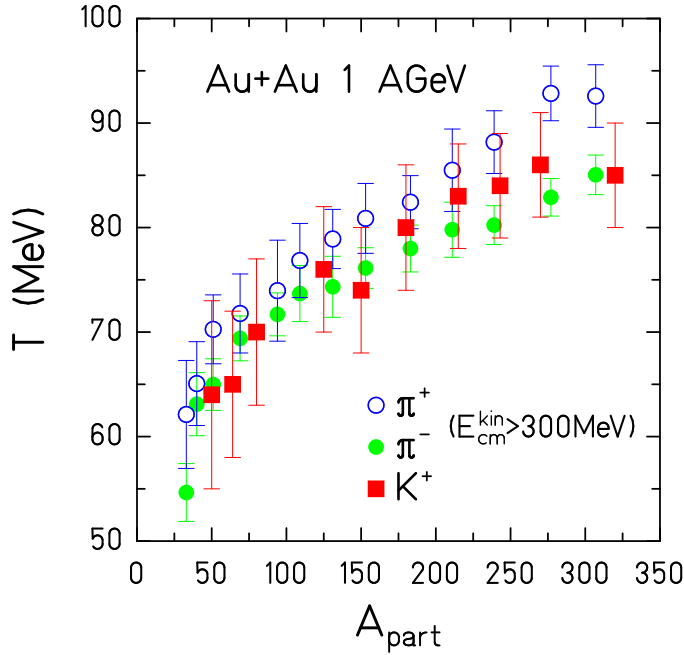,width=12.0cm}}}
\end{picture}
\caption{Measured inverse slope parameters for high-energy 
pions and $K^+$ mesons 
in Au+Au reactions at 1 AGeV. The figure is taken from
\protect\cite{senger99}. 
}
\label{fig_slope_1}
\end{figure}

$K^+$ mesons are produced in the early and dense phase of a heavy ion
reaction. E.g. at an incident energy of 1 AGeV they reach their
chemical freeze-out values already after 15 fm/c which is the phase
were maximal compression is reached. One should, however, keep in mind
that at such time scales the surrounding nuclear environment is still
in a pronounced non-equilibrium state \cite{pal01,fuchs03}. The same
is in principle also true for $K^-$ mesons. However, due to high
absorption rates through strangeness exchange reactions 
$K^- N \leftrightarrow Y\pi$ they have the tendency to reach chemical
equilibrium much faster. We will come back to this point in subsection 3.4. 

Experimental evidence for an early $K^+$ freeze-out is provided by the
fact that $K^+$ mesons and high-energy pions behave qualitatively
similar. This fact is demonstrated by Fig. \ref{fig_slope_1} where 
the experimentally extracted inverse slope parameters for kaons and 
high-energy pions ($E^{\rm kin}_{\rm cm} >300$ MeV) 
are compared. Also rapidity distributions of $K^+$ and high $p_T$
pions are similar and clearly distinct from low $p_T$ pions
\cite{best97}. At SIS energies both, kaons and pions 
with $p_T \gtrsim 0.5$ GeV are produced at equivalent ``subthreshold'' 
energies and identical slopes of the spectra imply that these 
particles freeze out under similar conditions. 
High-$p_T$ particles probe in general the early stage of a
heavy ion reactions and this feature is 
independent of the considered energy range. For pions at SIS energies 
this behavior was predicted by transport models \cite{bass95,Teis,uma97}
and experimentally verified by the simultaneous measurement of pion in-plane 
and out-off-plane flow \cite{wagner98,wagner00}. The energy dependence of the
pion flow is caused by spectator matter shadowing which can be used 
 as a time clock for the reaction
\cite{wagner98,senger99}. 
\begin{figure}[h]
\unitlength1cm
\begin{picture}(9.,8.5)
\put(3.5,0){\makebox{\epsfig{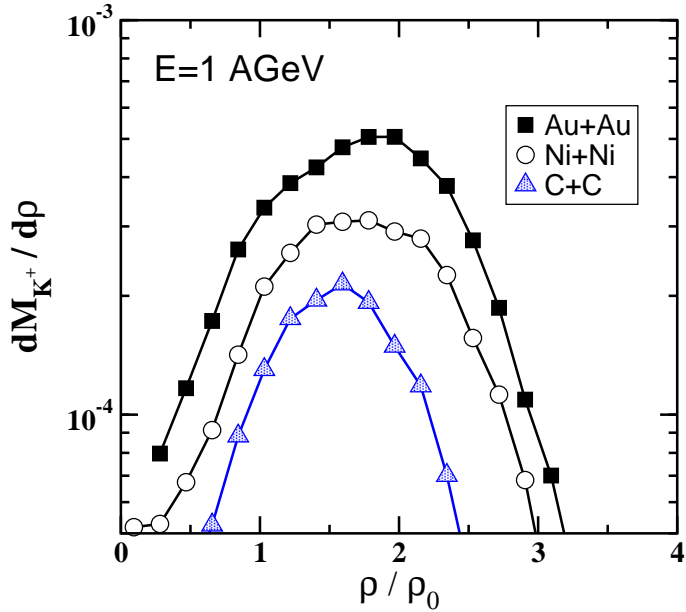}}}
\end{picture}
\caption{Nuclear density at the production of a $K^+$ meson. 
Results from transport 
calculations for central C+C, Ni+Ni and Au+Au collisions at 1 AGeV, 
normalized to the mass numbers, are shown.
}
\label{fig_rho_1}
\end{figure}
To obtain a quantitative picture of the density range probed 
by $K^+$ production, in Fig. \ref{fig_rho_1} baryon densities are shown at 
the space-time coordinates where $K^+$ mesons are created. 
The figure shows the densities in central collisions (b=0 fm) for 
three typical mass systems, Au+Au, Ni+Ni and C+C at 1.0 AGeV.  
The density distribution  $dM_{K^+}/d\rho$ 
is defined as 
\beq
dM_{K^+}/d\rho = \sum_{i}^{N_{K^+}} 
\frac{d P_i }{ d\rho_B ({\bf x}_i, t_i)}
\label{densdist}
\eeq 
where $\rho_B$ is the baryon density at which the kaon $i$ was created 
and  $P_i$ is the corresponding production probability. For 
a better comparison of the different reaction systems 
the curves are normalized to the 
corresponding mass numbers. Fig. \ref{fig_rho_1} illustrates 
the fact that nuclear compression is probed by $K^+$ production. Most 
kaons are produced at supra-normal densities between one and three
times $\rho_0$. The maximal densities are, on the other hand, 
correlated with the system size. A dependence on the nuclear EOS 
will therefore be most pronounced in large systems as 
discussed in detail in chapter 6.
\begin{figure}[h]
\unitlength1cm
\begin{picture}(10.,9.5)
\put(3.5,0){\makebox{\epsfig{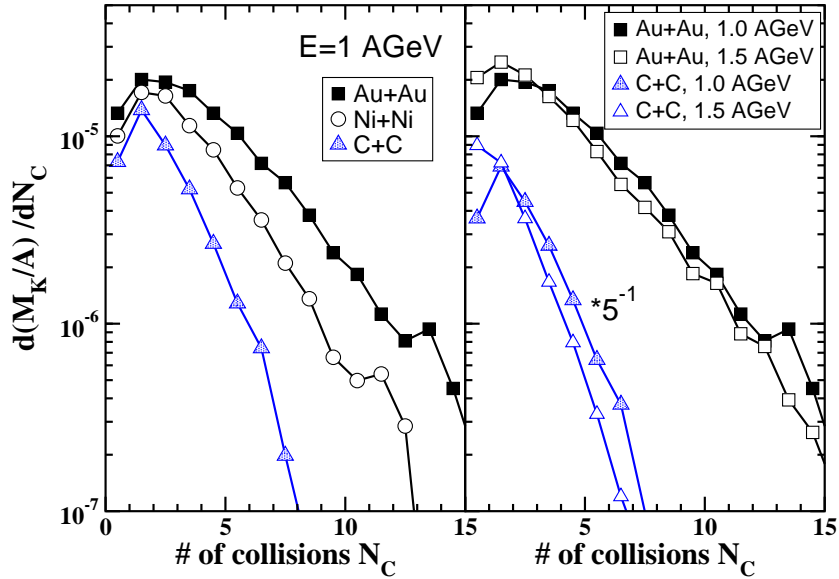}}}
\end{picture}
\caption{System size and energy dependence of the 
number of binary collisions which particles encountered
prior to the production of a $K^+$ meson. Left: results from transport 
calculations for central  Au+Au, Ni+Ni and  C+C reactions at 1 AGeV.  
Right: Same for Au+Au and  C+C at 1 and 1.5 AGeV (arbitrary
normalization of the curves). 
}
\label{fig_ncoll_1}
\end{figure}
Closely connected to the densities is the average number of collisions 
$N_C$ which particles 
encountered prior to the production of a $K^+$ meson.
The distributions shown in Fig. \ref{fig_ncoll_1} are obtained in 
analogous way as the density distributions, i.e.  
$dM/d N_C = \sum_{i}~ d P_i / d N_{C}^i$ 
with $N_{C}^i  = \frac{1}{2} (N_{C_{1}}^i +N_{C_{2}}^i) $ the average 
number of collisions which two hadrons ($1,2=N, \Delta, \pi$) 
experienced before they produced kaon $i$ with production probability 
$P_i$. As before the curves are normalized to the mass numbers of the 
different reaction systems. 
This quantity is a suitable measure for the 
collectivity probed by $K^+$ production. Fig. \ref{fig_ncoll_1} (left
panel) demonstrates that subthreshold kaon production is indeed a multi-step 
process. At 1 AGeV the average number of hadron-hadron collisions
prior to the production of a kaon is larger than one and 
increases with the system size. This feature is also 
reflected in the $A_{\rm part}$ dependence 
of the experimentally extracted inverse slope parameters of the kaon
spectra shown in 
Fig. \ref{fig_slope_1}. In more central reactions kaons seem to 
originate from an environment with apparently higher temperature which
must, however, not be thermally equilibrated. The inverse slope 
parameters rise also with system size, Tab. \ref{tab_temp1}, which is 
consistent with the collision 
history shown in Fig. \ref{fig_ncoll_1}. The inverse slope 
parameters can also simply be considered as a measure of the collectivity 
experienced by the kaons. The question if equilibrium has been reached 
or not is a subtle one and cannot be decided exclusively from 
experiment. 
\begin{table}
\begin{center}
\begin{tabular}{|c|c|c|c|}
\hline
 &  C+C  & Ni+Ni  & Au+Au\\ 
\hline\hline
  $K^+$ (1.0 AGeV) & $58\pm 6$ & $75\pm 6 $ & $85 \pm 6$ \\
\hline
  $K^-$ (1.8 AGeV) & $55\pm 6$ & $90\pm 5 $ & -- \\
\hline
\end{tabular}
\end{center}
\caption{\label{tab_temp1}
Experimental inverse slope parameters for $K^+$ and  $K^-$ at 
comparable energies above threshold for different mass systems
(from \protect\cite{senger99,laue00,barth}). 
}
\end{table}
With increasing energy the number of multi-step processes 
involved in  the kaon production decreases which can be seen from the 
right panel of Fig. \ref{fig_ncoll_1}. There the number of collisions 
are shown at different energies 1 and 1.5 AGeV. For better comparison 
the curves are normalized to each other. In the light C+C system the  
decrease is more pronounced than in Au+Au 
and similar to p+A reactions most kaons 
are now produced in primary reactions.  

\subsubsection{Energy and mass dependence}
\begin{figure}[h]
\unitlength1cm
\begin{picture}(8.,8.5)
\put(3.5,0){\makebox{\epsfig{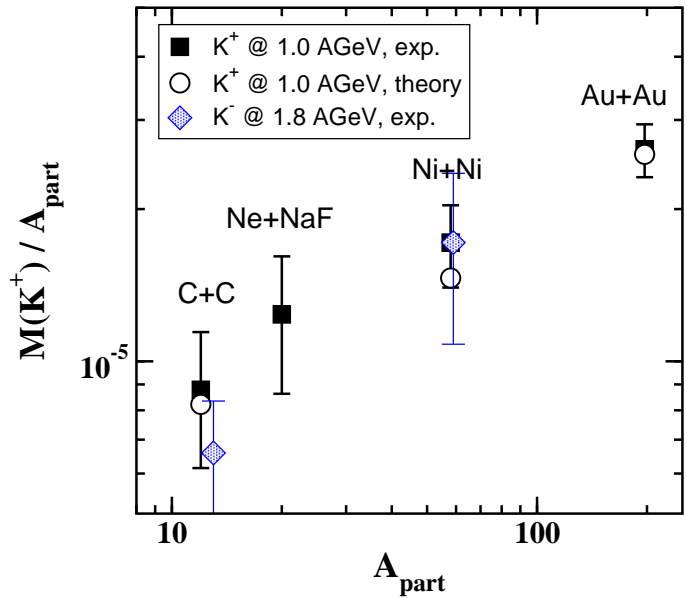}}}
\end{picture}
\caption{$K^+$ multiplicities per $A_{\rm part}$ for different 
colliding mass systems at 1 AGeV. Data from KaoS 
\protect\cite{barth,sturm01} are compared to theoretical 
QMD transport calculations. For C+C and Ni+Ni also $K^-$ data 
\protect\cite{barth,laue00}  at equivalent 
energy above threshold (1.8 AGeV) are shown. 
}
\label{fig_apart_1}
\end{figure}
The fact that subthreshold kaon production is a collective phenomenon
is clearly reflected in the dependence of the kaon yield on the number
of participating nucleons. $A_{\rm part}$ is defined as
the number of nucleons contained in the overlapping region of the two
colliding nuclei. For a given impact parameter the overlap in a geometrical
model is that of two hard spheres with radius $r_0 A^\frac{1}{3}$
($r_0=1.2$ fm). In symmetric A+A collisions the mean $A_{\rm part}$ is
given by $\langle A_{\rm part}\rangle = A/2$, in exclusive reactions 
  $A_{\rm part}$ is directly related to the centrality.

However, before turning to heavy ions it is useful to consider 
proton-nucleus reactions. A survey on the experimental
situation of the mass dependence of $K^+$ production in p+A can be
found in \cite{buescher01}. If the kaons were dominantly produced
through primary reactions the scaling law of the 
$\sigma^{p+A\longrightarrow K^+X}$ cross section should follow 
that of the inelastic
proton-nucleus cross section. For proton energies $T_p \leq 2$ GeV
the inelastic cross section is given by $\sigma_{\rm inel} \sim
A^{0.69\pm 0.03}$ which is close to that of an opaque nucleus 
($\sigma \sim A^\frac{2}{3}$). At $T_p= 1.5$ GeV, i.e. close to
threshold, a scaling law $\sigma_{K^+} \sim A^\alpha$ with $\alpha =
0.73 \pm 0.04$ has been extracted at SATURNE \cite{saturne96} which
supports the primary reaction mechanism. Deep below threshold the 
missing mass of the kaons has to be provided by the Fermi motion 
inside the nucleus or the high momentum tails of the nucleon spectral
functions. Such collective effects are not supposed to depend strongly
on the system size. On the other hand, deep subthreshold
energies favor the secondary production mechanism since, e.g., the
Fermi motion can be utilized several times in multi-step processes. 
This would lead to a proportionality $\sigma_{K^+} \sim A$. Indeed,
at PNPI a coefficient $\alpha = 1.04 \pm 0.01$ has been extracted at 
$T_p = 0.8\div 1.0$ GeV \cite{pnpi}. Such an exponent is in line with
the predictions from transport models which indicate that subthreshold kaon
production in p+A is governed by secondary reactions \cite{rudy02}. 
More recently ANKE extracted values of 
$\alpha = 0.74 \pm 0.05$ at $T_p = 1.0$ GeV \cite{anke01,buescher01} in 
line with SATURNE. 
The scaling depends, however, strongly on the experimantal acceptence 
and thus experimental situation is not fully conclusive.

\begin{figure}[h]
\unitlength1cm
\begin{picture}(8.,8.5)
\put(3.5,0){\makebox{\epsfig{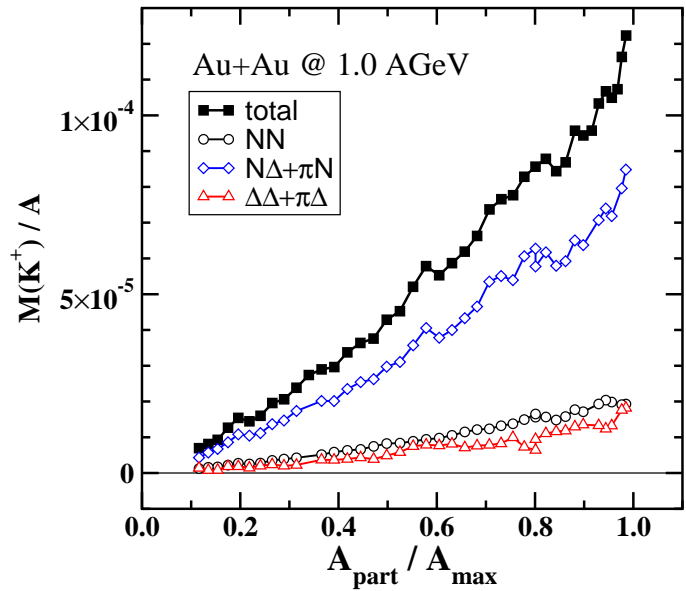}}}
\end{picture}
\caption{$K^+$ multiplicities as a function of $A_{\rm part}$ in Au+Au 
at 1 AGeV for the different production channels $NN$ (primary), 
$N\Delta + \pi N$ (secondary) and $\Delta\Delta + \pi \Delta$
(secondary). 
}
\label{fig_apart_3}
\end{figure}

In summary, p+A reactions reveal a scaling law of the kaon multiplicity  
$N_{K^+} = \sigma_{K^+}/\sigma_{\rm inel}\sim A^\alpha$ with 
$1.05 \leq \alpha \leq 1.48$. In a Glauber picture an A+A collision 
can be understood as the superposition of independent NN collisions or
as A$\times$(N+A). Assuming the Glauber
picture to hold, the kaon multiplicity $N_{K^+}$ should be linear
proportional to $A_{\rm part}$. This is evidently not the case. 
Fig. \ref{fig_apart_1}\footnote{Although in minimal bias reactions the mean $A_{\rm part}$
is given by A/2, we scale the yields by $\langle A_{\rm part}\rangle
=A/2$ which accounts for the fact that kaons are dominantly produced
in central and semi-central reactions.}  shows KaoS 
data \cite{senger99,sturm01} for different mass systems at 1 AGeV 
which can be fitted by $N_{K^+} \propto  A^\alpha$ with $\alpha = 1.36$. 
The  observed non-linear increase as a function of system size 
is thus a clear indication for a
collective production mechanism through multi-step scattering
processes. The system size dependence of the measured
multiplicities can be reproduced by theoretical transport
calculations. In Fig.  \ref{fig_apart_1} we show results from QMD
calculations which include a repulsive in-medium potential (see also
chapters 4 and 5). 
\begin{figure}[h]
\unitlength1cm
\begin{picture}(8.,8.5)
\put(3.5,0){\makebox{\epsfig{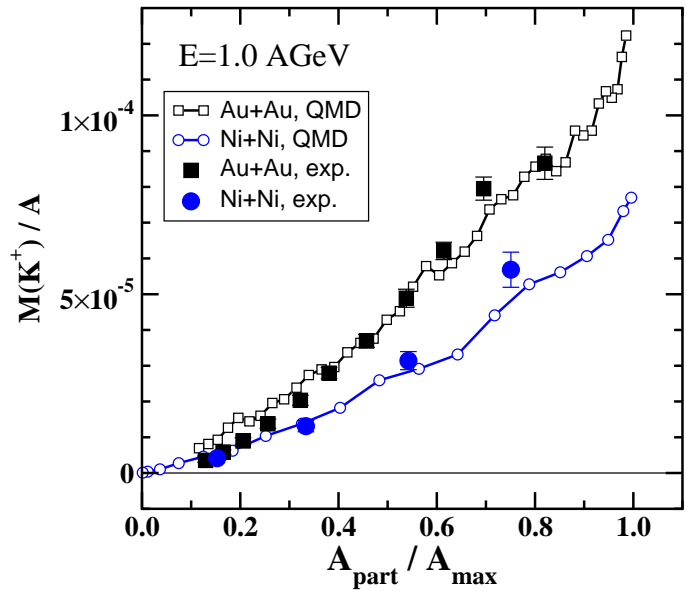}}}
\end{picture}
\caption{$K^+$ multiplicities as a function of $A_{\rm part}$ in Au+Au
and Ni+Ni reactions at 1 AGeV. Data from KaoS 
\protect\cite{senger99,barth} are compared to 
QMD transport calculations. 
}
\label{fig_apart_2}
\end{figure}
The same non-linear scaling with  $A_{\rm part}$ occurs as a function
of centrality, i.e. the $K^+$ yield in central reactions is
strongly enhanced relative to peripheral reactions. The scaling of 
$N_{K^+} \propto A_{\rm part}^\alpha$ with $\alpha = 1.8 \pm 0.15$
\cite{kaos94} is even more pronounced than the average A dependence of
the kaon multiplicity of different A+A systems
(Fig. \ref{fig_apart_1}). Since the bulk of kaons originates from 
semi-central collisions (e.g. in Au+Au from $b\sim 5$ fm which
corresponds to $A_{\rm part}/A_{\rm max} \sim 0.7$ in
Fig. \ref{fig_apart_2}) the pronounced centrality dependence is to
some extent washed out in minimal bias reactions. As a function of
centrality $\alpha$ is significantly larger than the value of 
$\alpha =1.48$ which can maximally be deduced from p+A within a
Glauber picture. This demonstrates that heavy ion collisions show
qualitatively novel aspects compared to p+A. Fig. \ref{fig_apart_2}
compares the experimental  $A_{\rm part}$ dependence in Ni+Ni and 
Au+Au reactions at 1 AGeV \cite{barth,senger99} to QMD
calculations. In both cases the data are well reproduced. The
contributions of the different production mechanisms can be seen from 
Fig. \ref{fig_apart_3}. As expected, the primary $NN \longrightarrow NYK^+$
channels show a linear $A_{\rm part}$ dependence while the strong
non-linear increase of the total yield is caused by secondary type
reaction mechanisms. Here the processes  $N\Delta \longrightarrow
NYK^+$ and $ \pi N \longrightarrow YK^+$ dominate while the higher order 
multi-step processes $\Delta \Delta \longrightarrow
NYK^+$ and $ \pi \Delta \longrightarrow YK^+$ are of minor importance.

For Ni+Ni the
scaling parameter $\alpha$ has also been measured as a function of the
beam energy \cite{barth}. A slight decrease of $\alpha = 1.9 \pm 0.25$
at 0.8 AGeV to $\alpha = 1.65 \pm 0.05$ at 1.8 AGeV indicates the
decreasing importance of collective effects with rising beam
energy. The experimental facts are consistent with the theoretical
findings discussed above, namely that the average number of collisions
which the nucleons encountered prior to the production of a $K^+$
meson increases a) with system size and b) going deeper below
threshold. The particles have to accumulate the necessary energy by
multiple scattering which is easier to achieve in central reactions
and/or larger systems where higher densities are reached. The
conditions are optimal in the early phase of the reaction where baryon
as well as energy density are maximal. 
Corresponding data for antikaons are rare. 
However, existing $K^-$ data indicate that the behavior of antikaons 
follows closely that of the
kaons. In Ni+Ni at 1.8 AGeV a scaling law of $N_{K^-} \propto A_{\rm
part}^\alpha$ with $\alpha = 1.8 \pm 0.3$ has been observed 
\cite{laue00,barth}.  At a first glance this may be astonishing since one could
expect that the strong $K^-$ absorption leads to a weaker  $A_{\rm
part}$ dependence than for $K^+$. On the other hand, strangeness
exchange is the dominant source for  $K^-$ production at subthreshold
energies and primary hyperons are produced in association with
the $K^+$'s. Since these primary hyperons have the same $A_{\rm part}$ as
the $K^+$'s the $K^-$'s are forced to follow them. 
\subsubsection{Chemical freeze-out}
\begin{figure}[h]
\unitlength1cm
\begin{picture}(9.,9.5)
\put(3.5,0){\makebox{\epsfig{file=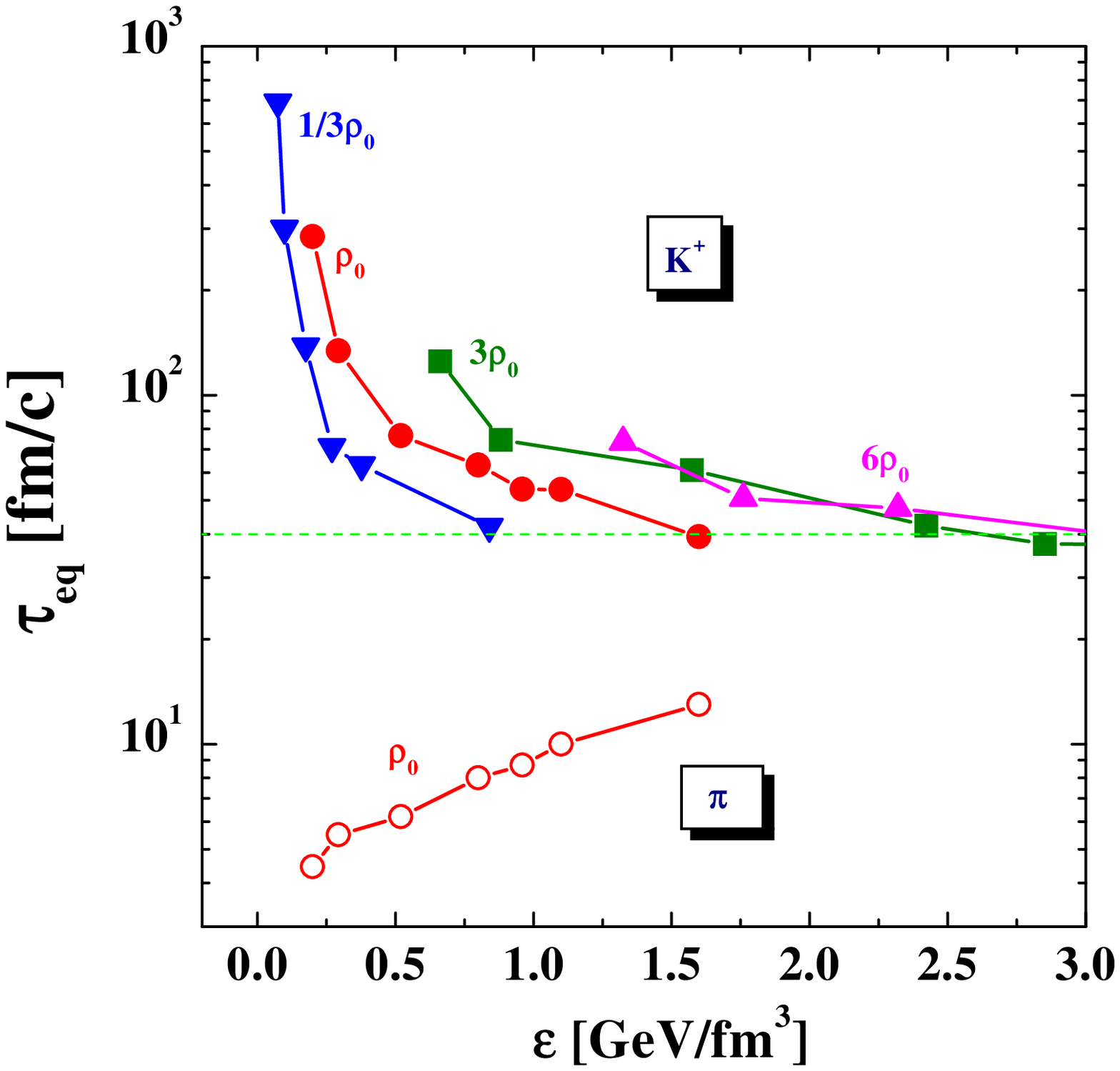,width=9.0cm}}}
\end{picture}
\vspace*{-2.5cm}
\caption{Chemical equilibration time $\tau_{eq}$ for pions and kaons in
infinite hadronic matter as a function of baryon density and energy 
density $\epsilon$. Results from a BUU cascade are shown. The figure
is taken from \protect\cite{brat00}.   
}
\label{fig_tfreeze1}
\end{figure}
The question of strangeness equilibration has a long history. Already
in the classic paper by Koch, M\"uller and Rafelski \cite{koch86} it
was argued that at CERN energies strangeness saturation requires 
equilibration times $\tau_{eq}\sim 80$ fm/c much longer than the 
reaction time and that the fact that particle spectra can nonetheless
be described by a common Boltzmann factor $T\sim 170$ MeV hints
towards a sudden hadronization where the system crosses the phase
boundary from the quark-gluon plasma to the hadronic phase. At SIS
energies there is of course no such phase transition but the question
remains whether the observed particle abundances have purely dynamical origin
or can be understood by means of statistical concepts.

Transport calculations indicate that kaons as well as antikaons are
dominantly produced in the early and dense non-equilibrium
phase. Since kaons interact with the surrounding medium almost
exclusively by elastic reactions $KN\longrightarrow NK$ (here we 
consider charge exchange as elastic) they have a long
mean free path $\sim 7$ fm and chemical freeze-out occurs early. 
This picture is supported by the fact that subthreshold kaons show similar features
like high energy pions, see Fig. \ref{fig_slope_1}. High energy pions were
experimentally proven to originate from the early phase
\cite{best97,wagner98,wagner00}. The influence of strangeness absorption
which might drive kaons to equilibrium has been studied in
\cite{pal01}. This process has extremely low probability since at
subthreshold energies no more than one $YK^+$ pair per nucleus-nucleus
collision is created. This
means that the same pair of particles has to meet again in the course
of the reaction for the kaon to get annihilated. As long as kaon
production is active the annihilation (loss) rate $dN_L /dt$ is about one to two orders
of magnitude smaller \cite{pal01} and can be safely neglected in
corresponding transport calculations. Hence the kaon yield 
freezes out early {\it and} at supra-normal nuclear density
\cite{pal01}. The conditions are not much different from 
those where the kaons are produced, see Fig. \ref{fig_rho_1}. 
The same conclusion can be
drawn from the investigations of Bratkovskaya et al. \cite{brat00}
where chemical equilibration times in ``infinite'' hadron matter,
i.e. a box with periodic boundary conditions, were studied using a BUU
cascade model. The resulting equilibration times for pions and kaons
as functions of the baryon density and the energy density $\epsilon$
are shown in Fig. \ref{fig_tfreeze1}. In contrast to pions, kaons need
at least $\tau_{eq}\gtrsim 40$ fm/c to reach chemical equilibrium. 
Equilibration takes even longer at
lower baryon and/or energy density. At SIS conditions, corresponding to
the lower limit of $\epsilon$ in Fig. \ref{fig_rho_1}, $\tau_{eq}$ is
above 100 fm/c which is at SIS energies about twice the duration of an heavy
ion reaction like Au+Au. 

\begin{figure}[h]
\begin{minipage}[h]{185mm}
\unitlength1cm
\begin{picture}(18.5,7.0)
\put(0.,0.){\makebox{\epsfig{file=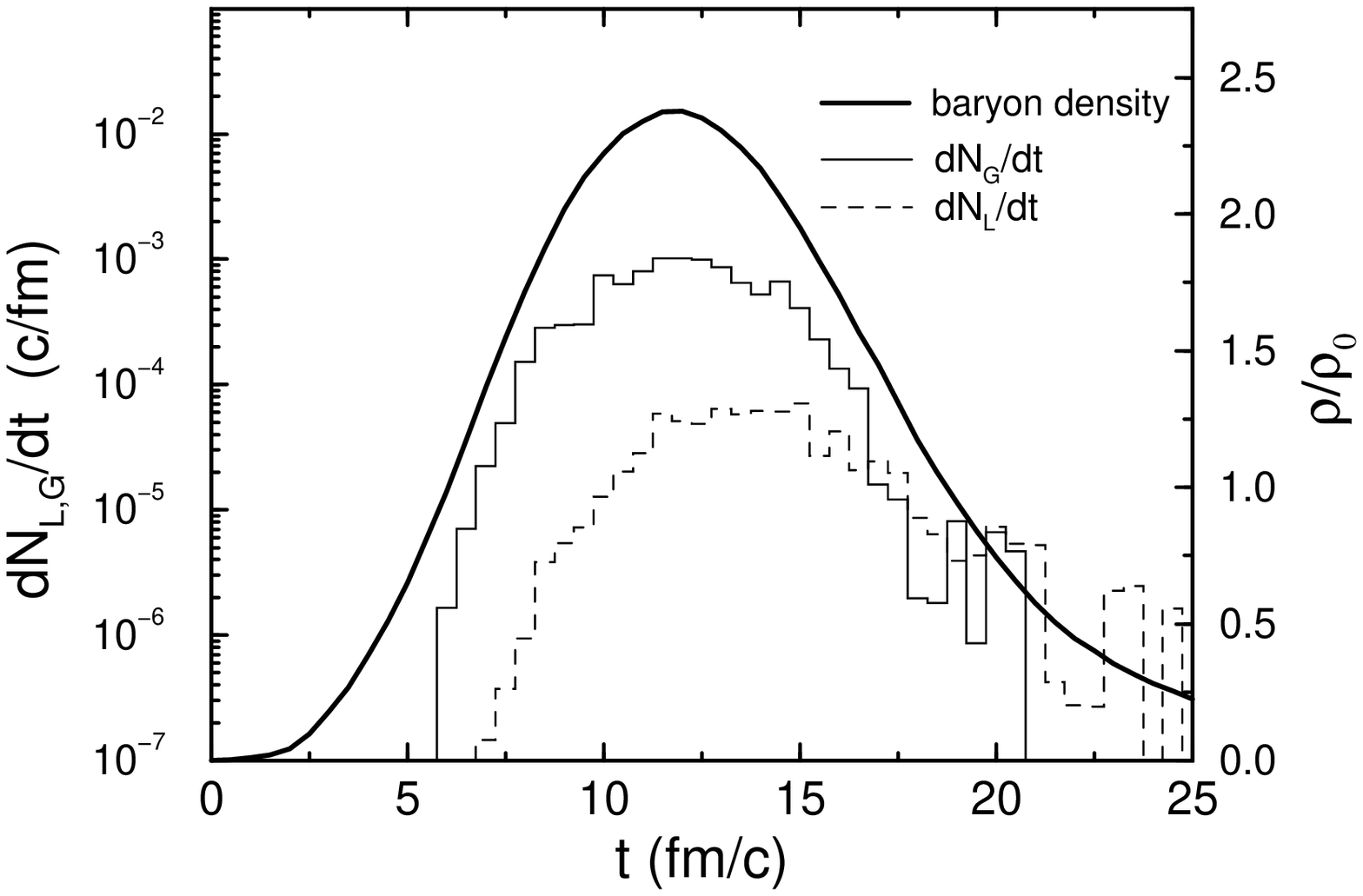,width=8.5cm}}}
\put(9.5,0.0){\makebox{\epsfig{file=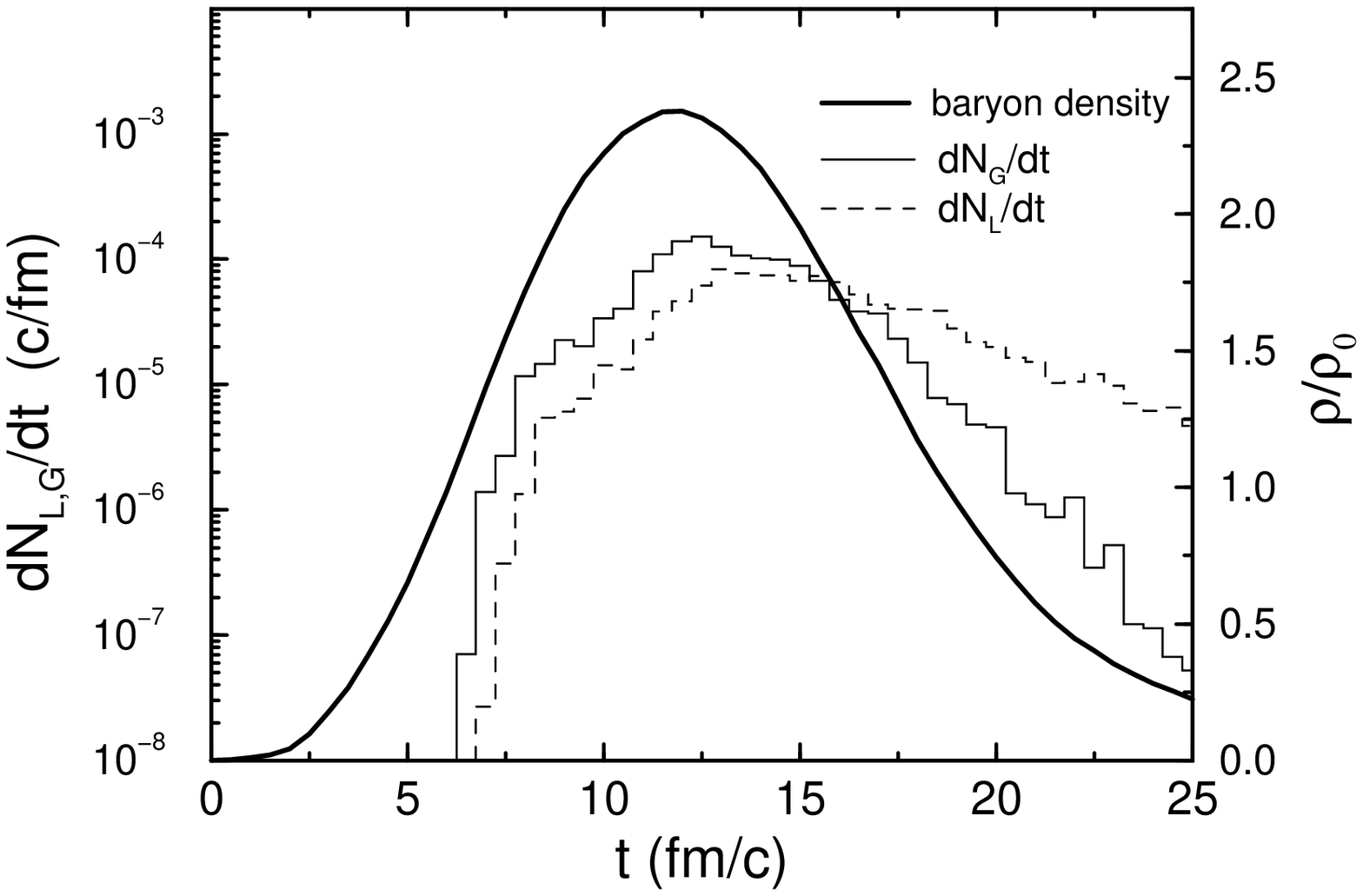,width=8.5cm}}}
\end{picture}
\end{minipage}
\caption{Kaon (left) and antikaon (right) production $dN_G/dt$ and 
annihilation $dN_L/dt$ rates in a central Ni+Ni reaction at 1.0 AGeV. 
The solid line shows the time evolution of the central 
density. The figure is taken from \protect\cite{pal01}.
}
\label{Kpm_eq_fig}
\end{figure}

For antikaons the situation is more complex. Though 
primary antikaon production 
takes also place in the initial and dense phase, strangeness can be
redistributed between hyperons and antikaons. 
Strangeness exchange leads
to annihilation rates which exceed the production rates in the 
expansion phase of a heavy ion collision \cite{pal01,aichelin03}. When 
the fireball becomes dilute the endothermic $K^-$  production  
process dies out while the exothermic inverse reaction can still 
continue. Due to such a
cross over there exists a situation where the equilibrium condition, 
namely  production rate = annihilation rate, is formally fulfilled. However,
this is not a stationary state since the system is expanding. 
The situation is illustrated by Fig. \ref{Kpm_eq_fig} where kaon and 
antikaon production (gain) rates $dN_G/dt$ and annihilation (loss) 
rates  $dN_L/dt$ in a central (b=0 fm) Ni+Ni reaction at 1.0 AGeV are 
shown. The results were obtained by Pal et al. \cite{pal01}.
Though closer, also $K^-$'s seem not to reach chemical equilibrium in a 
thermodynamical sense. In thermal equilibrium the yields do exclusively 
depend on the chemical potentials and the temperature. 
As has been demonstrated by Hartnack at al. \cite{aichelin03} this is 
not the case even when the equilibrium condition $dN_G/dt =  dN_L/dt$ is 
fulfilled: In the expanding system an artificial scaling of the strangeness exchange 
cross section $\pi Y \leftrightarrow NK^-$ is directly reflected in the 
total yield.

In view of these facts it is astonishing that many features can also 
be understood by means of purely statistical concepts. The statistical
model of an ideal hadron gas assumes that particles are 
emitted from a thermalized,
eventually radially expanding fireball. It turned out to describe hadron
yields and ratios successfully from AGS to RHIC energies (for a
recent review see \cite{pbm03}). 

In the framework of  a grand canonical ensemble the 
particle number density $n_i$ of species $i$ is given by a 
Boltzmann distribution \footnote{In most actual calculations 
the more accurate Fermi and Bose distributions are used. The
usage of Boltzmann distributions makes the present discussion,
however, more transparent.}
\beq  
n_i =\frac{N_i}{V}=g_i \int_0^{\infty} \frac{d^3 k}{(2\pi)^3} 
e^{-\left( E_i -\mu_B B_i - \mu_S S_i\right) /T } ~~.
\label{sm1} 
\eeq
$E_i = \sqrt{{\bf k}^2 + m_{i}^2}$ is the single particle energy of hadron 
species $i$, $T$ is the temperature, $\mu_B$ and $\mu_S$ are the 
baryon chemical potential and strangeness chemical potential. $B_i$ 
and $S_i$ are the baryon charge and strange charge and 
$g_i$ the degeneracy factor of hadron $i$. 
$V$ is the volume of the system. The characteristics of the 
fireball can be described by means of
four independent parameters, such as volume $V$, total energy $E$,
net baryon density $\rho_B$ and net strangeness density $\rho_S$. These
determine the fireball temperature $T$,
the strangeness and baryon chemical potentials $\mu_S$ and $\mu_B$.
Since the total strangeness is zero the number of free parameters 
reduces to three, namely, $V$, $E$ and $\rho_B = N_B / V$. If one is
only interested in particle ratios instead of total yields the volume
dependence drops out and the number of parameters is reduced to
two. They are usually chosen as the temperature and the baryon chemical
potential and are determined by fitting experimental hadron yields.

Such a grand canonical description fails at SIS energies. However, as
shown by Cleymans et al. \cite{cleymans98} the statistical model works
also at SIS if one accounts for exact strangeness
conservation which leads to the following modification
of Eq. (\ref{sm1}) for kaons and antikaons 
\beqa  
n_{K^+} &=& g_{K^+}\int_0^{\infty} \frac{d^3 k}{(2\pi)^3} 
e^{-E_{K^+}/T } 
\left[ g_{K^-}V\int_0^{\infty} \frac{d^3 k}{(2\pi)^3} e^{-E_{K^-}/T } 
 + g_{\Lambda}V\int_0^{\infty} \frac{d^3 k}{(2\pi)^3} e^{-(E_{\Lambda} -\mu_B)    /T } 
\right] \label{sm2}\\
 n_{K^-} &=& g_{K^-}\int_0^{\infty} \frac{d^3 k}{(2\pi)^3} 
e^{-E_{K^-}/T } 
\left[ g_{K^+}V\int_0^{\infty} \frac{d^3 k}{(2\pi)^3} e^{-E_{K^+}/T } 
\right] ~~.\label{sm3}
\eeqa 
With this modification the statistical model delivers an accurate
description of particle yields and ratios also at SIS energies (except
of the $\eta$ yield which is systematically underestimated) 
\cite{cleymans98,cleymans99}. The assumption of common freeze-out
parameters for all hadron species stands, however, in clear
contradiction to the dynamical transport calculations. In particular a
freeze-out density of about $\rho_0 /4$ \cite{cleymans98,cleymans99}
as obtained by the statistical model fit to the hadron multiplicities
 is much lower than the corresponding $K^+$ freeze-out density obtained
from the dynamical models. 

The statistical model reproduces nonetheless several basic experimental
features of the strangeness production at SIS. The canonical treatment
leads to a quadratic $A_{\rm part}$ dependence of the kaon
multiplicities $N_{K^+} \sim V^2 \sim A^{2}_{\rm part}$ which is close 
to the observed $A_{\rm part}$ dependence \cite{kaos94,barth}. 
Since pions are treated grand-canonical they
scale linearly with $V$ and $A_{\rm part}$. Hence, the observed
centrality dependence of the $K^+/\pi^+$ ratio \cite{kaos94} is
obtained for free. The same holds for the  $A_{\rm part}$ dependence of the
$K^-/K^+$ ratio which is predicted to be flat in agreement with the
Ni+Ni data. A fine tuning can be achieved by a variation of the
temperature as a function of centrality \cite{cleymans00}. 
This enforces, however,
in a slight decrease of $T$ for central reactions which contradicts the 
intuition of a larger energy deposit and a hotter fireball. In the
larger Au+Au system the $K^-/K^+$ ratio has even be found to slightly
decrease with centrality \cite{foerster03} which is at variance with a
statistical interpretation but easy to understand from a microscopic
point of view due to the large $K^-$ absorption cross section.

However, here appears another contradiction to the dynamical
approaches: In contrast to the statistical model based on bare hadron
masses, the transport models  need to introduce in-medium kaon masses 
in order to reproduce the experimental $K^-/K^+$ ratio
\cite{li97,wis00}. This apparent contradiction between
dynamical and statistical approaches has been addressed by Brown et al. 
\cite{brown02}. From eqs. (\ref{sm2}) and  (\ref{sm3}) the $K^+/K^-$
ratio is approximately given by
\beq
\frac{N_{K^+}}{N_{K^-}}\sim \left( \frac{m_\Lambda}{m_K}\right)^\frac{2}{3} 
\frac{ e^{-(E_{\Lambda} -\mu_B)/T } }{e^{-E_{K^-}/T}}~~.
\label{sm4}
\eeq
The experimental value 
\footnote{In Refs. \protect\cite{cleymans99,cleymans00} the earlier
and smaller ratio $N_{K^+}/N_{K^-} = 21\pm 9$ from
\protect\cite{barth} has been used.} of $N_{K^+}/N_{K^-} = 30$ in Ni+Ni at 1.8
AGeV \cite{menzel00} enforces a chemical potential which is significantly
smaller than the $\Lambda$ mass in order to punish hyperon production
relative to that of $K^-$'s. A dropping $K^-$ mass $m^*_{K^-} = m_K
-\alpha\rho_B$ compensates the $\mu_B$ dependence in (\ref{sm4}) to
large extent \cite{brown02} and enables one to reproduce the 
experimental ratio with much larger freeze-out densities $\rho\sim
1\div 2\rho_0$, depending on the strength of the mass reduction. 
Such values for the  freeze-out density are now in agreement with the
estimates from transport models. 

However, the problem is not
completely resolved since a more recent analysis \cite{tolos03} comes
to the conclusion that the usage of in-medium masses obtained from
coupled channel G-matrix calculations requires extremely low values for
$\rho$ and $T$, i.e. $T\leq 34$ MeV and $\rho \leq 0.02\rho_0$. Such 
freeze-out parameters seem, however, to be far from reality. When
the same in-medium potentials are, on the other hand, applied within
dynamical transport calculations \cite{cassing03} the $ K^-/K^+$ ratio
is fairly well reproduced. From this analysis one has  
to conclude that the statistical approach is at SIS energies not applicable to 
this observable, at least if one accounts also for in-medium effects.

\subsubsection{Thermal freeze-out}
In contrast to the chemical kaon freeze-out which takes place early,
the kaon momentum distributions are strongly influenced by final state
interactions. Both, elastic $KN$ scattering as well as the propagation
in a repulsive mean field from Coulomb and strong interactions
makes the spectra harder \cite{fang94b,wang98c} and influences 
the collective in-plane \cite{ko95,fuchs98,brat97,wang97,li98c} 
and out-of-plane flow \cite{wang97a,shin98}. In the case of the
$K^-$'s thermal and chemical freeze-out are not clearly separated
since elastic and strangeness exchange cross sections are of the same
order of magnitude. Propagation in an attractive mean field tends here
to make the spectra softer.  

The influence of the mean field on the shape 
of the spectra has e.g. 
been studied in Refs. \cite{cassing97a,cassing03,wang98c}. 
Both,  a  repulsive $K^+$ potential as well as an  
attractive $K^-$ potential lead to deviations from a thermal 
spectrum which can be characterized by a radial flow component.  
In \cite{wang98c} the transverse mass 
($m_T = \sqrt{{p_T}^2 + m_{\rm K}^2}$) 
spectra from QMD transport calculations were 
fitted by a radially expanding thermal source 
\beq
\frac{{d^3}N}{d{\phi}dy{m_T}d{m_T}} \sim e^{-(\frac{{\gamma}E}{T}-{\alpha})}
\{ {\gamma}^{2}E - {\gamma}{\alpha}T(\frac{E^2}{p^2}+1) + ({\alpha}T)^{2}\frac{
E^2}{p^2} \}\frac{\sqrt{({\gamma}E+{\alpha}T)^{2}-m^{2}}}{p}
\label{reswidth}
\label{blast}
\eeq
with a common radial kaon velocity $\beta$ = $\upsilon$/c. 
In (\ref{blast}) $E$ = $m_T~{\rm cosh}y$, 
$p = \sqrt{{p_T}^2+{m_T}^2~{\rm sinh}^2y}$,
$\alpha= \gamma\beta p/T$ and $\gamma$ is the Lorentz factor. 
The repulsive potential leads to a depletion of the 
low mass $m_T$ spectrum and creates a characteristic 'shoulder-arm' shape
which gives rise to a radial flow component. E.g. in 
typical Au+Au reactions at 1 AGeV the collective motion 
of the kaons due to the mean field has been found to 
be about 20$\%$ of the thermal motion ($\beta_{\rm coll.}\simeq 0.1$) 
\cite{wang98c}.   $K^-$ mesons exhibit an anlogous 
collective motion in the radial direction. However, 
the attractive potential leads here 
to a characteristic 'concave' structure 
in the transverse mass spectrum \cite{wang98c,cassing03} 
which was attributed to a 'virtual' radial flow in \cite{wang98c}. 
Thus a collective motion could in principle be distinguished from 
the thermal motion which would, however, requires to measure 
low mass spectra with high precision.

\begin{figure}[h]
\begin{minipage}[h]{185mm}
\unitlength1cm
\begin{picture}(8.5,5.0)
\put(4.5,0.){\makebox{\epsfig{file=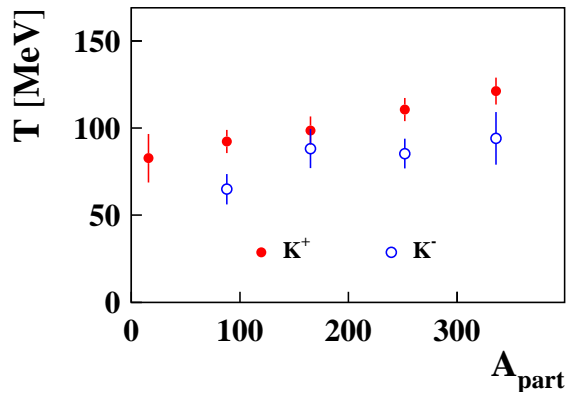,width=8.5cm}}}
\end{picture}
\end{minipage}
\caption{Inverse slope parameters $T$ of $K^\pm$ spectra in Au+Au reactions 
at 1.5 AGeV measured by KaoS \protect\cite{foerster03}. The Figure is 
taken from \protect\cite{foerster05}. 
}
\label{Tkaos_fig}
\end{figure}
Experimental evidence for
different kinetic freeze-out of kaons and antikaons has been reported
in \cite{foerster03} where in the same reaction (Au+Au at 1.5 AGeV) 
$K^-$ spectra were found to be systematically softer than those of the 
$K^+$'s (see Fig. \ref{Tkaos_fig}).  
Also the angular distributions are qualitatively different, 
i.e. strongly forward-backward peaked for $K^+$ and much more isotropic
for $K^-$ \cite{foerster03}. 
Both facts indicate that kaons decouple early from the
reactions dynamics while antikaons suffer strong final state
interactions.


\section{Strangeness production in transport models}

\subsection{Mean field dynamics for strange particles}
Semi-classical transport equations of a Boltzmann type can be derived
from first principle quantum field theory. The starting point is the
non-equilibrium real time Green's functions formalism \cite{ms59}
which leads  after
truncation of the  Green's functions hierarchy at the two-body level, to the
famous Kadanoff-Baym equations \cite{kb62} for the one-body Green 
function. After a gradient expansion of Wigner transforms one arrives
finally at the semi-classical kinetic equation. The kinetic equation
is composed by a drift term, the Vlasov part, which propagates the
one-body density and a collision term which  accounts for two-body
scattering. Derivations of the complete 
kinetic equation in full beauty and complexity are involved 
and can be found in the literature
\cite{btm90,cassing00,kb62,glw80,henning95,greiner98} for both, fermions and bosons. 
To set the context  the relevant steps are briefly reviewed:

Starting from the real time 
Schwinger--Keldysh formalism one obtains a 
matrix ${\underline G}$ of Green functions, (anti-) 
chronologically ordered Green functions 
and correlation functions $G^{>,<}$. The correlation functions 
are defined as 
$G^>(1,1^\prime)= -i <\Psi(1)\overline{\Psi} (1^\prime)>$ and 
$G^<(1,1^\prime) = i <\overline{\Psi}(1)\Psi(1^\prime)>$ using the notation 
$1=(t_{1},{\bf x}_{1})$. The quantity of interest 
is $G^<$ because in the limit $t_1 = t_1'$ it corresponds to a density. 
The Dyson equation in non-equilibrium is given as a matrix 
equation  
\begin{equation}
D(1,1^\prime) {\underline G}(1,1^\prime) 
= {\underline \delta}(1-1^\prime)+\int d2 
{\underline\Sigma}(1,2){\underline G}(2,1^\prime)
\quad ,
\label{Ndyson1}
\eeq
with $D(1,1^\prime) \equiv {\underline \delta}(1-1^\prime) 
(i \gamma_{\mu} \partial_{1}^{\mu} -m) $ for spin-1/2 fermions and 
$D(1,1^\prime) \equiv - {\underline \delta}(1-1^\prime)
(\partial_{1\mu} \partial_{1}^{\mu} +m^2)$ for bosons. The matrix of self-energies 
${\underline\Sigma}$\footnote{In this subsection 
fermions or bosons are not distinguished in the 
notation for their self-energy. Otherwise a bosonic self-energy is denoted 
by $\Pi$.} in Eq. (\ref{Ndyson1}) contains all higher order 
correlations originating from the 
higher order Green functions of the Schwinger--Keldysh hierarchy. When 
the hierarchy is truncated at the two--body level this leads to 
the  T-matrix approximation summing all two--body ladder correlations. 
The kinetic equation for the correlation function $G^<$ is obtained 
by subtracting from the Dyson equation (\ref{Ndyson1}) its adjoint. A 
Wigner transformation then allows to represent the kinetic 
equation in phase space, i.e. $x-k$--space, rather than in coordinate 
space. An essential step is the truncation of the gradient 
expansion of the Wigner transform of products retaining 
only terms of first order in $\hbar$, which neglect memory terms. 
The self-energy, Eq. (\ref{Ndyson1}), is decomposed into scalar 
and vector parts 
\beq
\Sigma^+ = \Sigma^{+}_s - \gamma^\mu \Sigma^{+}_\mu
\label{sigma1}
\eeq
and the real part of $\Sigma$ is used to define effective 
masses and kinetic momenta 
\beqa
m^* = M +  Re\Sigma^{+}_s (x,k)
\quad , \quad
k^{*}_\mu = k_\mu + Re\Sigma^{+}_\mu (x,k)
\label{pstar}
\eeqa
of the dressed particles in the nuclear medium. 
$\Sigma^{+}$ is the retarded self-energy constructed by the 
difference of the corresponding correlation functions 
$G^{\pm}(1,1^\prime) = \theta(\pm(t_1-t_{1^\prime}))\left[G^>(1,1^\prime)
-G^<(1,1^\prime)\right]$ \cite{bm90}. The 
Dirac structure of the correlation functions $G^{>,<}$ can  
be separated off by a decomposition into a scalar spectral 
function $a$, a scalar distribution function $f$ and the projector 
onto positive energy states $\Lambda^+$ 
\beqa
G^< (x,k) &\propto & i \Lambda^+~ a(x,k) f(x,k) 
\label{NGF5}
\\
G^> (x,k) &\propto & -i \Lambda^+~ a(x,k) \left[ 1\mp f(x,k)\right] 
\quad .
\label{NGF6}
\eeqa
The $\mp$ sign in Eq. (\ref{NGF6}) stands for fermions or bosons, 
respectivly. In the case of nucleons the projector is given 
by $\Lambda^+ = \left( \not k^* + m^* \right)/2 m^*  $.  
In an essential approximation which concerns the   
the spectral properties of the hadrons is 
the quasiparticle approximation. It is valid in the limit of a 
small imaginary part of the self energy ($Im\Sigma^+ << Re\Sigma^+$). 
The spectral 
function then reduces to the mass shell constraint 
$a (x,k) = 2\pi \delta \ls k^{*2} - m^{*2}\rs  2 \Theta (k_{0}^*)$ 
which sets the energy on the mass shell 
$k_{0}^* = E^* (x,{\bf k}) =\sqrt{ {\bf k}^{*2} + m^{*2} }$. Thus, the number of 
variables of the distribution function $f(x,k)$ is reduced from eight to seven
\beq
a(x,k) f(x,k) = 2\pi \delta [k^{*2} - m^{*2}] 2\Theta (k_{0}^* ) f(x,{\bf k}) 
\label{dist2}
\eeq
which simplifies considerably practical implementations. As can be 
seen from Fig. \ref{spectral_fig} the quasiparticle approximation is 
well justified for kaons but less applicable to antikaons.

In many cases the kinetic equation is further treated in the Hartree approximation 
which implies to neglect the explicit momentum dependence 
of the mean field,  i.e. $Re\Sigma^{+} = Re\Sigma^{+}_H (x)$. 
Then the resulting kinetic equation can completely be 
formulated in terms of kinetic momenta 
instead of canonical momenta  \cite{bm90}
\beqa
&&  \left[k^{*\mu} \partial_{\mu}^x  + \left( k^{*}_{\nu} F^{\mu\nu}     
+ m^* \partial^{\mu}_x m^* \right) 
\partial^{k^*}_{\mu} \right] (af) (x,k^* ) 
\nonumber\\
&=&  \frac{1}{2} \int \frac{d^4 k_{2}}{(2\pi)^4} \frac{d^4 k_{3}}{(2\pi)^4}
             \frac{d^4 k_{4}}{(2\pi)^4} 
             a(x,k)  a(x,k_2)  a(x,k_3)  a(x,k_4) W(kk_2|k_3 k_4) 
\nonumber \\    
&\times& (2\pi)^4 \delta^4 \left(k + k_{2} -k_{3} - k_{4} \right)    
\left[ f(x,k_3) f(x,k_4) \ls 1\mp  f(x,k) \rs \ls 1\mp f(x,k_2) \rs \right.  
\nonumber \\  
&-&   \left. f(x,k) f(x,k_2) 
\ls 1\mp  f(x,k_3) \rs \ls 1\mp f(x,k_4) \rs \right]    
\quad .
\label{TP5}
\eeqa
Eq. (\ref{TP5}) resembles the well known transport equation of a 
Boltzmann--Uehling--Uhlenbeck type. The left hand side is a drift term driven  
by the mean field via the kinetic momenta $k^{*}$, the field strength tensor 
$F^{\mu\nu} (x) = \p^{\nu}_x  Re\Sigma^{+\mu}_H (x) 
                -\p^{\mu}_x  Re\Sigma^{+\nu}_H (x) $, 
and the effective mass $m^*$. The right hand side is a collision integral which 
contains the transition rate $W$ or equally the in--medium cross section given by 
$(k^* + k_{2}^*)^2 d\sigma /d\Omega(k,k_2) = W(kk_2|k_3 k_4)$. The 
collision term contains Pauli-blocking, respectively Bose enhancement 
factors $(1\mp f)$ for the final states.  

As soon as one has to deal with different particle species which are 
interacting, e.g. via production and absorption processes, the 
transport problem becomes a coupled channel problem. This means that one has 
to solve a seperate transport equation for each degree of freedom, 
but these are coupled to each other by their collisions intergrals 
and the mean fields. 

\subsubsection{Transport equation for kaons}
A full transport equation for kaons can in principle be derived as 
outlined above. However, since we are only interested in 
the structure of the relativistic equations of motion,  
we restrict the discussion in the 
following to the mean field level and the left hand side 
of the transport equation (\ref{TP5}), i.e. the Vlasov equation. 

The Vlasov equation can
easily be derived from the Klein-Gordon equation for the kaon field. 
For simplicity we restrict the discussion to $K^+$ mesons since the 
derivation for $K^-$'s is fully analogous. 
Starting from the Klein-Gordon equation (\ref{kg2}) the field 
equations for the kaon field and its adjoint read
\beqa
&&\left[ \left( \partial_\mu + i V_\mu \right)^2  + \msq \right] 
\phi_{\mathrm{K} } (x_1) = 0 
\label{A1}\\
&&\left[ \left( \partial_\mu - i V_\mu \right)^2  + \msq \right] 
\phi_{\mathrm{K} }^* (x_2) = 0 
\quad . 
\label{A2}
\eeqa
In mean field approximation to the chiral $SU(3)$ Lagrangian as discussed 
in Chap. 2 a vector potential with alternating sign for kaons and 
antikaons, i.e. $V_\mu = \pm \frac{3}{8\fps} j_\mu $ for $K^\pm$ and 
an attractive scalar part occur. The latter can be absorbed 
into the effective mass 
$\ms = \sqrt{ \mks - \frac{\Sigma_{\mathrm{KN}}}{\fps} \rhos 
     + V_\mu V^\mu } $ which is equal for kaons and antikaons. 
Here one has already made use of the quasiparticle approximation since 
Eqs. (\ref{A1},\ref{A2}) contain only the real part of the kaon 
self-energy $\Pi_K$ (\ref{uopt3}). 
Multiplying equations (\ref{A1}) and (\ref{A2}) by the respective 
adjoint field and subtracting them, yields
\beqa
\left[ \partial^{x_1}_{\mu}\partial_{x_1}^{\mu} - 
 \partial^{x_2}_{\mu}\partial_{x_2}^{\mu} + 
2i \left( V_\mu (x_1) \partial_{x_1}^{\mu} + 
          V_\mu (x_2) \partial_{x_2}^{\mu} \right) 
 - V_\mu (x_1 )V^\mu (x_1 ) \right.
\nonumber \\
\left.
+ V_\mu (x_2 ) V^\mu (x_2 ) 
+ \msq (x_1 ) -\msq (x_2 )  \right] 
\phi (x_1) \phi^* (x_2)= 0  ~~~.
\label{A3}
\eeqa
Introducing center-of-mass and relative coordinates 
\bdm
x = \frac{1}{2}\left( x_1 + x_2 \right) \quad , \quad r =  x_1 - x_2
\edm 
one has 
\bdm  
\partial^{x_1}_{\mu}\partial_{x_1}^{\mu} - 
 \partial^{x_2}_{\mu}\partial_{x_2}^{\mu} = 
2 \partial^{x}_{\mu}\partial_{r}^{\mu}
\quad . 
\edm
The fields are now expanded to first order in gradients 
\beqa
V_\mu (x_1) \partial_{x_1}^{\mu} + V_\mu (x_2) \partial_{x_2}^{\mu}
&=& \cosh\left( \frac{r}{2} \cdot \partial_x \right) 
V_\mu (x) \partial_{x}^{\mu} + 
2 \sinh\left( \frac{r}{2} \cdot \partial_x \right) 
V_\mu (x) \partial_{r}^{\mu}
\nonumber\\
&\approx&  
V_\mu (x) \partial_{x}^{\mu} + 
\left(r \cdot \partial_x V_\mu (x) \right) \partial^{\mu}_r
\label{A4}
\eeqa 
where $r \cdot \partial_x$ denotes a four-vector product. 
Similar one obtains 
\beqa
\msq (x_1 ) -\msq (x_2 ) &\approx& 2 \ms (x) 
\left( r \cdot \partial_x \ms (x)\right)
\nonumber \\
V_\mu (x_1 )V^\mu (x_1 ) - V_\mu (x_2 ) V^\mu (x_2 ) &\approx&
2 V_\mu (x)\left( r \cdot \partial_R V^\mu (x) \right)
\label{A5}
\quad .
\eeqa
Inserting Eqs. (\ref{A4}), (\ref{A5}) into (\ref{A2}) gives 
\beqa
&&\biggl[ \partial_{\mu}^x \partial^{\mu}_r 
+ i \biggl( V_\mu(x) \partial^{\mu}_x  + 
\left( r \cdot \partial_x V_\mu (x)\right)\partial_{r}^\mu \biggr) 
+ \ms (x) \left(r \cdot \partial_x \ms (x)\right) 
\biggr.
\nonumber \\
&& \biggl. 
- V_\mu (x) \left(r \cdot \partial_x V^\mu (x)\right) \biggr]
\phi \left(x+\frac{r}{2}\right) \phi^* \left(x-\frac{r}{2}\right) =0~~.
\label{A6}
\eeqa
Integration of Eq. (\ref{A6}) over $\int d^4 r e^{ik\cdot r}$ 
yields the Vlasov equation 
\beq
\left[ k^{*\mu} \partial_{\mu}^x + 
\left( k_\nu \partial^{\mu}_x V^\nu 
- V_\nu \partial^{\mu}_x V^\nu +  \ms \partial^{\mu}_x \ms \right)
\partial_{\mu}^k
\right] (af)(x,k) = 0
\label{A7}
\eeq
for the scalar phase-space 
distribution $f(x,k)$. The phase-space distribution is defined as the 
Wigner transform of the density matrix $\phi\phi^*$
\beq
f(x,k) = \int d^4 r e^{ik\cdot r}
\phi \left(x+\frac{r}{2}\right) \phi^* \left(x-\frac{r}{2}\right)
\quad .
\eeq
Introducing the field strength tensor 
$F^{\mu\nu} = \partial^{\mu} V^\nu -\partial^{\nu} V^\mu $ and 
kinetic momenta $k^{*}_{\mu} = k_{\mu} \mp V_\mu $, Eq. (\ref{A7}) 
can now be written as 
\beq 
\left[ k^{*\mu} \partial_{\mu}^x + 
\left( k_{\nu}^* F^{\mu\nu} +  \ms \partial^{\mu}_x \ms \right)
\partial_{\mu}^{k^*}
\right] f(x,k^* ) = 0
\quad .
\label{A8}
\eeq
The corresponding equation for $K^-$ mesons is obtained by 
replacing in Eq. (\ref{A8}) the field strength tensor $F^{\mu\nu}$ 
with $F^{\nu\mu}=-F^{\mu\nu}$.

Equation  (\ref{A8}) has exactly the same structure 
as the relativistic Vlasov equation for nucleons 
 which is obtained from the projection 
onto the scalar component of the  Wigner density matrix for
spin-$\frac{1}{2}$ fermions in the 
spinor representation. Thus, in the semi-classical limit the information 
of the fermionic/bosonic character of the particles is lost at the
mean field level. In the collision integral differences are retained,
e.g. fermionic Pauli-blocking factors are replaced by Bose-enhancement
factors.

\subsubsection{Equations of motion}
From the Vlasov eq.  (\ref{A8}) classical equations of motion are 
obtained  in the quasi-particle limit which implies to  put the 
phase space distribution on mass shell  
\beq
 f(x,k^* ) \equiv
 f({\bf q},t;{\bf k}^* ) ~ \delta\left( k_{0}^* -  E^* ({\bf q},{\bf k})\right)~~,
\label{quasi1}
\eeq
with 
\beq
E^*  = \sqrt{ {\bf k}^{*2} + \msq } ~~~.
\eeq
Inserting (\ref{quasi1}) into (\ref{A8}) yields the covariant 
equations of motion 
\beqa
\frac{ d  q^\mu}{ d\tau} &=& \frac{k^{*\mu}}{\ms} = u^\mu
\\
\frac{ d  k^{*\mu}}{ d\tau} &=& \frac{k^{*}_{\nu}}{\ms} F^{\mu\nu} 
+\partial^\mu \ms ~~.
\label{como}
\eeqa
Again the classical equations of motion are identical 
to those for nucleons,  obtained  
e.g., in relativistic mean field theory \cite{btm90}. 
Only the different structure of the 
effective mass reflects the bosonic character of the kaons. 

Alternatively the equations of motion can also be obtained from the dispersion 
relation $E = k_0 = E^* \pm V_0 $ through the 
Hamilton equations 
\beqa
\frac{ d{\bf q}}{ d  t} = \frac{\partial E}{\partial {\bf k}} 
\quad, \quad  
\frac{d {\bf k}}{ d t} = -\frac{\partial E}{\partial {\bf q}} ~~.
\label{ham1}
\eeqa
Thus one obtains
\beqa 
\frac{ d {\bf q}}{ d t} &=& \frac{{\bf k}^*}{E^*} \\
\frac{d {\bf k}}{d t}   &=& - \frac{\ms}{E^*} 
\frac{\partial \ms }{\partial {\bf q}} \mp 
\frac{\partial V^0 }{\partial {\bf q}} 
\pm \frac{ {\bf k}_{i}^*}{E^*} 
\frac{\partial  {\bf V}_{i}}{\partial {\bf q}} 
\quad .
\label{ham2} 
\eeqa
Eq. (\ref{ham2}) provide the non-covariant version of (\ref{como}).  
The term proportional to the space-like component of the vector 
potential gives rise to a momentum dependence in Eq. (\ref{ham2}) 
which can be attributed to a Lorentz force term. This structure 
becomes more evident when Eq. (\ref{ham2}) is rewritten in terms 
of kinetic momenta
\beqa
\frac{ d{\bf k}^*}{d t} &=& - \frac{E^*}{\ms} 
\frac{\partial \ms }{\partial {\bf q}} \mp 
\frac{\partial V^0 }{\partial {\bf q}} 
\pm \frac{{\bf k}^*}{E^*} \times 
\left( \frac{\partial}{\partial {\bf q}} \times {\bf V} \right)
 = -\frac{\partial U_K }{\partial {\bf q}}
\pm {\bf v}_{i} \frac{\partial  {\bf V}_{i}}{\partial {\bf q}} 
\label{lorentz}
\eeqa
with ${\bf v} = {\bf k}^* / E^*$ the kaon velocity. 
\begin{figure}[h]
\unitlength1cm
\begin{picture}(8.,4.5)
\put(4.5,0){\makebox{\epsfig{file=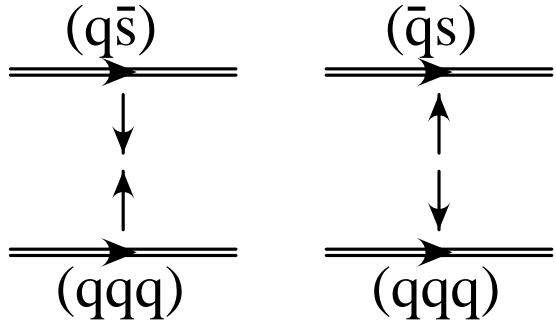,width=8.0cm}}}
\end{picture}
\caption{Action of the Lorentz force between nucleons $(qqq)$ 
and kaons. The force acts opposite for $K^- (s{\bar q})$ and 
$K^+ ({\bar s}q)$.
}
\label{fig_lorentz}
\end{figure}
The appearance of the velocity dependent 
(${\bf v} = {\bf k}^* / E^*$) Lorentz force in 
Eqs. (\ref{como})-(\ref{lorentz}) is a genuine feature 
of the relativistic kaon dynamics as soon as a vector field is involved. 
In addition to the trivial $|\bf k |$ momentum dependence of the 
optical potential the $\bf k $ dependent part of the 
interaction appears in reference frames where the spatial components  
${\bf V}$ and  $\bf k $ do not vanish. Hence, in 
nuclear matter at rest or p+A reactions no Lorentz 
force is present. The importance of this interaction for the nucleon 
dynamics in heavy ion collisions is well known 
since a long time, see e.g. \cite{blaettel93}. For the 
kaon dynamics it was first discussed by Fuchs et al. 
\cite{fuchs98}. The physical picture behind is completely 
analogous to electrodynamics: assuming that the 
interaction between kaons and nucleons is determined by the 
non-strange quark content of the hadrons, two parallel 
currents are attracted, two anti-parallel currents are repelled. 
Hence the Lorentz force acts opposite for $K^- (s{\bar q})$ and 
$K^+ ({\bar s}q)$ as illustrated in Fig.\ref{fig_lorentz}.  
\begin{figure}[h]
\unitlength1cm
\begin{picture}(9.,7.5)
\put(3.5,0){\makebox{\epsfig{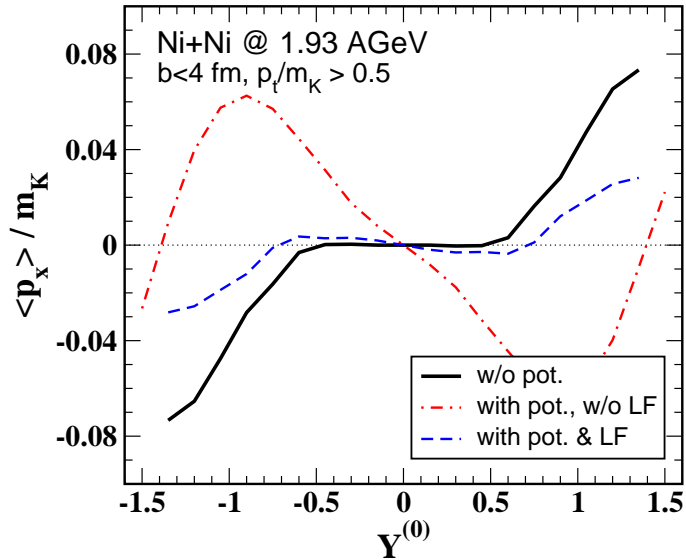}}}
\end{picture}
\caption{Transverse $K^+$ flow 
in 1.93 AGeV $^{58}$Ni + $^{58}$Ni reactions 
 at impact parameter b$\leq 4$ fm. 
 The calculations are performed without and including the kaon mean
 field dynamics. In the latter case results without and including 
the contribution from the  Lorentz-force (LF) are shown.
}
\label{fig_kflow1}
\end{figure}
In heavy ion reactions kaons are primordially produced in the 
early non-equilibrium phase. In contrast to p+A reactions, 
there the baryon vector current is in general locally non-zero 
even in the c.m. frame of the colliding nuclei. Thus a 
Lorentz force will be acting on the kaon. As discussed in
\cite{fuchs98} the Lorentz force counterbalances to large 
extent the influence of the time-like $V_0$ component of the vector 
field acting on the $K^+$ mesons. This cancellation can easily 
be understood from Eq. (\ref{ham2}). 
The vector field is proportional to the baryon current 
$ j_\mu = (\rho_B, {\bf u}\rho_B) $ where ${\bf u}$ denotes the 
streaming velocity of the surrounding nucleons. If 
${\bf u}$ is locally constant, then the 
total contribution of the vector field in Eq. (\ref{ham2}) 
can be written as 
$ \mp \frac{3}{8\fps} 
\left( 1 - |{\bf v}| |{\bf u}| \cos\Theta \right) 
\partial \rho_B /\partial {\bf q}$. 
Now the angle $\Theta$ between the kaon and the baryon 
streaming velocities determines the influence of the 
Lorentz force. For  $\Theta =0$ one obtains an attraction (repulsion) 
for $K^+$ ($K^-$) and the opposite in the case of $\Theta =180^o$. 

Such dynamics have e.g. consequences for the kaon transverse flow. 
This is illustrated in Fig. \ref{fig_kflow1} where 
the $K^+$ transverse flow in 
Ni+Ni reactions at 1.93 AGeV is shown for various cases. 
It is seen that the strongly repulsive 
static $K^+$ potential $U_K$ in (\ref{lorentz}) 
tends to push the kaons away from the spectator matter, leading to a 
strong anti-flow around mid-rapidity. 
The Lorentz force pulls the kaons back to 
the spectator matter, resulting finally in a flow pattern 
which is close to that obtained without any in-medium potential. 
However, the magnitude of the cancellation effect and the final flow 
pattern depends also on the strength of the vector potential 
and the interplay between repulsive vector and the 
attractive scalar potential. 
In the early works on kaon flow \cite{ko95,wang97} the Lorentz force
has been disregarded which led to an overestimation of the vector
repulsion. One has, however, to keep in mind that correlations beyond 
mean field lead to an explicit momentum dependence of the scalar 
and vector potentials. Such an explicit momentum 
dependence reduces e.g. in the case of the nucleons the size of 
the scalar and vector potentials and, correspondingly 
the  Lorentz force which softens the nucleon optical potential. 
A detailed discussion of the kaon flow and a 
comparison to data will be
performed in chapter 5.  

\subsection{Off-shell transport}

Essential for the validity of the classical equations of motion is the 
quasi-particle approximation (QPA) which assumes that the spectral
strength of a hadron 
is concentrated around its quasi-particle pole. Since kaons
have practically zero width this condition is readily fulfilled in
the vacuum. Particle widths can, however, dramatically change in a
dense hadronic environment. To first order in density the in-medium width 
of a hadron in nuclear matter can be estimated by the collision width 
$\Gamma^{\rm tot}=\Gamma^{\rm vac} + \Gamma^{\rm coll}$, 
\beq
\Gamma^{\rm coll} = \gamma v \sigma \rho_B~~,
\label{gcoll}
\eeq
with $v$ the hadron velocity relative to the surrounding matter and $\sigma$
the total hadron-nucleon cross section. Due to their long mean free path the 
collisional broadening of kaons is small. The kaon-nucleon cross section is
dominated by its elastic part which is about 10 mb and the majority of kaons
is produced with relatively small velocity. Thus for $K^+$'s in-medium widths 
can be expected to be small and the quasi-particle-approximation
 appears to be justified. For antikaons the situation is 
completely different. The strong coupling
of the $K^- N$ system to the $\Lambda(1405)$ and $\Sigma(1385)$
hyperons leads not only to a shift of the quasi-particle pole but also
to a strong distortion of the $K^-$ spectral function \cite{oset01,lutz02}  
(see Fig. \ref{spectral_fig}). The application of mean
field dynamics is therefore questionable. 

First attempts to account for in-medium spectral properties in a more
consistent way within the framework of semi-classical 
transport models were already made some time ago. A fully consistent 
treatment of the off-shell dynamics, i.e. a solution of the full 
Kadanoff-Baym equations has up to now only been performed for 
toy models and simplified geometries. Danielewicz \cite{dani84} was 
the first to solve the quantum evolution equations for the correlation 
functions $G^{<,>}$ (\ref{NGF5},\ref{NGF6}) for an infinite cylindrically 
symmetric system using a simplified two-body potential of Gaussian form. 
The work of K\"ohler \cite{koehler95} 
who solved the Kadanoff-Baym equations under the same conditions on a 
Cartesian momentum space grid can be viewed as the first step 
towards a realization of quantum transport on a lattice. To develop 
a lattice quantum transport for non-uniform systems and realistic 
interactions will be one of the future challenges in theoretical 
heavy ion physics. 
\begin{figure}[h]
\begin{minipage}[h]{185mm}
\unitlength1cm
\begin{picture}(18.5,7.0)
\put(0.,0.){\makebox{\epsfig{file=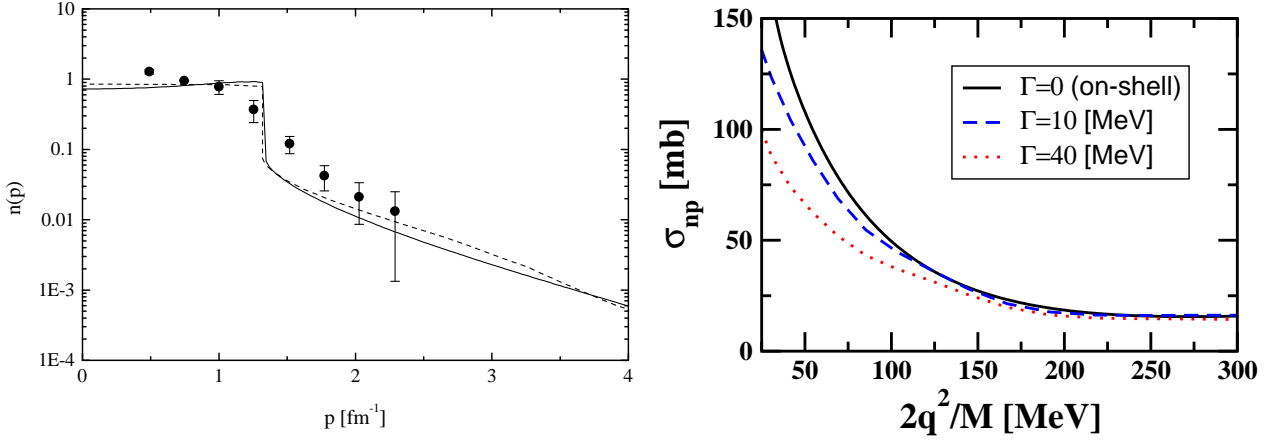,width=9.5cm}}}
\put(9.0,0.0){\makebox{\epsfig{file=sig_offNP.eps,width=8.0cm}}}
\end{picture}
\end{minipage}
\caption{Left: momentum distribution in nuclear matter at $\rho_0$ 
from a self-consistent off-shell transport (solid) is compared to 
a many-body calculation \protect\cite{benhar} (dotted) 
and to data from electron 
scattering \protect\cite{ciofi91}. The figure is taken from 
\protect\cite{lehr00}. Right: Off-shell dependence of the 
elastic in-medium $np$ cross section at $\rho_0$. Off-shell DBHF 
matrix elements are folded over nucleon spectral functions of 
different width. Results are from \protect\cite{offcs}.   
}
\label{offshell_fig1} 
\end{figure}

On the other hand, substantial progress has been  made in the recent 
years to map 
part of the off-shell dynamics on a modified test-particle formalism 
\cite{cassing00,greiner98,lehr00,cassing00b}. This allows to apply 
off-shell dynamics, although in a simplified form, to the complex 
space time evolution of a heavy ion reaction. That such type 
of approach is able to describe essential features of nuclear correlations 
beyond mean field has e.g. been demonstrated in Ref. \cite{lehr00}:  
in the left part of Fig. \ref{offshell_fig1} 
the nucleon phase space distribution $f({\bf k})$ in nuclear 
matter resulting from a self-consistent off-shell calculation is 
compared to a microscopic many-body calculation \cite{benhar} and to 
electron scattering data \cite{ciofi91}. Although these results were 
obtained with adjusted on-shell scattering matrix elements the 
agreement with the realistic distribution function is remarkably 
good. 

The present knowledge of off-shell matrix elements is rather 
limited. Except of particular cases of interest, e.g 
$\pi Y \leftrightarrow K^- N$ matrix elements \cite{tolos01,tolos02}, 
theoretical 
investigations are scarce. In Ref. \cite{offcs} in-medium half-off-shell 
matrix elements for elastic nucleon-nucleon scattering have 
been determined from relativistic DBHF many-body calculations based 
on realistic meson exchange potentials. 
The results indicate that a smooth transition from on-shell to 
off-shell matrix elements is possible and the usage of 
on-shell cross sections is justified for $NN$ scattering. 
The right part of Fig. \ref{offshell_fig1}  displays  the elastic off-shell 
in-medium neutron-proton cross section at $\rho_0$, folded over nucleon 
spectral functions of different width, as a function of the c.m. momentum. 
The mean $NN$ cross section is in general 
only little affected which  justifies the standard on-shell 
transport approach for the overall reaction dynamics. 
\begin{figure}[h]
\unitlength1cm
\begin{picture}(9.,9.5)
\put(3.5,0){\makebox{\epsfig{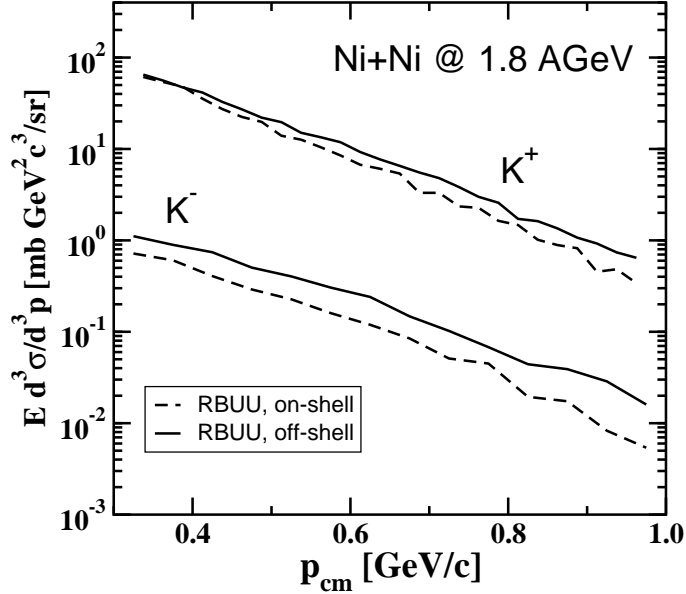}}}
\end{picture}
\caption{Influence of hadron off-shell propagation on $K^\pm$ spectra 
in Ni+Ni reactions at 1.8 AGeV. On-shell cross sections have been 
used for the strangeness production and exchange reactions in the 
RBUU calculation of \protect\cite{cassing00b}.
}
\label{offshell_fig2}
\end{figure}

The question to what degree a depletion of the Fermi surface due to 
particle-hole excitations and the high momentum tails of the nuclear 
spectral functions will affect subthreshold particle 
production is, however, not so obvious to answer. The high momentum tails 
correspond to deeply bound states which are off-shell and to treat 
such states in a standard transport 
approach like on-shell quasi-particles 
would violate energy-momentum conservation. The contribution 
of the nuclear short-range correlations to subthreshold $K^+$ production 
in p+A reactions have been estimated in \cite{saturne96}, based on the 
spectral distributions of Benhar et al. \cite{benhar} shown in 
Fig. \ref{offshell_fig1}. The removal energy 
for a high momentum state compensates the naively expected energy 
gain and the short-range 
correlations do therefore not significantly contribute to subthreshold 
particle production \cite{saturne96}. 
The situation changes, however, when the medium 
is heated up and high momentum particles become on-shell or when 
the spectral distributions of the produced hadrons themselves 
are broadened.   

  
Possible implications of off-shell dynamics connected to 
a spectral broadening of the kaons 
have first been studied by Cassing and Juchem 
\cite{cassing00b,cassing03}. These calculations were based on 
{\it on-shell} cross sections and hadronic spectral functions 
($N,\pi,\Delta,K^+,K^-$) which are exclusively determined by the 
collisional broadening (\ref{gcoll}) of these particles. Nucleons 
and pions were found to be practically unaffected by the 
off-shell propagation while the high momentum tails of 
the $K^+$ spectra are enhanced. But even at subthreshold energies 
the effect on the $K^+$ yield was found to be moderate which 
justifies the standard quasi-particle approach and is in line with the 
observations of Ref. \cite{saturne96}. As expected, 
the off-shell effects were found to be much stronger for 
antikaons. $K^-$'s aquire a large spectral width in the medium 
and it is energetically favorable to produce them off-shell in the 
low momentum tails of their spectral functions. During the propagation 
through the medium these particles become on-shell as they reach the 
detector. Obviously this transition must take place in a controlled 
way which is guaranteed within consistent off-shell dynamics 
\cite{cassing00b,cassing03}.  As can be seen from 
Fig. \ref{offshell_fig2} where the corresponding $K^\pm$ spectra 
in Ni+Ni reactions at 1.8 AGeV are shown, the 
collisional broadening leads to an enhancement of the 
$K^-$ yield by about a factor of 2-3 compared to the on-shell 
treatment. However, in particular for $K^-$ the off-shell 
behavior of the in-medium strangeness exchange cross sections $\pi Y 
\longleftrightarrow N K^-$ plays an important role \cite{cassing03}. 
Results based on off-shell cross sections will be discussed in the 
following chapter.  

\subsection{Collisions}
Transport models treat particle production at subthreshold energies
usually in a perturbative way. In each individual hadron-hadron
collision the incident energy $\sqrt{s}$ must be above the production 
threshold. However, such processes are rare and the corresponding
cross sections lie in the vicinity of thresholds usually several
orders of magnitude below the total cross sections. Since 
Monte-Carlo methods which are applied to solve the collision 
integral of the kinetic 
equation \cite{bg88} select the actual reaction channel $i$ with 
probability 
\beqa
P_i = \frac{\sigma_i (\sqrt{s})}{\sigma_{\rm tot} (\sqrt{s})}~,~~~
\sum_{\rm all~channels} P_i =1~~,
\eeqa
 meson production is at subthreshold energies an extremely rare
 process. To obtain nevertheless reasonable statistics, for such
 reactions a perturbative treatment is applied
 \cite{lang92,fang93,fang94}. In short, this means that the meson is produced
 if kinematically allowed. Phase space and Pauli-blocking
 factors for the final states are determined and the produced meson
 $k$ gets the probability 
\beq
P_k = \sum_X   \frac{\sigma^{Y\longrightarrow K^+ X}}{\sigma_{\rm tot}
} (\sqrt{s})
\eeq
assigned. $Y$ denotes the initial state and the sum runs over all 
possible final states. From the set of possible final states the
actual reaction channel $k$, i.e. $Y\longrightarrow K^+ X_k$, is selected by 
Monte-Carlo. Thus, e.g. the total $K^+$ multiplicity 
is given 
\beq
N_{K^+} = \sum_{{\rm all}~K^+} P_k~~.
\eeq
Finally the initial  particles are reset to their initial values 
and the cascade
continues as if no such reaction would have taken place. The
perturbatively produced mesons are propagated as 'virtual' particles
parallel to the ongoing reaction dynamics in such a way that they are
influenced by the bulk of particles, however, without any feed back to
the global reaction dynamics. The
perturbative treatment is justified as long as the considered particles are 
rare enough not to influence the average reaction dynamics.
\subsubsection{Elementary cross sections}
As discussed in the previous chapter, 
the elementary production mechanism for $K^+$ mesons can be divided into 
baryon induced reactions $BB\longrightarrow BYK^+ $ 
($B$ stands either for a nucleon or a $\Delta$--resonance and $Y$ for 
a $\Lambda$ or a $\Sigma$ hyperon, respectively) and processes 
$\pi B\longrightarrow YK^+ $ induced by pion absorption. 
Here and in the previous works of the T\"ubingen group 
\cite{fuchs98,fuchs97b,wang97a,wang97,wang99,wang99b,fuchs01,zheng03} 
for pion induced reactions the elementary cross sections of 
Tsushima et al. \cite{tuebingen} have been used. 
The cross sections are derived within the resonance 
model in Born approximation, including all baryonic resonances   
with masses below 2 GeV as intermediate states. In the 
meantime they are standardly 
and applied in most transport calculations 
\cite{cassing97a,brat97b,cassing99,ko95,li97,li98,nantes99,hartnack01}.
\begin{figure}[h]
\unitlength1cm
\begin{picture}(12.,12.0)
\put(2.0,0){\makebox{\epsfig{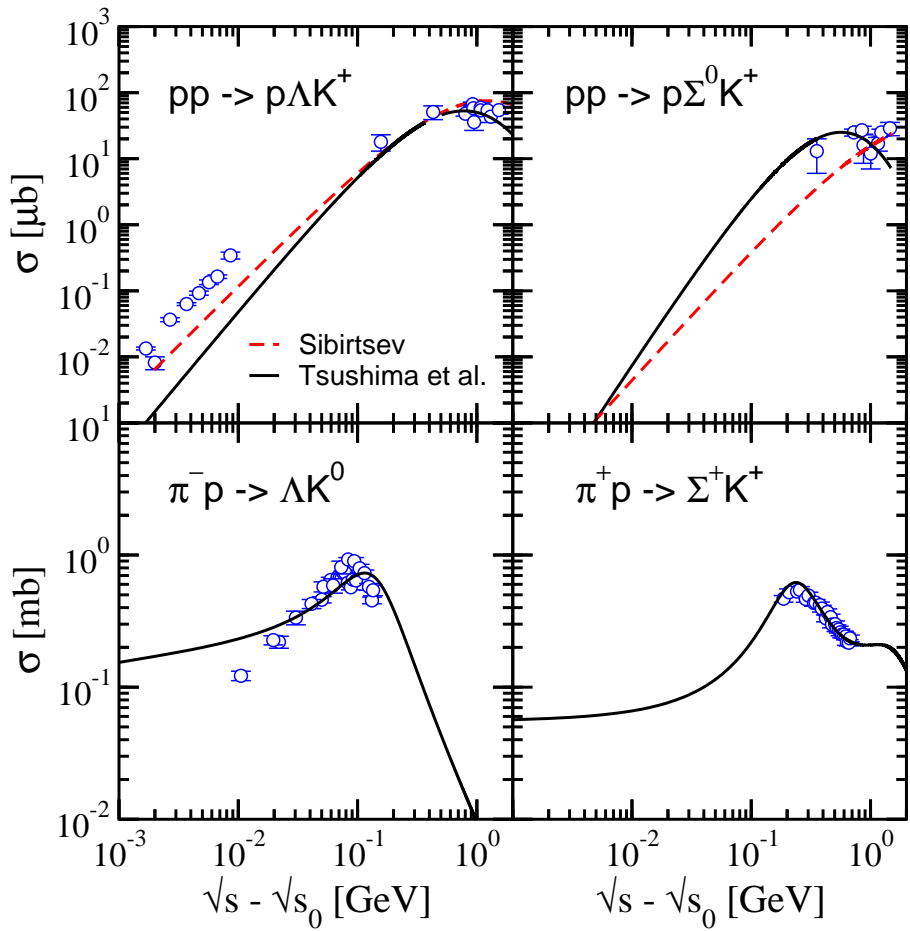}}}
\end{picture}
\caption{Elementary cross section for kaon production. 
Resonance model calculations from Tsushima et al. 
\protect\cite{tuebingen,tsushima99} and 
OBE  calculations from Sibirtsev \protect\cite{sibirtsev95} are 
compared to data \protect\cite{cosy11,flaminio,jones71,candlin83}.
}
\label{fig_cross1}
\end{figure}

For baryon induced channels there are presently several 
different parameterizations on the market: Those from 
Li and Ko \cite{li97} used by the Texas and Stony Brook groups;  
those given by Sibirtsev \cite{sibirtsev95} which are used by 
the Giessen group \cite{cassing97a,cassing99} 
and have been applied in some of our previous works 
\cite{fuchs97b,wang97,wang97a,fuchs98};  
and more recent calculations within the resonance model by 
Tsushima et al. \cite{tsushima97,sibirtsev98,tsushima99}. 
In principle it would be desirable to 
base the cross sections for both, pion and baryon induced reactions 
on the same model \cite{tuebingen,tsushima99}. However, the 
resonance model \cite{tsushima99} under-predicts the COSY-11 data 
\cite{cosy11} for $pp \longrightarrow p\Lambda K^+ $ at 
threshold. Thus, here and in \cite{fuchs01,zheng03}  
the cross sections from \cite{sibirtsev95} were used   
for $N N \longrightarrow N\Lambda(\Sigma)K^+ $ and those of ref. \cite{tsushima99} 
for reactions involving nucleon resonances 
($N N \longrightarrow \Delta YK^+,~ N \Delta \longrightarrow BYK^+ $ and 
$\Delta \Delta \longrightarrow BYK^+ $). Figure \ref{fig_cross1}
compares the model predictions for $pp$ and $\pi p$ channels 
to available data. In the vicinity of the threshold the nucleon-hyperon
final state interaction is attractive and strong which tends to increase the cross
section \cite{watson52}. Since both, the OBE as well as the resonance model 
calculations do not account for FSI effects, the COSY data are under-predicted 
near threshold \cite{cosy11}. As pointed out in \cite{hanhart98} the
modification of the energy dependence of the cross section due to FSI 
is determined by the on-shell T-matrix. Since in the semi-classical transport
models on-shell potentials enter into the evaluation of the scattering
amplitude, e.g. by shifts of thresholds, some FSI contribution is effectively 
included which justifies the usage of the cross sections from
\cite{sibirtsev95}.  

\begin{figure}[h]
\unitlength1cm
\begin{picture}(12.,12.0)
\put(2.0,0){\makebox{\epsfig{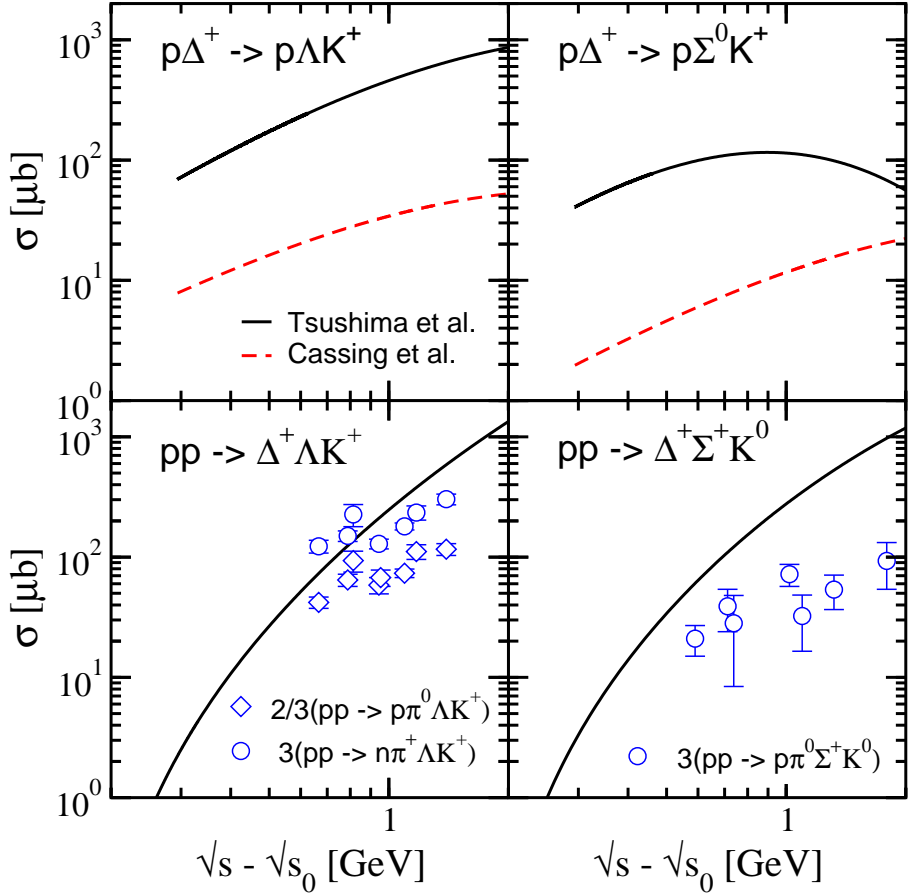}}}
\end{picture}
\caption{Cross sections involving $\Delta$ resonances. Upper part:  
for the experimentally unknown 
$p\Delta \longrightarrow p Y K^+$ cross sections resonance model calculations 
\protect\cite{tsushima99} are compared to the parameterizations  
of \protect\cite{cassing99}. Lower part: Resonance model calculations 
\protect\cite{tsushima99} for $pp \longrightarrow \Delta Y K$ are 
compared to $pp \longrightarrow N\pi Y K$ data \cite{flaminio}. 
}
\label{fig_cross2}
\end{figure}
In contrast to the $NN$ channel, cross sections with an incoming $\Delta$ 
resonance $N\Delta \longrightarrow BYK^+ $ are not constrained by data. 
Here the largest differences
between various parameterizations occur. Fig. \ref{fig_cross2}
compares the resonance model results \cite{tsushima99} with 
parameterizations based on isotopic relations for $pp$
\cite{cassing99,brat97}. The latter cross sections turn out to be
about one order of magnitude smaller. This difference is crucial since
the $N\Delta  \longrightarrow K^+ X$ is one of the primary production channels
for kaons. It explains why there existed a significant 
inconsistency between transport
calculations from different groups concerning the total kaon multiplicity in 
heavy ion reactions. Kaon yields obtained from the Giessen group
\cite{cassing99,brat97,cassing97a} turned out to be systematically smaller
than those of the Texas and Stony Brook \cite{ko96,li97}, the Nantes
\cite{hartnack01} and the T\"ubingen \cite{fuchs01,zheng03} groups. The source
of this discrepancy, which is crucial with respect of an interpretation of
corresponding data in terms of kaon in-medium potentials, could finally be
traced back to the usage of the different cross sections for kaon production 
in $N\Delta$ collisions \cite{hartnack01}.

Reactions with multi-particle final states, such as 
$NN \longrightarrow N\pi Y K$, $NN \longrightarrow N\pi\pi Y K$, 
are subdominant at subthreshold energies. If taken into account they
can either be treated by direct parameterizations of experimental data
\cite{li97} or by two-step processes via resonance production 
$NN \longrightarrow R Y K$ and subsequent decay 
$R\longrightarrow N(n\cdot\pi)$. Fig. \ref{fig_cross2} compares 
resonance model calculations  \cite{tsushima99} to data. The
experimental values in  Fig. \ref{fig_cross2} are normalized to the
iso-spin factors of the corresponding $\Delta \longrightarrow N\pi$
decays. The present calculations include reactions with a $\Delta$ resonance 
in the final state with cross sections taken from  \cite{tsushima99}.

\subsubsection{Shift of thresholds}

The cross sections are obtained for free 
scattering. The incorporation of medium effects in the scattering
process is a peculiar question, in particular since at SIS energies most 
kaons are created near threshold. A shift of the 
thresholds by in-medium potentials affects the production 
mechanism significantly. However,  
the treatment of the threshold conditions and the determination 
of the phase space of final states is in the medium a subtle 
problem. Since there exists up to now no unified description 
of relativistic and non-relativistic approaches we will discuss 
this point in more detail. In this context problems arise generally 
due to on-shell scattering of quasi-particles. In the presence of the 
medium mass-shell conditions are modified and phase space relations 
valid in free space have to modified as well. 

Let us for example consider baryon-baryon 
induced $K^+$ production $BB\longrightarrow BYK^+ $. 
In free space the momenta of the outgoing particles are 
distributed according to the 3--body phase space
\beq
d\Phi _{3}(\sqrt{s},m_{B},m_Y, m_{K}) = 
d\Phi_{2}(\sqrt{s},m_{B},M)dM^{2}
\Phi_{2}(M,m_Y, m_{K})
\label{phase1}
\eeq
with $\sqrt{s}$ the center--of--mass energy of the initial baryons.
The two-body phase space in Eq. (\ref{phase1}) has the well known form 
\beq
\Phi _{2}(\sqrt{s},m_{1},m_{2})
=\frac{\pi p^{*}(\sqrt{s},m_{1},m_{2})}{\sqrt{s}}
\label{phase2}
\eeq
where 
\beq
p^{*}(\sqrt{s},m_{1},m_{2})=
\frac{\sqrt{(s-(m_{1} + m_{2})^{2})
(s-(m_{1} - m_{2})^{2})}}{2\sqrt{s}}
\label{phase3}
\eeq
is the momentum of the particles $1$ and $2$ in their 
center--of--mass frame. 

Concerning the in-medium description differences arise also 
between non-relativistic and relativistic 
approaches. In the relativistic case the mean field is 
generally composed by scalar and vector parts, a feature 
which is absent in purely non-relativistic approaches. 
However, also the kaon mean field is of a 
scalar-vector type structure which makes a non-relativistic treatment 
somewhat ambiguous. The reason lies in the fact that in 
most non-relativistic approaches canonical momenta 
$k_\mu $ are propagated whereas in relativistic 
approaches usually kinetic momenta $k_{\mu}^*$ 
are used. To illustrate this effect we consider first nucleon-nucleon  
scattering. 

In relativistic dynamics, e.g. given by the $\sigma\omega$ model 
of Quantum Hadron Dynamics \cite{serot88} 
the nucleon mean field is also composed by a scalar $\Sigma_S$ 
and a vector $\Sigma_\mu$ part. 
Like for kaons the vector field enters into the 
kinetic momenta $k_{\mu}^* = k_\mu - \Sigma_\mu$ and the 
scalar field into the effective mass $m^* = m + \Sigma_S$ of the 
 particles. The dressed quasi-particles  fulfill the mass shell condition 
\beqa 
k_{\mu}^{*2} - m^{*2} = 0 \quad , \quad 
E^* = \sqrt{ {\bf k}^{*2} +m^{*2}} 
\label{mass1}
\eeqa
and thus transport models are usually formulated in terms of 
these kinetic quasi-particle quantities. 
If two nucleons are scattered, energy-momentum conservation requires 
\beq
k_{1\mu } + k_{2\mu } = k_{1\mu }^{'} + k_{2\mu }^{'}
\label{conserv1}
\eeq 
which is equivalent to the conservation of the kinetic quantities
\beqa
k_{1\mu }^{*} + k_{2\mu }^{*} = k_{1\mu }^{*'} + k_{2\mu }^{*'}
\label{conserv2}
\eeqa 
as long as the fields are density dependent but  
do not explicitely depend on momenta, i.e. 
$\Sigma_{\mu,S} = \Sigma_{\mu,S} (\rho)$. 
Then both conditions (\ref{conserv1}) and (\ref{conserv2}) 
can be fulfilled simultaneously. This means that the phase space 
relations (\ref{phase1}--\ref{phase3}) can be used by replacing 
the bare quantities $s, m_B, m_Y, m_K$ by the effective 
quantities $s^*, m^{*}_{B}, m^{*}_{Y}, m^{*}_{K}$.

However, $K^+$ mesons are created by associated strangeness production 
which leads automatically to a shift of the corresponding 
production thresholds in the medium, even when no modifications of the 
kaon properties are taken into account. The reason is the
creation of the associated hyperon. Due to its reduced non-strange quark 
content the mean field of the hyperon should 
scale -- at least in a simple $SU(3)$ flavor picture -- with about 
a factor of 2/3 compared to the nucleon field. Such a scaling is 
in rough agreement with mean field calculations for hyper nuclei 
\cite{hyper1,hyper2,lenske00}.  Now the baryon 
fields are no more conserved and therefore it is no more possible to conserve 
both, kinetic and canonical momenta simultaneously. From the 
derivation of kinetic equations \cite{dani84,btm90} it is clear that 
in this case single particle energies 
\beqa
E = k_0 = E^* + \Sigma_{0}
\label{esp}
\eeqa
 and canonical 
momenta $k_\mu = (E, {\bf k})$ are the quantities 
which have to be conserved. 
Consequently, the usage of on-shell  phase space 
relations (\ref{phase1}--\ref{phase3}) with the 
quantities $s^*, m^{*}_{B}, m^{*}_{Y},m^{*}_{K}$ will lead to a violation 
of energy conservation. 

To overcome this problem and to make 
nonetheless use of on-shell relations it is useful 
to formulate the mass-shell conditions in terms of canonical momenta.  
This can be achieved with the help of the optical potential 
defined in (\ref{uopt}) which allows to rewrite the in-medium 
dispersion relation as 
$0= k_{\mu}^{*2} - \msq = k_{\mu}^{2} - \mks - 2\mk U_{\rm opt}$. 
Since $U_{\rm opt}$ is a Lorentz scalar 
it can be absorbed into an newly defined effective mass 
${\tilde m}_{\mathrm K}$  
\beq
{\tilde m}_{\mathrm K} (\rho ,{\bf k}) 
= \sqrt{ \mks + 2\mk U_{\rm opt}(\rho ,{\bf k}) }
\label{effmass}
\eeq
which sets the canonical momenta on mass-shell
\beqa
0= k_{\mu}^{*2} - \msq =  k_{\mu}^{2} - {\tilde m}_{\mathrm K}^{2}
\quad .
\label{mass3}
\eeqa
The single particle energy follows from the 
dispersion relation (\ref{disp2}) written now as 
\beq
E = \sqrt{ {\bf k}^2 + {\tilde m}_{\mathrm K}^2 }
\label{disp}
\quad .
\eeq 
The threshold condition for $K^+$ production in baryon induced 
reactions reads now
\beq
\sqrt{s} \ge {\tilde m}_B + {\tilde m}_Y + {\tilde m}_K
\label{tresh}
\eeq
with $\sqrt{s}$ the center--of--mass energy of the colliding baryons. 
The momenta of the outgoing particles are distributed according 
to the 3--body phase space
\beq
d\Phi _{3}(\sqrt{s},{\tilde m}_{B},{\tilde m}_Y, {\tilde m}_{K}) = 
d\Phi_{2}(\sqrt{s},{\tilde m}_{B},M)dM^{2}
\Phi_{2}(M,{\tilde m}_Y, {\tilde m}_{K})
\quad .
\label{phase4}
\eeq
The integration over the mass distribution 
of the ${\tilde m}_B,{\tilde m}_Y, {\tilde m}_{K}$ system in Eqs. 
(\ref{phase3}) and (\ref{phase4}) has to be performed numerically. 
The introduction of ${\tilde m}_{\mathrm K}$ 
is thereby of practical use. However, in contrast to the 
quasi-particle mass $\ms$ which is in mean field approximation 
only density dependent, ${\tilde m}_{\mathrm K}$ 
depends on density and momentum. This means that 
3-body-phase-space of the final states and the corresponding 
final state ${\tilde m}$ masses for the final state hadrons 
have to determined self-consistently by iteration. 
Thus the production threshold 
depend also on the final momentum distribution. The usage 
of a simply shifted in-medium mass 
${\tilde m}_{\mathrm K} = \mk \pm \alpha \rho$ in eq. 
(\ref{disp}) neglects such a momentum dependence. 

The potential of the final state hyperons is usually determined under the 
assumption of SU(3) symmetry. The scaling with the non-strange quarks 
leads than to the reduction by 2/3 compared to the nucleon potential.  
Such a scaling has been found to be in reasonable agreement with 
the $\Lambda$ dynamics in heavy ion reactions, in particular the  
$\Lambda$ flow \cite{wang99,li98} and is also close 
to the value extracted from hyper nuclei \cite{hyper1,hyper2,lenske00}.
Thus one obtains an additional shift of the thresholds by 
$U_Y - U_B = -\frac{1}{3}U_B$. In the 
QMD calculations discussed in the work the shifts of the thresholds 
due to the different initial and final state in-medium potentials 
have been treated within a relativistic framework, i.e. 
$U_Y - U_B = -\frac{1}{3}(\Sigma_s - \Sigma_0)$.  
This  allows well defined Lorentz transformations of the kaon 
and baryon mean fields concerning their scalar-vector structure. 
In correspondence with the soft/hard Skyrme forces a soft/hard version of the 
non-linear $\sigma\omega$--model  \cite{li95b} 
with K=200 MeV and  K=380 MeV are used to determine the reaction threshold. 

Besides the  scaling of the hyperon optical potential also the 
momentum dependence of baryon optical potentials influences the 
production thresholds. The nucleon 
optical potential is repulsive at high momenta which are necessary for 
the incoming states to overcome the production threshold. The momenta 
of the final states are significantly smaller. Due to the less 
repulsive optical potential experienced by the final states 
additional energy is gained to overcome the production threshold. 
At subthreshold energies all these effects are essential and have to be 
taken into account in the transport calculations.

\subsubsection{Angular distributions}
For most reactions elementary cross sections are assumed to be
isotropic. In some cases there exists experimental evidence for
anisotropic angular distributions. However, from heavy ion spectra it
is difficult to disentangle the possible sources for anisotropic kaon
emission. One source are final state interactions such as $KN$
rescattering and the mean field propagation
\cite{fang94b,wang97a,brat97}. For a reproduction of the strongly 
forward-backward peaked $K^+$ pattern reported in 
\cite{elmer96} these sources turned out to be not
sufficient and an additional forward-backward anisotropy was introduced
into the pion induced $\pi N\longrightarrow YK^+$ channel in \cite{wang97a}.   
\begin{figure}[h]
\unitlength1cm
\begin{picture}(9.,7.0)
\put(3.5,0){\makebox{\epsfig{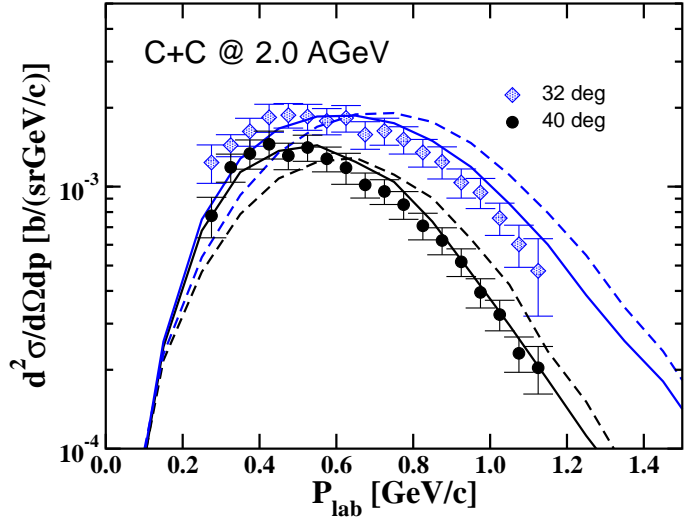}}}
\end{picture}
\caption{Inclusive $K^+$ spectra at $\theta_{\rm lab}=32^o,~40^o$ 
in C+C reactions at 2.0 AGeV. The calculations are performed with 
different $BB\rightarrow BYK^+$ final state 3-body 
phase space distributions and compared to KaoS data
\protect\cite{sturm02}. The dotted curves refer to an isotropic 3-body 
phase space while the solid curves are obtained using the 
parameterization of Eq. (19) with an additional empirical 
c.m. angular anisotropy (see text). The figure is taken from 
 \protect\cite{zheng03}. 
}
\label{fig_spec1}
\end{figure}
Instead of an ideal 3-body final state phase space in the baryon
induced  reactions $NN\rightarrow NYK^+$ 
Li et al. \cite{li97b} proposed an empirical parameterization of the form 
\beq
d\Phi _{3}(\sqrt{s},{\tilde m}_{B},{\tilde m}_Y, {\tilde m}_{K}) = 
dW_{K}(\sqrt{s},{\tilde m}_{B},{\tilde m}_Y,M_K)dM^{2}_K
\Phi_{2}(\sqrt{s}-M_K,{\tilde m}_Y, {\tilde m}_{B})
\label{phase6}
\eeq
where the kaon momentum $p$ is distributed according to 
\beq
dW_K \simeq \left(\frac{k}{k_{\rm max}}\right)^3 
\left(1-\frac{k}{k_{\rm max}}\right)^2 ~~, 
\label{phase5}
\eeq 
with $k_{\rm max}=p^*(\sqrt{s},{\tilde m}_{B}+{\tilde m}_Y, {\tilde m}_K )$ 
the maximal kaon momentum in the $BB$ c.m. frame. 
$M_K = \sqrt{k^2 +{\tilde m}_{K}^2}$ in Eq. (\ref{phase6}). 
The parameterization of Eq. (\ref{phase6}) has been motivated 
by analyzing corresponding $pp\rightarrow p\Lambda K^+$ data
\cite{bland69}. Compared to the ideal 3-body phase space it 
shifts the kaon spectrum towards lower momenta. 
The influence of different treatments
is illustrated in Fig. \ref{fig_spec1} where $K^+$ spectra in C+C 
collisions at 2.0 AGeV are shown. KaoS measured inclusive $K^+$
spectra at various c.m. angles with high precision \cite{sturm02}.
From the comparison to data one sees that an 
isotropic 3-body phase space in the 
$BB\rightarrow BYK^+$ channel shifts the spectra to too high momenta.
The empirical parameterization of Eq. (\ref{phase5})  
improves the situation but is still not fully sufficient to account for the 
angular asymmetry of the data \cite{sturm02}. 
A relatively good agreement can be achieved introducing 
an empirical angular dependence
\beq 
d\sigma\propto (1+a\cos^2\theta_{\rm c.m.})d\cos\theta_{\rm c.m.} 
\eeq 
in the elementary cross sections. An asymmetry parameter of $a=1.2$ 
leads to slightly forward/backward peaked elementary 
$BB\rightarrow BYK^+$ cross sections  and the corresponding 
spectra shown in Fig. \ref{fig_spec1} are then well reproduced. 

\subsection{Comparison of different transport models}
This subsection addresses the question how consistent the results
of present transport models are. Differences may occur due to the use of
different physical input such as e.g. elementary cross sections. The type of
the model, i.e. BUU or QMD should not be of relevance. However, the
corresponding simulation codes are complex and sometimes based on different
numerical and methodical solution techniques. Although physical observables
should be independent on such technical questions one has to exclude them as possible 
sources of uncertainties as far as possible. This was the major goal of two
workshops held in Trento 2001 and 2003 where all major groups 
doing transport model 
calculations in the SIS energy range participated. In a first round of 
homeworks the default versions of the codes were compared, in a second round 
further specifications were made for the comparison. The results of the 
second round have been published in \cite{trento03}. 

\begin{figure}[h]
\unitlength1cm
\begin{picture}(16.,10.0)
\put(1.0,0){\makebox{\epsfig{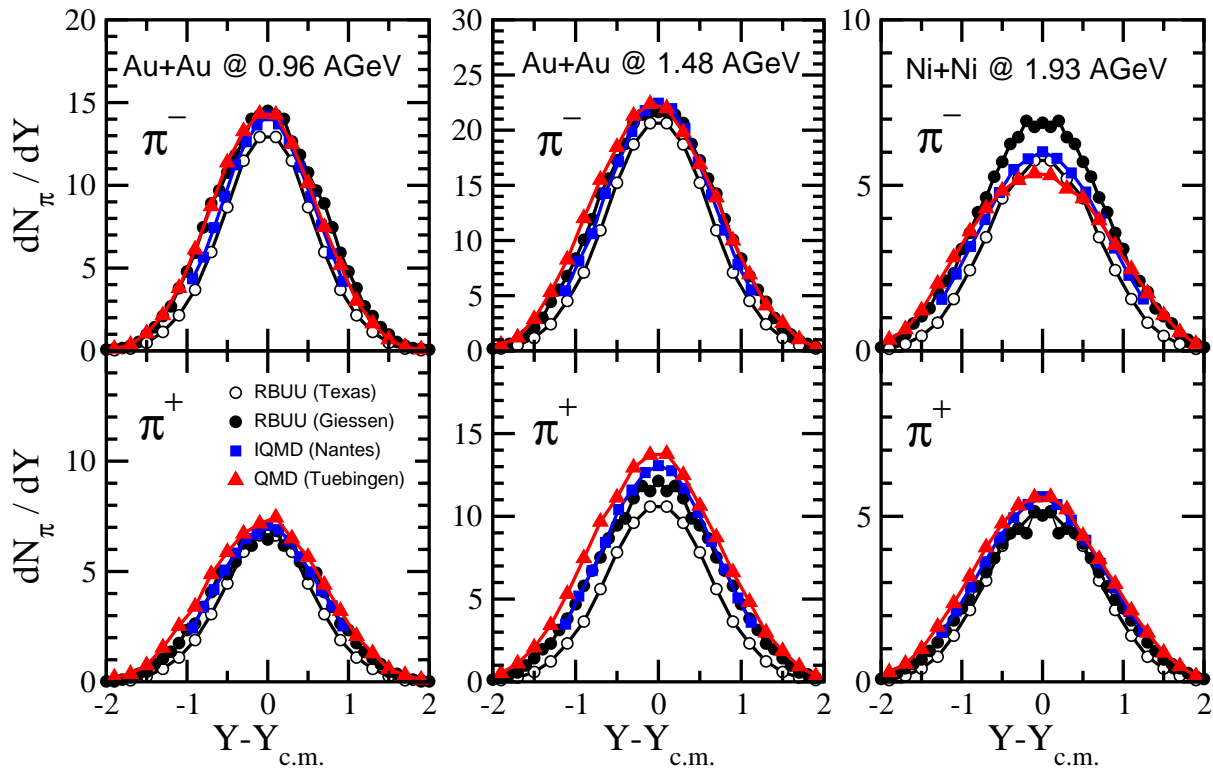}}}
\end{picture}
\caption{$\pi^\pm$ rapidity distributions in central (b=1 fm) 
Au+Au reactions at 0.96 and 1.48 AGeV and Ni+Ni reactions at 1.93 AGeV 
from various transport models: 
RBUU (Texas, open circles), RBUU (Giessen, full circles), 
IQMD (Nantes, full squares) and 
QMD (T\"ubingen, full triangles). 
}
\label{fig_pimult}
\end{figure}
In the following I compare results for pion and kaon production from four state of the art
transport models which participated in that benchmark test. This selection represents only 
a subset of the models compared in \cite{trento03} and is motivated the fact 
that most of the publications on subthreshold $K^+$ production over the 
last five to ten years are based on one of these codes. Of course each 
of the models experienced a steady devellopment during this period but 
the selection represents well established transport models used to 
describe $K^+$ production at subthreshold energies:   
the RBUU model of the Texas A\&M group \cite{ko96,chen04}, 
the RBUU model of the Giessen group \cite{cassing99}, the
IQMD model of the Nantes group \cite{hartnack01} and the QMD model of the
T\"ubingen group \cite{fuchs01,zheng03}. 
Additional reasons why the present discussion is restricted to the 
four mentioned approaches instead of all models presented in 
\cite{trento03} are the following ones: the BUU model of Danielewicz 
\cite{dani00} does not contain explicit kaonic degrees of freedom, 
UrQMD (Reiter) \cite{bass98} treats kaon production non-perturbatively 
through resonance excitations and is not well suited for an application 
at subthreshold energies, the Munich/Catania RBUU model (Gaitanos) 
\cite{ferini05} uses the T\"ubingen kaon package and gives therefore 
similar results. The status of the Rossendorf/Budapest BUU model (Barz/Wolf) 
is hard to estimate since there exist no publications on $K^+$ production 
based on this model from recent years. Finally the Giessen 
BUU model (Larionov) included for the benchmark test all baryonic 
resonances with masses below 2 GeV under the assumption 
that the excited $\Delta^*$ and $N^*$ resonances contribute to the 
$K^+$ production with the same cross sections as the $\Delta$ and 
the nucleon, respectively. Hence a comparision to the four 
models which include only the $\Delta(1232)$ and $N^*(1440)$ for 
kaon production can be missleading, in particular since the 
contributions from higher excitated nucleon resonances to 
the kaon production will be suppressed by a quenching of the elementary 
resonance production cross sections at finite density. 
The resonance quenching is taken into account in \cite{larionov04} where 
the Giessen BUU model shows a much better agreement with the four selected 
models than the benchmark test from \cite{trento03} would suggest. For the 
benchmark results of this model we refer the reader to \cite{trento03}. 
However, the model is included in Fig. \ref{fig_dndy_2} where the 
default versions of the various codes are compared.

The bechmark tests were performed for three
different systems, Au+Au at 0.96 and 1.48 AGeV and Ni+Ni at 1.93 AGeV, all at
impact parameter b=1 fm. For this comparison a soft nuclear EOS with momentum
dependent forces was applied. Except of the Giessen RBUU
model\footnote{Private communication with E.L. Bratkovskaya.}
 the results were obtained with 
a constant $\Delta$ width $\Gamma_\Delta = 120$ MeV, respectively a constant 
lifetime $\tau = 1/\Gamma_\Delta$. A constant 
resonance life time is unphysical and not used in the default versions of the 
codes but it simplifies such a comparison to some extent, in particular 
when predictions for pion production are considered. In QMD\footnote{An
  extended version of the QMD model which includes all nucleon resonances with
  masses below 2 GeV was used for the description of dilepton production at SIS 
energies \protect\cite{shekhter03}.} and IQMD 
only the $\Delta(1232)$ and $N^*(1440)$ resonances
have been included, while the Giessen RBUU includes additionally 
the $N^*(1535)$. 
Fig. \ref{fig_pimult} shows the resulting $\pi^-$ and  $\pi^+$ rapidity
distributions: For Au+Au at 0.96 and 1.48 AGeV the agreement between the
different codes is quite satisfactory, i.e. within a 5-10\% level. Only at the 
highest considered energy of 1.93 stronger deviations 
are visuable since the Giessen calculations includes one higher lying resonance. 
In this context it should be mentioned that the $\pi^\pm$ yields from 
Texas RBBU shown in Fig. \ref{fig_pimult} differ from those of 
\cite{trento03}. There isospin averaged rapidity distributions 
are shown while here the isospin dependence is taken into account 
by the isobar model.  
The default versions of the codes use different descriptions of the 
energy dependence of the resonance life times. As discussed in \cite{trento03} 
the two RBUU models under consideration use $\tau \propto 1/\Gamma_\Delta (|{\bf k}|)$ 
with an energy  dependent width. The QMD \cite{shekhter03} 
and IQMD \cite{aichelin97} models apply the time delay 
description \cite{pratt96} where the lifetime is obtained from the 
phase shift $\tau = 2d\delta(E)/dE$ which results in a Breit-Wigner form 
$ \tau (\mu)  = 4\pi \mu dW_{\Delta}(\mu)/d\mu^2$ with 
\begin{equation}
dW_\Delta (\mu) = 
\frac{1}{\pi} \frac{\mu \Gamma_\Delta (\mu) d\mu^2 }
{(\mu^2 - m_{\Delta}^2)^2 +(\mu\Gamma_\Delta (\mu))^2}
\label{BW}
\end{equation}
In Eq. (\ref{BW}) $ m_{\Delta}$ is the resonance pole mass and $\mu$ the running mass.  
The pion yields obtained by such a description are lower 
than those shown in Fig. \ref{fig_pimult} (15-20\% for QMD and about 30\% 
for IQMD) while they do not much change when the constant width is replaced 
by an energy dependent width in the RBUU calculations.  

\begin{figure}[h]
\unitlength1cm
\begin{picture}(16.,10.0)
\put(1.0,0){\makebox{\epsfig{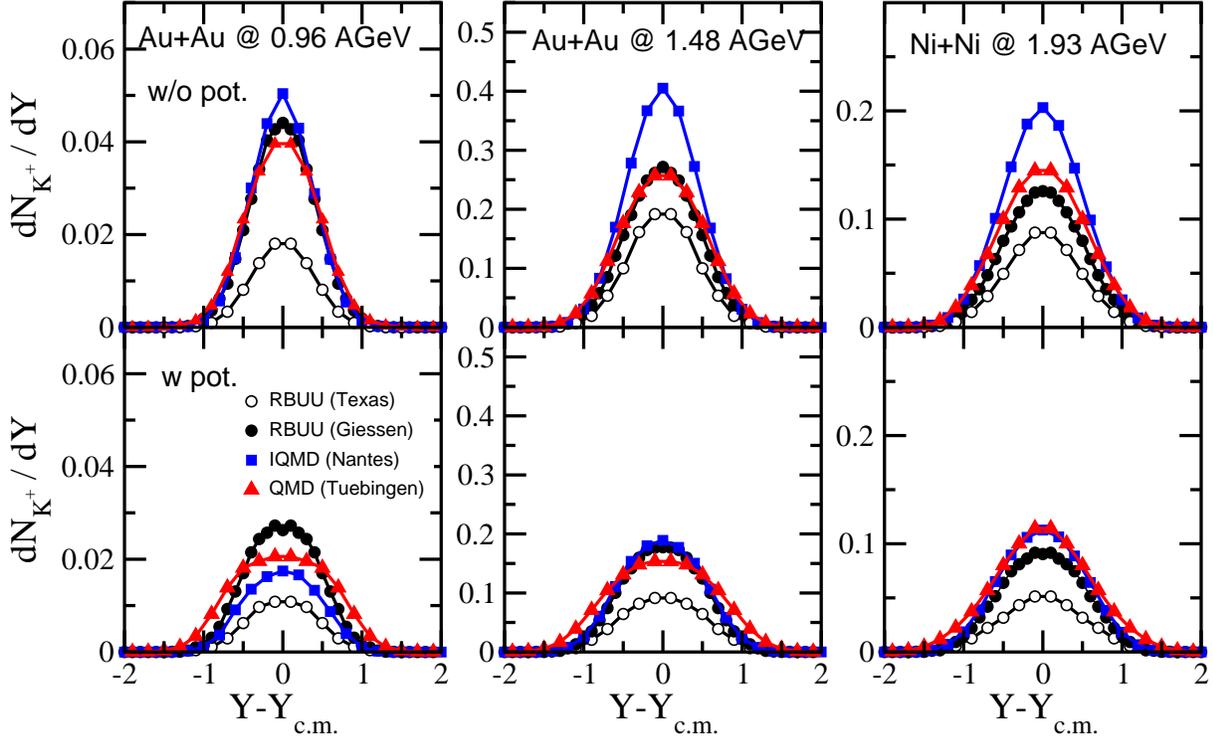}}}
\end{picture}
\caption{$K^+$ rapidity distributions in central (b=1 fm) 
Au+Au reactions at 0.96 and 1.48 AGeV and Ni+Ni reactions at 1.93 AGeV 
from various transport models 
\protect\cite{homework}: RBUU (Texas, open circles), 
RBUU (Giessen, full circles), 
IQMD (Nantes, full squares) and 
QMD (T\"ubingen, full triangles). The upper curves show  
results without kaon in-medium potentials  while  
the lower curves include  potentials. 
}
\label{fig_Kmult}
\end{figure}

Though pion yields agree well the situation is not yet as clear concerning
the kaon production. Fig. \ref{fig_Kmult} shows the $K^+$ rapidity
distributions obtained without in-medium effects and including an in-medium
kaon potential. First we will discuss the results without 
medium effects: 

The results are based on comparable, however, not completely
identical input. The Giessen RBUU, IQMD and QMD 
calculations are based on exactly the same
set of elementary cross sections, i.e. the parameterizations of 
\cite{sibirtsev95} for $NN \longrightarrow N Y K^+$ and those of
\cite{tsushima99} for $N\Delta \longrightarrow N Y K^+$ (in contrast to
previous publications of the Giessen group). Pion induced reactions are based
on the cross sections of \cite{tuebingen}. Hence the RBUU results from Giessen 
agree relatively well with those of the T\"ubingen QMD, i.e. on a 10-15\% 
level. Compared to these two models the
Nantes IQMD leads to an about 30\% higher kaon yield, in particular at higher
energies. This discrepancy can probably not be traced back to the level of 
elementary input and further clarification is needed.  
The Texas RBUU yields lie significantly below the other ones. This 
discrepancy is, however, understandable since the Tsushima 
cross sections \cite{tsushima99} have been used for the 
$NN \longrightarrow N Y K^+$ channel which are at threshold 
about one order of magnitude smaller than those of \cite{sibirtsev95}, 
see Fig. \ref{fig_cross1}. Consequently, the suppression of the 
RBUU Texas yields is most pronounced at the lowest energy of 0.96 AGeV. 
Here the impact of the elementary input is clearly reflected. 

All models show the same qualitative in-medium effect, namely a 
sizable reduction of the $K^+$ yield.  
One has, however, to be aware that the different models use 
 repulsive in-medium potentials of different strength and treat their 
momentum in different ways. When the kaon potential is parameterized in the 
form of an effective mass 
\beq
m^*_{\rm K} = m_{\rm K} (1+\alpha\rho/\rho_0)
\label{mparam}
\eeq
the coefficients for the results shown in Fig. \ref{fig_Kmult} are 
$\alpha = 0.07$ (QMD; MFT ChPT+corr. potential), 0.075 (IQMD), 0.04 (RBUU Giessen) and 0.04 
(RBUU Texas) \cite{trento03}. This leads  
automatically to different predictions for the kaon 
yields. According to the weakest 
potential the relative suppression of the kaon yields is least  
pronounced in the RBUU calculations. There exist, however, not only
differences in the size of the potentials but also in the way how the 
kaon mean field is treated. The present QMD \cite{fuchs98,fuchs01} 
calculations (and those of \cite{larionov04}) are the only ones which 
account for the full covariant structure including the Lorentz force 
from the spatial components of the vector field (\ref{lorentz}). RBUU Texas
and IQMD neglect these contributions and in the RBUU Giessen model the total 
kaon potential accroding to expression (\ref{mparam}). 

\begin{figure}[h]
\unitlength1cm
\begin{picture}(14.,8.5)
\put(1.0,0){\makebox{\epsfig{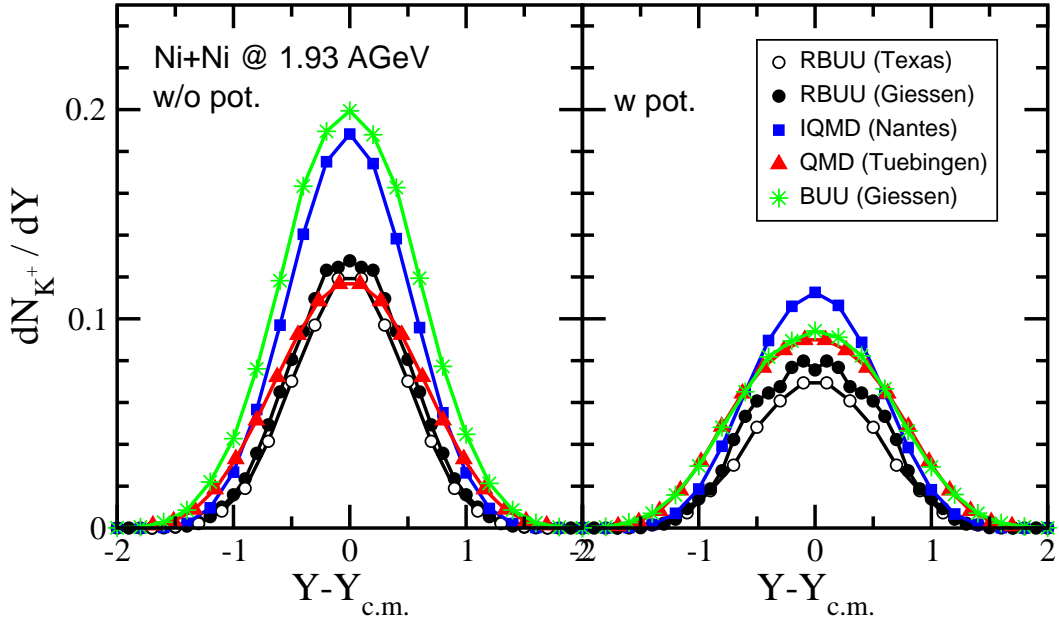}}}
\end{picture}
\caption{$K^+$ rapidity distributions in central (b=1 fm) 
Ni+Ni reactions at 1.93 AGeV 
from various transport models in their default versions: 
RBUU (Texas, open circles), 
RBUU (Giessen, full circles), 
IQMD (Nantes, full squares),  
QMD (T\"ubingen, full triangles) and BUU (Giessen). The left figure shows   
results without kaon in-medium potentials  while  
the right one includes  potentials. 
}
\label{fig_dndy_2}
\end{figure}

From the above comparison it seems surprising that all four models are 
able to fit the experimental $K^+$ data for Ni+Ni at 1.93 AGeV 
\cite{zheng03,hartnack01,mishra04,chen04} 
(see also Fig. \ref{fig_dndy_1}). Therefore in Fig. \ref{fig_dndy_2} the 
predictions of the default versions of the codes are 
shown, again for Ni+Ni at 1.93 AGeV at impact parameter b=1 fm. 
This figure contains in addition the predictions of the BUU model 
of Larionov and Mosel (BUU Giessen) which is also able to reproduce 
experimental kaon multiplicities when a quenching of higher nucleon 
resonances is taken into account \cite{larionov04}. This model 
uses exactly the same in-medium $K^+$ potential as is used in the 
 present QMD calculations.   
From Fig.  \ref{fig_dndy_2} it becomes clear that the unphysical constraint of  
a constant $\Delta$ width (which violates detailed balance in the 
codes if pion absorption is not modified accordingly) has a strong 
influence on the kaon yields. Similar to the pion yields the kaon yields 
are smaller in the QMD/IQMD default versions, in particular 
when in-medium potentials are used. For the Texas RBUU the behavior 
is opposite, i.e. the kaon yields are now enhanced. Considering the 
case without potential one sees that the models group into two fractions: 
(IQMD, BUU (Giessen)) and (QMD, RBUU (Texas), RBUU (Giessen)). The 
latter ones yield an about 30-40\% smaller kaon yield. The reason for this 
discrepancy is still an open question which has to be settled in 
future. 

When in-medium potentials are taken into account, the QMD and BUU rapidity 
distribution are slightly broader than the other ones.  
This is due to the momentum dependence originating from the 
relativistic Lorentz force which is only accounted for in these 
two models. The relative 
potential effect is now larger for the Giessen RBUU compared to 
Fig. \ref{fig_dndy_1} since the  stronger chiral RHA potential of 
\cite{mishra04} is applied, corresponding to $\alpha\simeq 0.055$ in the 
above parameterization. When in-medium potentials are applied,  
the uncertainty of the present model calculations, 
is of similar magnitude than experimental error bars. However, 
as mentioned above, the agreement of the codes in their bare 
versions is still not completely satisfactory and requires 
further efforts to improve on this.

For a compilation of the predictions for $K^-$ production, which turned 
out to be still burdened with much higher uncertainties, we refer the 
reader to Ref.  \cite{trento03}.
\setcounter{footnote}{0}
\section{Probing in-medium kaon potentials}
In-medium kaon potentials shift the single particle energies 
(\ref{disp1}) and the thresholds for kaon production 
(\ref{effmass}-\ref{tresh}). The measurement of kaon multiplicities 
should therefore provide a direct access to such in-medium 
potentials and, consequently probe an expected partial restoration of 
chiral symmetry at supra-normal nuclear densities \cite{fang94b}. 
For $K^+$ mesons the 
repulsive mean field reduces the yields while the situation is opposite 
for $K^-$  where the yields should be significantly enhanced by the 
attractive potential. The hope is to extract information about the 
existence and size of such potentials from heavy ion reactions by 
transport calculations. 
However, the situation turned out to be more complex than originally 
expected: 

Despite the fact that the kaon mean field has indeed a strong influence 
on the multiplicities their absolute values depend as well on the 
elementary cross sections. Uncertainties due to an incomplete knowledge 
of the cross sections turned out to be of the same order as the potential 
effects. However, in the meantime a more or less 
consistent picture has emerged concerning the $K^+$ mesons. 
For antikaons the situation is less satisfying since 
cross sections are known with 
less precision and the $K^-$ yield itself is strongly coupled to the 
$K^+$ production rate via strangeness exchange reactions. 

As a way out of this dilemma there were also strong attempts to 
extract the potential from dynamical observables which are altered 
by the in-medium optical potential while uncertainties in the 
total production rates drop out. Promising probes are, e.g., 
collective flow pattern.   
\subsection{Total Yields}
For $K^+$ mesons the various transport models now provide  a 
relatively consistent picture what concerns the net potential effect 
on the kaon multiplicities. 
The repulsive mean field leads to a reduction of the yields by about 
30-50\%, depending on the actual strength of the potential, the  
system size and the energy of the reaction. The magnitude of the reduction 
within different transport model realizations can be read off from Fig. 
\ref{fig_Kmult}. Figures. \ref{fig_ex_1} to \ref{fig_dndy_1} demonstrate the 
energy and system size dependence and the conclusion which can be 
drawn from the comparison to data.
\begin{figure}[h]
\unitlength1cm
\begin{picture}(9.,8.5)
\put(3.5,0){\makebox{\epsfig{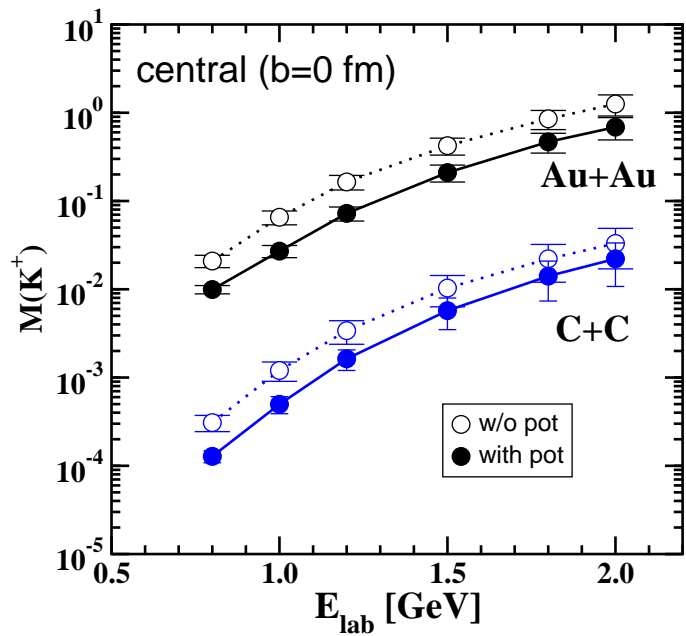}}}
\end{picture}
\caption{Influence of $K^+$ in-medium potential on the kaon yields. 
Multiplicities obtained with and w/o in-medium potential are shown 
for central (b=0) Au+Au and C+C reactions.
} 
\label{fig_ex_1}
\end{figure}
Fig. \ref{fig_ex_1} shows the potential effect in central 
Au+Au and C+C reactions as a function of beam energy. Throughout this 
work the calculations which include an in-medium potential 
are based on the $K^+$ 
mean field proposed by Brown and Rho \cite{brown96b}, denoted in Figure 
\ref{kmass1_fig} as MFT ChPT+corr., which has been derived from ChPT 
and includes effectively higher order corrections beyond mean field. The 
reduction of the $K^+$ yield due to the repulsive potential is, as expected, 
slightly larger in heavy systems than in light systems and most pronounced at 
energies far below threshold. 
\begin{figure}[h]
\unitlength1cm
\begin{picture}(9.,8.5)
\put(3.5,0){\makebox{\epsfig{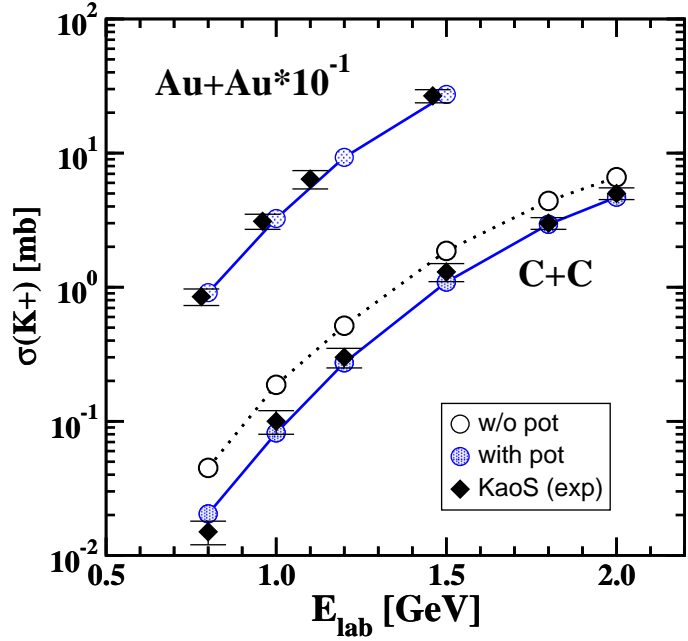}}}
\end{picture}
\caption{The $K^+$ excitation functions in Au+Au (scaled by $10^{-1}$) 
and C+C reactions are compared to data from KaoS 
\protect\cite{kaos99,sturm01}. Calculations include an 
in-medium kaon potential. For C+C also results w/o in-medium potential 
are shown.
} 
\label{fig_ex_2}
\end{figure}
The excitation functions of the $K^+$ cross sections in inclusive 
Au+Au and C+C  are shown in Fig \ref{fig_ex_2}.  Calculations 
were performed with $b_{{\rm max}}=11$ fm 
for Au+Au and $b_{{\rm max}}=5$ fm for C+C and are normalized to 
the experimental reaction cross sections. 
The comparison to data from KaoS  \cite{sturm01,kaos99} clearly 
supports the existence of such a repulsive $K^+$ potential. 

The same conclusion has been obtained by other groups, e.g. 
the Texas and Stony Brook groups \cite{ko96,li97,li97b,li98d} 
and the Nantes group \cite{hartnack01}. For 
some time these findings were in contradiction to calculations from 
the Giessen group \cite{brat97,cassing97a,cassing99,cassing03} 
where the in-medium suppression of the $K^+$ yield was found to be of 
similar size but corresponding 
data where then underpredicted. The reason for this discrepancy could 
be traced back to the usage of different elementary cross sections, 
in particular for the $N\Delta\mapsto NYK^+$ channels 
(see also Fig. \ref{fig_cross2}). The cross sections used by the 
Giessen group have been derived from $pp\mapsto p\Lambda K^+$ by isotopic 
relations while the calculations of the present work (T\"ubingen group) 
are based on the cross sections of Tsushima et al. \cite{tsushima00} for 
the $N\Delta$ channel. The latter ones have a better 
theoretical foundation and are now standardly 
used in transport calculations. However, in earlier calculations 
the uncertainty due to this channel (not constrained by 
data) was of the same order as the net potential effect. 
When model calculations 
are based on comparable sets of elementary cross sections 
the results of the various transport models are rather consistent, 
as can be seen from Figs. \ref{fig_Kmult} and \ref{fig_dndy_1}. 

Fig. \ref{fig_dndy_1} compares 
 the $K^+$ rapidity distributions in Ni+Ni reactions at 1.93 AGeV 
to data from FOPI \protect\cite{best97} and KaoS \protect\cite{menzel00} 
\footnote{$Y^{(0)}$ denotes the center-of-mass rapidity normalized to 
the projectile rapidity $Y^{(0)} = (Y/Y_{\rm proj})_{\rm c.m.} = 2Y_{\rm lab}/Y_{\rm beam} -1$.}.  
Again the description of the data requires the repulsive mean field. 
The same conclusions are obtained from 
IQMD \cite{hartnack01} and the RBUU calculations of Mishra et al. 
\cite{mishra04}, now with the
 $N\Delta\mapsto XK^+$ cross sections of Tsushima et al. \cite{tsushima00}.
The RBUU calculations shown here are based on a chiral mean field evaluated in 
relativistic Hartree approximation with $\Sigma_{\rm KN} = 450$ MeV 
\cite{mishra04}. The range term is included and thus the mean field 
is close that one used in the present QMD calculations. The IQMD 
calculations are based on the RMF kaon optical potential of 
Schaffner et al. \cite{schaffner97} which is also of similar 
strength than the chiral mean field. 
\begin{figure}[h]
\unitlength1cm
\begin{picture}(11.,9.0)
\put(2.5,0){\makebox{\epsfig{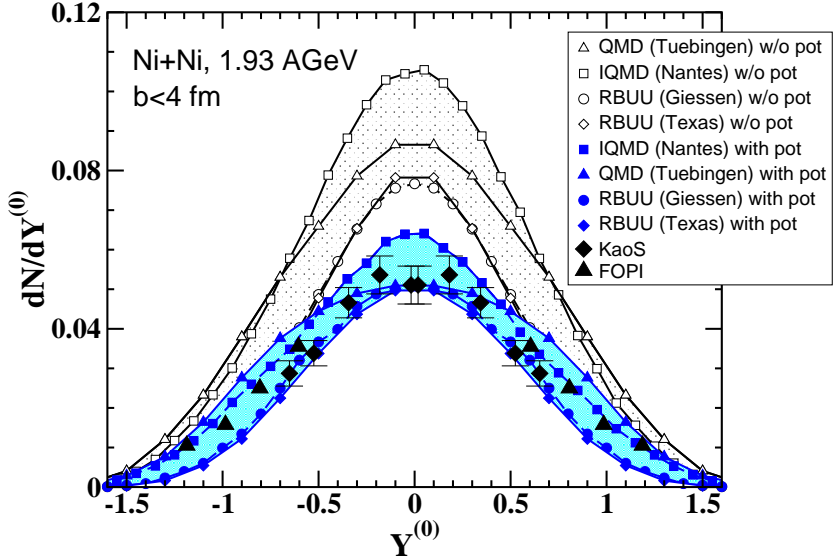}}}
\end{picture}
\caption{$K^+$ rapidity distributions in Ni+Ni reactions at 1.93 AGeV 
obtained including  and w/o an in-medium kaon potential. QMD, IQMD 
\protect\cite{hartnack01} and 
RBUU (Giessen from \protect\cite{mishra04}, 
Texas from \protect\cite{chen04}) and  calculations are compared to data  
from FOPI \protect\cite{best97} and KaoS \protect\cite{menzel00}. 
} 
\label{fig_dndy_1}
\end{figure}
The suppression of the kaon production by the repulsive in-medium potential 
is most pronounced at mid-rapidity which is understandable from kinematical 
reasons. Since subthreshold particle production takes dominantly place around
mid-rapidity the kaons are produced close to threshold where 
in-medium shifts have maximal impact. As already reflected in 
Fig. \ref{fig_Kmult}  the in-medium effect is slightly stronger in 
the IQMD calculations \cite{hartnack01} compared 
to QMD and RBUU. Both RBUU calculations use a slightly weaker 
in-medium potential (RBUU Giessen is based on the chiral RHA potential of 
\cite{mishra04} and RBUU Texas on the empirical potential of \cite{chen04}).  
In the QMD calculation the in-medium potential 
leads not only to a suppression but also to a slight broadening 
of the rapidity distribution which is due to the covariant dynamics 
including the Lorentz force. This contribution is absent in the 
IQMD and RBUU calculations.  Although the 
theoretical descriptions show still same variance they allow nevertheless 
to distinguish clearly between the two scenarios with and without in-medium 
effects. Data support the first one. 

Another parameter which has influence 
on the kaon multiplicities is the stiffness of the nuclear equation of state. 
The EOS dependence will in detail be discussed in the next chapter. The 
usage of a stiffer nuclear EOS reduces the $K^+$ yields to some extent, 
in particular in  heavy systems and thus one might be worried that the 
combination of a hard EOS w/o kaon potential might also be able to 
match with existing data.  But here one can take advantage from the 
fact that light systems like C+C show almost no nuclear EOS dependence while 
the influence of the kaon potential is still present. Hence the two effects 
can be disentangled and the scenario without in-medium kaon potential 
can be ruled out from the light reaction systems. 

Concerning the $K^-$'s mesons the situation is more complex and less clear. 
This is to large part due to 
the interplay of two different productions mechanisms, i.e. strangeness 
production and strangeness exchange reactions. Here transport model 
calculations do not yet deliver a consistent picture. 
On a qualitative level the in-medium energy shifts can 
easily be understood within the mean field picture. In $K^+K^-$ pair 
production reactions the vector potentials $\pm V_0$ cancel due to 
alternating signs for $K^+$ and $K^-$ and only the attractive scalar parts 
$\Sigma_S$ (\ref{sigs}) lead to a shift $\Delta Q$ of the 
production thresholds
\beq
BB \longrightarrow BBK^+K^-~~,~~\Delta Q = -2\Sigma_S~~.
\label{shift1}
\eeq
The same holds for the combination of strangeness production and strangeness 
exchange which involves the excitation of an intermediate hyperon 
\beqa
\left. 
\begin{array}{ccc}
BB \longrightarrow BYK^+    & ~~,~~& \Delta Q = V_0 - \Sigma_S \\
\pi Y \longrightarrow BK^-   & ~~,~~& \Delta Q = -V_0 - \Sigma_S \\
\end{array}
\right\} \Delta Q = - 2\Sigma_S~~.
\label{shift2}
\eeqa
The latter can be viewed as a three-body process 
$ \pi BB \longrightarrow BBK^+ K^-$ which is energetically favored 
compared to the two-body process $ BB \longrightarrow BBK^+ K^-$. 
Exactly the same arguments can be applied to 
the pion induced pair production $\pi B \longrightarrow BK^+ K^-$ and 
effective three-body process $\pi \pi B \longrightarrow BK^+ K^-$ which 
runs over strangeness exchange via an intermediate hyperon. If there 
are enough pions in the system these processes are the dominant sources 
for $K^-$ production \cite{cassing97a,aichelin03}. 
From these considerations one would conclude that the $K^-$ production 
threshold is generally lowered by $\Delta Q = - 2\Sigma_S$ and the 
corresponding $K^-$ should be significantly enhanced by the presence 
of the in-medium potentials.

\begin{figure}[h]
\unitlength1cm
\begin{picture}(11.,9.0)
\put(2.5,0){\makebox{\epsfig{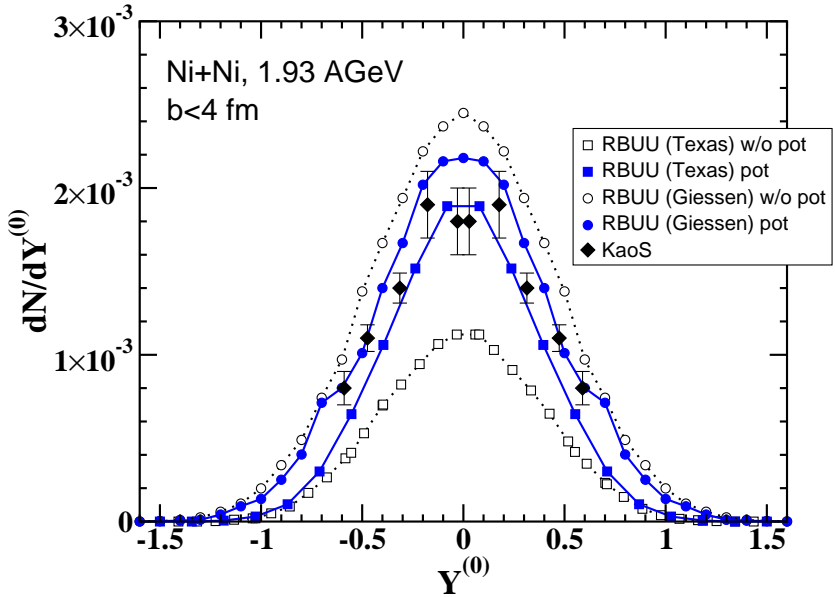}}}
\end{picture}
\caption{$K^-$ rapidity distributions in Ni+Ni reactions at 1.93 AGeV 
obtained including  and without in-medium kaon and antikaon potentials. 
RBUU calculations (Giessen from \protect\cite{mishra04}, 
Texas from \protect\cite{chen04}) are compared to data  
from KaoS \protect\cite{menzel00}. 
} 
\label{fig_dndy_3}
\end{figure}
However, these arguments are only valid if strangeness production and 
strangeness exchange reactions take place at equal nuclear densities 
where the corresponding potentials are of equal size. This is generally 
not the case (see also Chap. 3). While strangeness 
production takes predominantly place in the early high density phase, 
strangeness exchange reactions are the driving processes for $K^-$ production  
and absorption at later stages and lower nuclear densities 
\cite{pal01,aichelin03}. Now 
the balance between attractive and repulsive potential shifts is violated. 
The same holds when less attractive $K^-$ potentials, e.g. from coupled 
channel calculations are used instead of mean fields potentials. Hence 
the net effect of the in-medium potentials on the $K^-$ yield may be 
small \cite{aichelin03,cassing03,mishra04}. 
Also the data situation is much less 
satisfying than for $K^+$. Some older transport calculations explain the 
available $K^-$ yields better using in-medium potentials 
\cite{li97,cassing99}. The comparison to old Ni+Ni in 1997 data 
clearly needed strong $K^-$ potentials. 
In more recent measurements the $K^-$ yield was found to be about 
a factor of two lower \cite{menzel00}. Fig. \ref{fig_dndy_3} compares recent  
RBUU results from Giessen \cite{mishra04} and Texas \cite{chen04}  
to $K^-$ multiplicities measured by KaoS in semi-central 
Ni+Ni reactions at 1.93 AGeV. Both models 
support the $K^-$ in-medium scenario, but on the basis of a qualitatively 
different behavior. While the results of Chen at al. \cite{chen04}
follow the argumentation above, leading to an enhanced $K^-$ yield 
when potentials are switched on, the results from Mishra et al. 
\cite{mishra04} show the opposite behavior. The $K^-$ yields depend 
not only on the strength of the attractive $K^-$ potential but, due to 
strangeness exchange, also on the strength of the repulsive  $K^+$ 
potential. Around threshold the number of  $K^+$ mesons and, 
correspondingly, that of $\Lambda$'s in the system is about one 
order of magnitude larger than that of primary  $K^-$ mesons. Small 
relative changes in the $\Lambda$ abundances can have large impact on 
the final $K^-$ yields. This complicated interplay can even 
lead to a reversed potential dependence of the $K^-$ yield as 
has first been pointed out by Hartnack et al. \cite{aichelin03}. 
The results from  Ref. \cite{mishra04} shown in  Fig. \ref{fig_dndy_3} 
are based on the chiral RHA potential, i.e. the same model as applied in 
Fig. \ref{fig_dndy_1} for $K^+$, which is able to reproduce both sets 
of data. Slightly different mean fields which reproduce the $K^+$ data 
as well were found to fail for $K^-$  \cite{mishra04}. Hence, the 
measured $K^-$ yields allow at present no definite conclusions
on the strength of the attractive $K^-$ potential. 
The situation is further complicated by a possible strong medium 
dependence of the strangeness exchange reactions 
$\pi Y \longleftrightarrow N K^-$ which is, however, theoretically 
not yet completely settled. The predictions 
obtained within coupled channel calculations range from a moderate 
enhancement close to threshold \cite{schaffner00,lutz02} to 
a strong suppression \cite{cassing03}. A consistent treatment of 
these effects requires in any case to account for the  off-shell dynamics 
within the transport approach. The off-shell calculations 
of \cite{cassing03} support the scenario 
of in-medium potential shifts for $K^-$.  
\begin{figure}[h]
\unitlength1cm
\begin{picture}(14.,7.0)
\put(1.0,0){\makebox{\epsfig{file=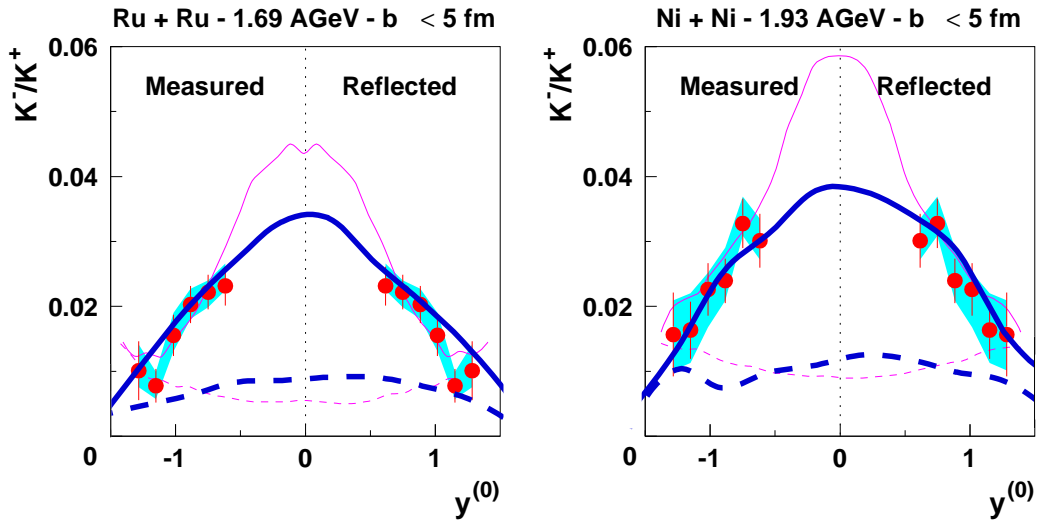,width=14.0cm}}}
\end{picture}
\caption{$K^-/K^+$ ratio as a function of rapidity in Ru+Ru reactions at 
1.69 AGeV and Ni+Ni reactions at 1.93 AGeV. RBUU calculations 
(thick lines: Ref. \protect\cite{cassing99}, 
thin lines: Ref. \protect\cite{li98d}) are compared to FOPI 
data \protect\cite{wis00}. In both cases solid lines include in-medium potentials, 
dashed lines refer to calculations without  in-medium potentials.  
The figure is taken from  \protect\cite{wis00}.
}
\label{fig_Kpmratio}
\end{figure}

The situation should become clear if one considers the $K^-/K^+$ ratio, in 
particular its phase space dependence. This is done in Fig. 
\ref{fig_Kpmratio} where FOPI data \cite{wis00} for the 
$K^-/K^+$ ratio as a function of rapidity are compared to transport 
results from \cite{li98d} and \cite{cassing99}. 
Without further in-medium effects the distributions are 
predicted to be flat as also expected within a statistical approach. 
The presence of the potentials pushes the kaons outwards to higher 
rapidities while the attractive antikaon potential binds $K^-$'s 
at mid-rapidity. Both effects lead to an increase of the $K^-/K^+$ ratio 
around mid-rapidity as also seen in the data. Supplemenatry data 
from KaoS \cite{menzel00} show that $K^-/K^+$ ratio reaches 
in Ni+Ni at 1.93 AGeV a value of about 0.04 at mid-rapidity which 
is in good agreement with the predictions from \cite{cassing99} 
(with pot.) but in contrast to those of \cite{li98d} were the in-medium 
effects are over estimated.  
However, more recent calculations which include 
in-medium modifications of the pion induced $K^-$ 
productions cross sections $\pi Y \longrightarrow NK^-$ and the 
corresponding absorption cross sections  do no more deliver such 
a clear picture \cite{schaffner00,tolos01}. The $K^-$ chemistry and the 
freeze-out time depends crucially on the magnitude of the 
strangeness exchange cross sections and this seems also to be reflected in 
the corresponding rapidity distributions. As already mentioned, 
the medium modifications 
of the cross sections are still a matter of current debate.  
\subsection{Dynamical Observables}
Dynamical observables such as collective flow patterns are to large extent 
free from uncertainties in the total production rates. They depend on the 
phase space pattern of the primordial sources and the final state 
interaction. For $K^+$ mesons the final state interaction is well 
under control since only elastic (and charge exchange) 
reactions occur, the total elastic cross section is of the order of 
$10\div 15$ mb. 
\subsubsection{In-plane flow}
\begin{figure}[h]
\unitlength1cm
\begin{picture}(10.5,7.5)
\put(2.5,0){\makebox{\epsfig{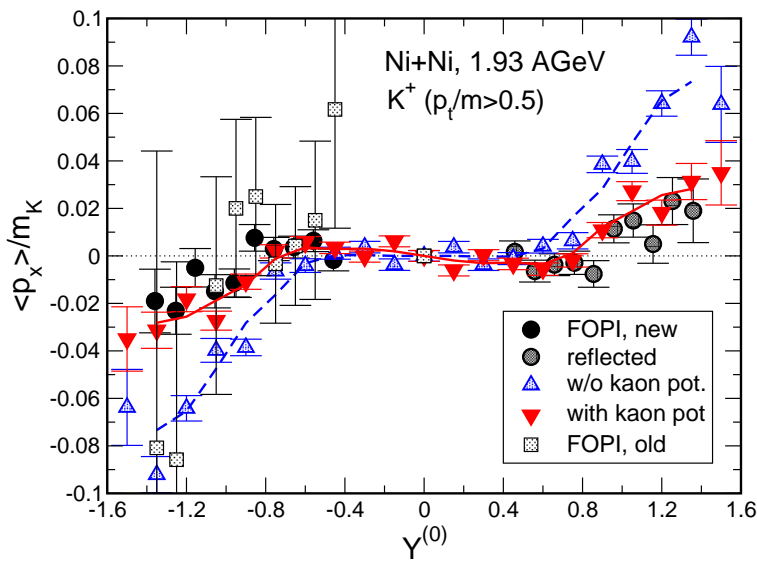}}}
\end{picture}
\caption{Transverse $K^+$ flow 
in 1.93 AGeV $^{58}$Ni + $^{58}$Ni reactions 
 at impact parameter b$\leq 4$ fm. 
 Calculations with and w/o kaon in-medium potential are 
compared to FOPI data \protect\cite{fopi95} (old) and 
\protect\cite{fopi99} (new). The MFT ChPT+corr. potential 
is used   \protect{brown96b} and the Lorentz force is included.
}
\label{fig_Ni_flow1}
\end{figure}
The transverse or in-plane  
flow of $K^+$ mesons is a particularly attractive observable. 
It was first proposed by Ko and Li \cite{ko95} 
that the kaon flow pattern should provide a sensitive probe for the in-medium 
kaon potential. The repulsive potential should push the kaons 
away from the nuclear matter and produce slight anti-flow at spectator 
rapidities and a zero flow signal around mid-rapidity. This was found 
to be consistent with the first available flow 
data from FOPI \cite{fopi95}. Other theoretical studies 
predicted similar features for the 
kaon flow \cite{brat97,brat97b,wang97,li98c}. However, as pointed out 
by Fuchs et al. \cite{fuchs98} the 
scalar-vector type structure of the kaonic mean field implies 
the occurrence of a Lorentz-force (LF) in moving frames which has 
been disregarded in the previous investigations. The Lorentz-force 
from the vector field counterbalances the influence of the time-like 
vector potential on the $K^+$ in-plane flow to large extent 
(see discussion in Chap. 4) which makes it more difficult 
to draw definite conclusions from transverse flow pattern. 

In Refs. \cite{zheng03,larionov04} the $K^+$ flow in Ni+Ni reactions was 
re-investigated since in the meantime FOPI data with improved statistics 
became available \cite{fopi99}. As can be seen from Fig. \ref{fig_Ni_flow1} 
it is difficult to distinguish between the scenarios w/o in-medium 
potentials and full covariant in-medium dynamics around mid-rapidity. 
However, at spectator rapidities clear differences appear and the 
data favor again the in-medium scenario. The QMD calculations are based 
on the MFT ChPT+corr. potential \protect{brown96b} including the LF. 
Very similar results have been obtained in 
\cite{larionov04} where the same potential (including LF) has been used. 
In both, QMD \cite{zheng03} and BUU \cite{larionov04}, the data are 
best described by the relatively strong  MFT ChPT+corr. mean field 
while the weaker MFT ChPT potential or no potential at all lead to 
too strong flow at spectator rapidities.  The dependence of the kaon flow 
on the nuclear EOS and the Coulomb force is quite weak \cite{wang97,zheng03}. 
The present calculations are based on a soft nuclear EOS and the Coulomb force 
is included. 

Since the magnitude of the  Lorentz-force depends on the size of the vector 
potential an explicit momentum dependence of these potentials beyond 
mean field can reduce the Lorentz force. Such a momentum dependence 
reduces the nucleon flow in relativistic approaches compared to a 
mean field description, e.g. within QHD. An explicit momentum 
dependence is necessary in order to comply with the empirical optical 
nucleon-nucleus potential and nucleon flow data above 1 AGeV 
\cite{gaitanos01,hombach99,lca}. Similarly, spectra 
($p_{\rm lab},~ p_T~{\rm and}~ m_T$) imply that the $KN$ interaction 
is less repulsive at high $p_T$ which might be an indication for 
an explicit momentum dependence counterbalancing the Lorentz 
force to some extent. Slopes obtained -- in particular in central --  
Au+Au reactions are too hard while C+C spectra spectra are well 
described  \cite{larionov04} (see also Fig. \ref{fig_spec1}). 
The FOPI Collaboration measured also the 
$p_T$ dependence of $v_1$ in Ni+Ni and Ru+Ru reactions at 
spectator rapidities where a  transition from anti-flow to flow 
with rising $p_T$ was observed \cite{crochet00}. Also these 
data require a repulsive in-medium potential as found in RBUU  
\cite{crochet00}. Recent BUU calculations \cite{larionov04} 
indicate that the Lorentz force is thereby essential not to 
overestimate the data. Hence, for a precise determination of the 
density {\it and} momentum dependence of the $K^+$ potential more 
theoretical efforts are needed. 
\begin{figure}[h]
\unitlength1cm
\begin{picture}(10.5,7.5)
\put(2.5,0){\makebox{\epsfig{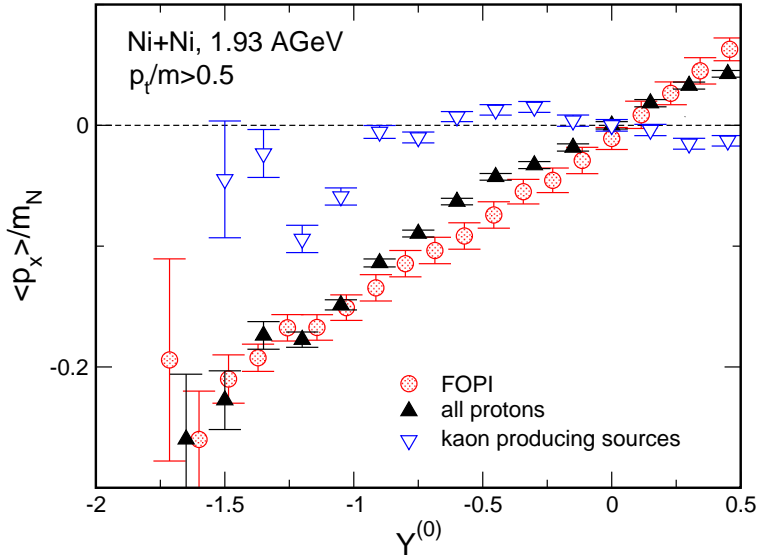}}}
\end{picture}
\caption{Transverse flow of the kaon production sources 
in 1.93AGeV $^{58}$Ni + $^{58}$Ni reactions. The total proton flow 
is compared to FOPI data \cite{fopi95}.
}
\label{fig_Ni_flow2}
\end{figure}

In this context one might worry about conclusions 
based on the small kaonic flow signal (about $10^{-3}$ of the 
initial kinetic energy) if it is not completely clear how 
well the flow of the primordial sources is under control. 
The proton flow, shown in Fig. \ref{fig_Ni_flow2}, is 
reasonably well described by present transport models 
but there exist still deviations from data which are generally 
on the 10\% level. The situation is, however, less severe since 
the final proton and kaon 
flow pattern are only very loosely connected \cite{nantes99}. 
The production sources carry a large in-plane flow while the 
kaons themselves carry a much smaller flow fraction since, at given 
rapidity, they are produced from two baryons originating from 
very different rapidity regions. 
Therefore baryon sources with positive and negative 
$ p_x$ add up to an almost vanishing net flow 
(see the results given by open down triangles 
in Fig. \ref{fig_Ni_flow2}. $\Lambda$'s which are produced in association 
with $K^+$'s have very similar flow pattern as protons, i.e. 
they show almost the same $p_x/m$ scaling \cite{fopi99}. 
In \cite{wang99b} the $\Lambda$ flow was 
investigated within the present model and the data were 
best reproduced including the  hyperon mean field
according to  SU(3) scaling 
$U_{\rm opt}^Y = \frac{2}{3} U_{\rm opt}^B$. As 
also observed in \cite{li98c} the primordial $\Lambda$ flow is 
moderate but strongly enhanced by the $\Lambda N$ final state 
interactions, i.e. rescattering and the $\Lambda$ mean field. 

\subsubsection{Out-off-plane flow}
\begin{figure}[h]
\unitlength1cm
\begin{picture}(16.,9.0)
\put(1.0,0){\makebox{\epsfig{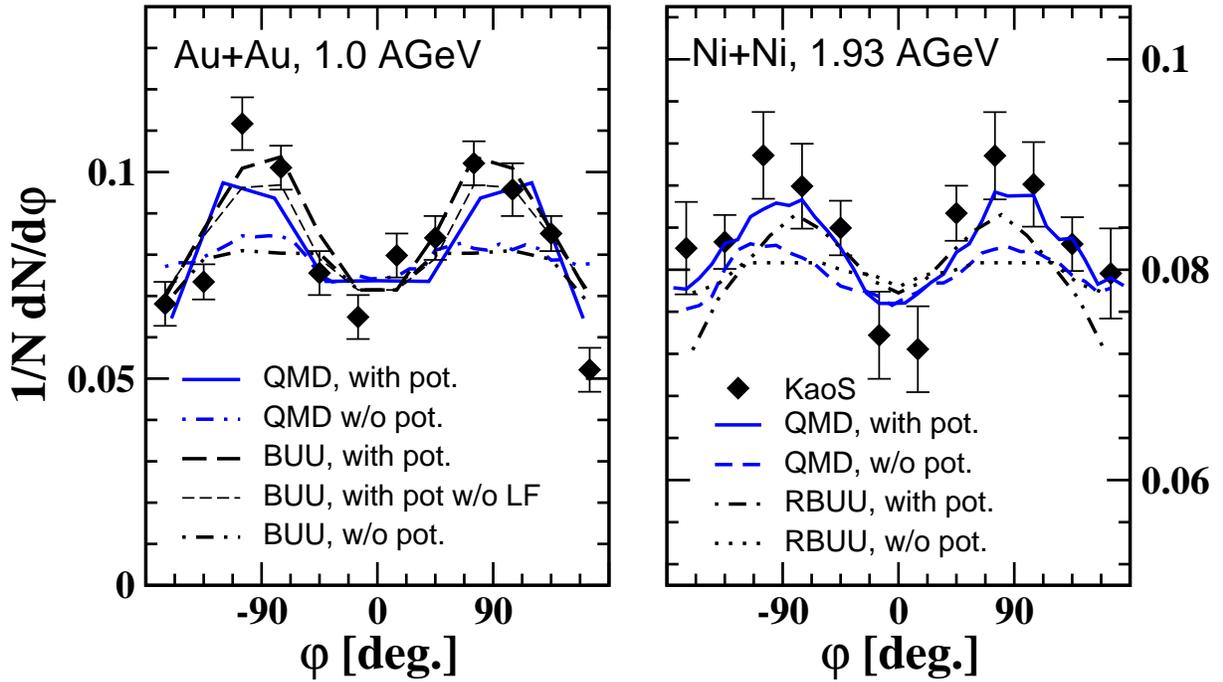}}}
\end{picture}
\caption{$K^+$ azimuthal angular distributions in semi-central 
Au+Au reactions at 1.0 AGeV and Ni+Ni reactions at 1.93 AGeV. 
QMD, BUU \protect\cite{larionov04} and RBUU 
\protect\cite{elena05} calculations without and 
with in-medium potential are compared to data from KaoS 
\protect\cite{shin98,kaos04}.
}
\label{fig_Ksqueeze}
\end{figure}
The phenomenom of collective flow can generally be 
characterized in terms of anisotropies of the azimuthal 
emission pattern, expressed
in terms of a Fourier series 
\beq
\frac{dN}{d\phi}(\phi) \propto  1+ 2v_1\cos(\phi) 
+2 v_2\cos(2\phi) + \dots 
\label{dndphi}
\eeq
which allows a transparent interpretation of the coefficients 
$v_1$ and $v_2$. The dipole term $v_1$ 
arises from a collective sideward deflection of 
the particles in the reaction plane and characterizes the 
transverse flow in the reaction plane. The second harmonics 
describes the emission pattern perpendicular to the reaction 
plane. For negative $v_2$ one has a preferential out-of-plane
emission, called {\it squeeze-out}. Pions exhibit a clear 
out-of-plane preference \cite{brill93,venem93} which is due 
to shadowing by spectator nucleons. The short mean free path 
of the pions hinders pions produced in the central reaction zone 
to traverse the spectator matter located in the reaction plane. 
Therefore it is easier for them to escape perpendicular to 
the reaction plane \cite{bass95}. Since the $K^-$ mean free 
path is comparable to that of the pions one might  
expect the same phenomenon for $K^-$ while the mean free path of $K^+$ 
mesons is large and no squeeze-out signal should be observed. These 
arguments hold when the final state interaction is exclusively 
determined by  scattering and absorption processes. 
However, the presence of a mean field changes the 
dynamics. Azimuthal anisotropies are therefore considered as 
a suitable tool to study medium effects.  

The azimuthal asymmetry of the $K^+$ production in heavy ion 
reactions has been first studied by Li et al. \cite{li96}. 
First data from KaoS \cite{shin98} for semi-central 
Au+Au at 1 AGeV showed a clear squeeze-out signal for 
midrapidity kaons. In corresponding transport calculations from 
the Texas/Stony Brook group \cite{li96,shin98} and the 
T\"ubingen group \cite{wang99} the data could only be reproduced by 
the presence of the repulsive $K^+$ mean field. Elastic 
rescattering of $K^+$ mesons was found to be too weak 
to create the observed squeeze-out signal.  
However,  if the repulsive potential is taken into account, 
the kaons are driven by potential gradients preferentially 
out-of-plane since gradients are larger perpendicular 
than parallel to the reaction plane. With other words, in the reaction 
plane the kaons are repelled by the spectator matter. Thus, the 
potential leads to an additional  
dynamical focusing out of the reaction plane.

Fig. \ref{fig_Ksqueeze} shows the azimuthal distributions 
for semi-central Au+Au reactions at  1 AGeV and Ni+Ni 
at 1.93 AGeV \cite{kaos04}. In both 
cases a transverse momentum cut of $0.2 < p_t < 0.8$ GeV/c 
and a mid-rapidity cut of $|Y^{(0)}|<0.2$ (Au+Au) and 
 $|Y^{(0)}|<0.4$ (Ni+Ni) has been applied. The 
corresponding impact parameters ranges are $5<b<10$ fm (Au+Au)  
and $3.8<b<6.5$ fm (Ni+Ni). In the Au+Au case we compare 
QMD calculations and recent BUU calculations 
from Larionov et al. 
\cite{larionov04} to the KaoS data \cite{shin98}. The results 
confirm the findings that the in-medium potential is needed in order 
to explain the experimental squeeze-out signal. 
Another interesting observation is the fact that 
the Lorentz force, present in covariant dynamics, has only a small 
influence on the out-of-plane flow, contrary to the in-plane 
flow discussed above \cite{larionov04} \footnote{In contrast 
to the earlier statement made in \protect\cite{wang99} we agree 
with \protect\cite{larionov04} on this point.}. The BUU calculations are 
based on the same kaon mean field as QMD, i.e. 
on Ref. \cite{brown96b}. For the  Ni+Ni system shown in the 
right panel of Fig. 
\ref{fig_Ksqueeze} also full experimental filter cuts 
are applied 
($0.267 < p_{\rm lab} < 1.182$ GeV/c and $28 < \Theta_{\rm lab} < 54$). 
The calculations with in-medium potential include the 
full covariant dynamics. For completeness the right panel of Fig. 
\ref{fig_Ksqueeze} shows also recent RBUU/HSD calculations 
\cite{elena05} based on the chiral RHA potential \cite{mishra04} which 
agree well with QMD. In summary, there exists a convergence 
of the various transport models on 
the conclusion that the azimuthal  $K^+$  emission pattern 
require a repulsive mean field.

\begin{figure}[h]
\unitlength1cm
\begin{picture}(16.,9.0)
\put(1.0,0){\makebox{\epsfig{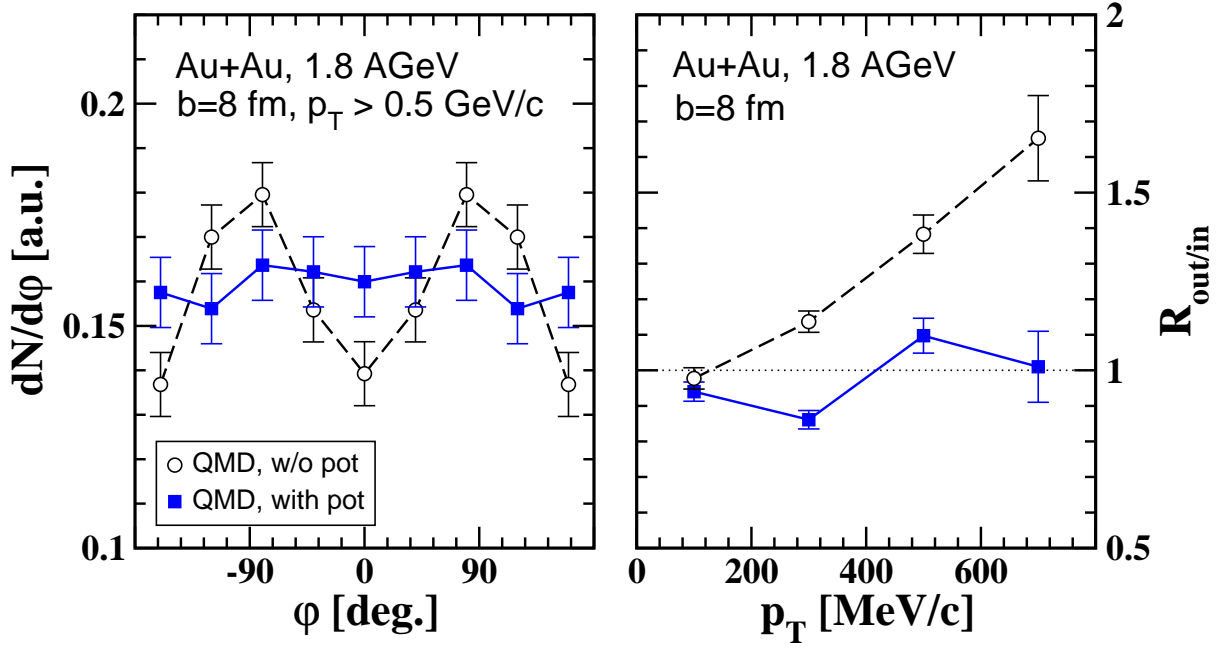}}}
\end{picture}
\caption{$K^-$ azimuthal angular distribution and the 
$R_{\rm out/in}$ ratio as a function of $p_T$ 
as predicted in Ref. \protect\cite{wang99} for semi-central 
Au+Au reactions at 1.8.  The calculations have been performed 
with in-medium $K^-$ potential. A mid-rapidity cut of 
$|Y^{(0)}|<0.2$ has been applied. 
}
\label{fig_Ksqueeze2}
\end{figure}

Turning to $K^-$ the situation is less clear. The first predictions for 
$K^-$ out-of-plane emission pattern were made by Wang et al. \cite{wang99} 
and are shown in Fig. \ref{fig_Ksqueeze2}. The energy of 1.8 AGeV 
has been chosen threshold equivalent to $K^+$ at 1 GeV. The results 
shown in Fig. \ref{fig_Ksqueeze2} were obtained at mid-rapidity 
($|Y^{(0)}|<0.2$) for a semi-central (b=8 fm) reaction. 
Standard strangeness exchange reactions 
$\pi Y\leftrightarrow K^- N~(Y=\Lambda,\Sigma)$ have been taken 
into account and the MFT ChPT mean field (see Fig. \ref{kmass1_fig}) 
has been used. These calculations 
concentrated on the dynamical mean field effect in the particle 
propagation and thus shifts of the $K^+$ and $K^-$ production 
threshold by the in-medium potentials have been disregarded. 

Under these assumptions a very transparent picture was predicted: 
due to the short mean free path $K^-$ behave similar like pions, 
i.e. they show a clear squeeze-out signal caused by 
absorption and rescattering. By the presence of a strongly attractive 
potential this signal is destroyed. In summary: 
without in-medium potential no squeeze-out signal for $K^+$ and a strong 
  squeeze-out signal for $K^-$, with in-medium potentials a sizable 
squeeze-out signal for $K^+$ and an isotropic emission pattern for  
$K^-$. However, already in 
\cite{wang99} a freeze-out time or $p_T$ correlation has been observed, 
reflected in the $p_T$ dependence of the $R_{\rm out/in}$ ratio. 
$R_{\rm out/in}$ quantifies the strength of the azimuthal asymmetry 
and is defined by the ratio
of the particle multiplicity emitted perpendicular to that emitted 
parallel to the reaction  plane
\beqa
R_{\rm out/in} = \frac{N(\phi=90^0) + N(\phi=270^0)}{N(\phi=0^0) +
 N(\phi=180^0)}  = \frac{1 - 2v_2}{1 + 2v_2}
\eeqa
$R_{\rm out/in}$$>$1 corresponds to a preferrential out-of-plane emission. 
The calculation without $K^-$ potential 
shows a steady rise of  $R_{\rm out/in}$ with $p_T$, reflected in 
the emission pattern (Fig. \ref{fig_Ksqueeze2} left part) where 
a $p_T > 0.5 GeV/c$ cut has been applied. When the potential is 
switched on  $R_{\rm out/in}$ is close to unity but 
high $p_T$ particles which are assumed to freeze out 
early carry still some squeeze. The low  $p_T$ $K^-$ mesons are, on 
the other hand, equilibrated and have flat emission pattern or even 
a slight in-plane flow. Such a transition from in-plane to out-of-plane 
flow has recently been observed by KaoS in Au+Au at 1.5 AGeV \cite{kaos04b} 
but the experimental signal is much larger than that 
obtained in \cite{wang99}. 

First data on $K^-$ azimuthal emission pattern have  
only recently been delivered by KaoS for the Ni+Ni system at 
1.93 AGeV \cite{kaos04}. The data agree with none of the two 
predicted scenarios but a preferred in-plane emission of the $K^-$ 
mesons at mid-rapidity has been observed. 
Although IQMD transport calculations from the 
Nantes group match the data when the in-medium potential is 
used it is not really obvious how such in-plane flow can develop. 
In  \cite{kaos04} it is argued that it might be due to late emission 
times caused by strangeness exchange reactions 
$K^- N\longrightarrow Y\pi \longrightarrow K^- N$. However, the 
same holds for pions which undergo several absorption cycles 
$\pi N\longrightarrow \Delta \longrightarrow \pi N$, have late 
emission times but show a clear squeeze-out signal. The 
main difference between 
the 1997 QMD calculations from Wang et al. \cite{wang99} and 
recent IQMD and RBUU results shown in Fig. \ref{fig_Ksqueeze3} lies in the 
fact that in \cite{wang99} {\it free} cross sections without  
in-medium shifts, both for $K^+$ and $K^-$ production and 
absorption, have been used. Shifts of the thresholds 
change the $K^+$ and $K^-$ multiplicities which, at a first glance, 
one would not expect to affect flow pattern significantly. However, 
reduced $K^+$ multiplicities go in line with a reduced number of hyperons 
in the system which reduces the $K^-$ absorption rate and 
the rate of strangeness exchange reactions 
$Y\pi \longrightarrow K^- N$.    
\begin{figure}[h]
\unitlength1cm
\begin{picture}(16.,9.0)
\put(1.0,0){\makebox{\epsfig{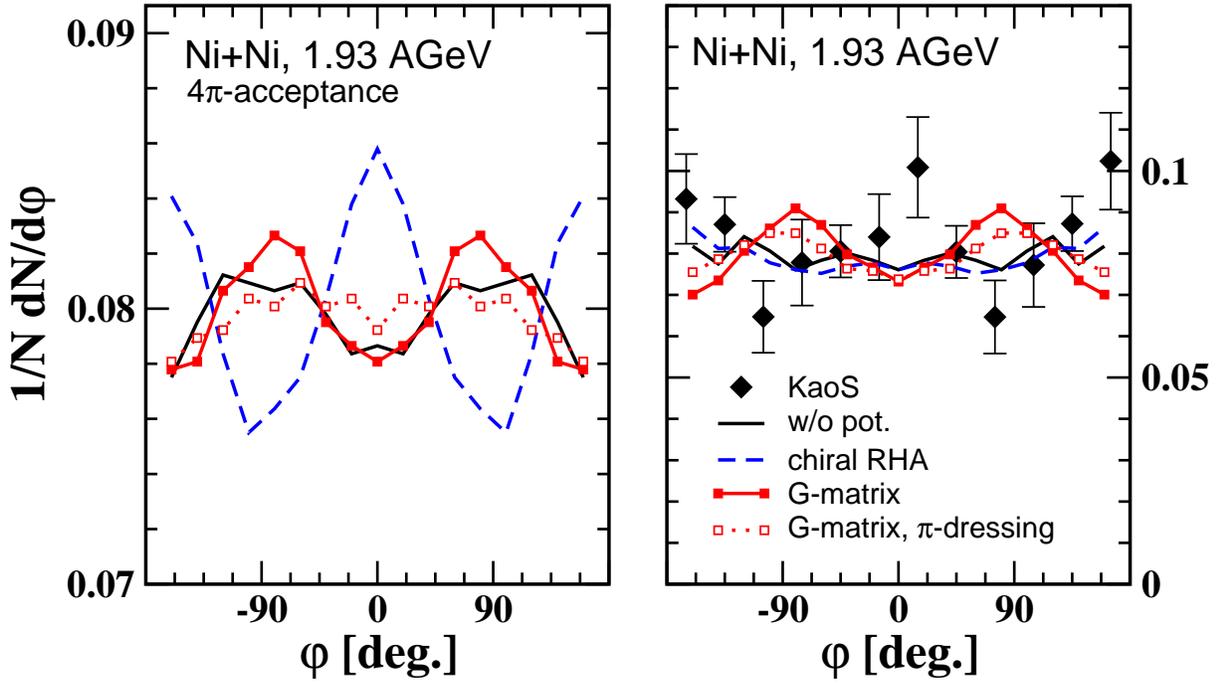}}}
\end{picture}
\caption{$K^-$ azimuthal angular distribution in Ni+Ni at 
1.93 AGeV. RBUU calculations \protect\cite{elena05} 
without and using $K^\pm$ in-medium 
potentials of different type are shown for $4\pi$-acceptance (left) 
and compared to the KaoS data \protect\cite{kaos04} with the 
corresponding acceptance cuts (right).
}
\label{fig_Ksqueeze3}
\end{figure}

Thus the $K^-$ flow pattern are determined by the interplay between 
mean field and in-medium cross sections. The medium dependence of the 
cross sections is significant, 
in particular when off-shell effects are included. 
The $4\pi$ pattern from recent RBUU calculations \cite{elena05} 
demonstrate the following: Without any medium 
effects a clear squeeze-out signal is observed, as predicted in 
\cite{wang99}. Using off-shell dynamics based coupled channel 
G-matrix potentials and in-medium cross sections 
of Tolos et al. \cite{tolos03,cassing03} 
this signal is even enhanced. This potential see Fig. \ref{kpot_fig1}, 
is significantly weaker than 
the chiral RHA mean field but the $K^-$ absorption 
cross sections are strongly enhanced at threshold. This means that the 
shadowing wins against the attractive potential in this scenario 
which leads to an even stronger squeeze-out than without in-medium 
effects. When, on the other hand, pion dressing is taken into account 
(G-matrix with $\pi$-dressing) the situation 
changes and the azimuthal pattern are almost flat. The corresponding 
potential is slightly less attractive than in the previous case but  
the absorption cross sections are now strongly reduced.  
The last case, namely the strongly attractive chiral RHA potential 
with free cross sections leads to an anti-squeeze-out  which 
is also seen in the data. However, the KaoS acceptance cuts lead to a 
strong distortion of the flow pattern and none of the discussed scenarios 
is able to explain the observed distribution shown in the right panel of 
Fig. \ref{fig_Ksqueeze3}. Thus the observed 
$K^-$ emission pattern are at present not really understood and 
further experimental and theoretical efforts are needed to clarify 
the picture.

\subsection{Consistency of the results}
\subsubsection{Consistency between transport predictions}
Now the question arises up to which degree a consistent picture 
has emerged after more than ten years of intensive experimental and 
theoretical efforts to understand  kaon production in heavy ion reactions 
at intermediate energies. Since one of the major goals was to extract 
information on the existence and size of in-medium potentials we summarize in 
Table \ref{tab_model1} the answers which are provided by present transport 
calculations. This summary is organized in the following way: for the most 
important measured observables the agreement or disagreement of transport 
calculations of different groups, using independent simulation codes and 
models, is classified by three classes: 

A filled bullet ({\Large $\bullet$}) 
denotes a good agreement within error bars 
with the bulk of existing data for this observable, an open 
bullet ({\Large $\circ$}), on the other hand, 
denotes a clear disagreement. When the situation is unclear, i.e. 
when calculations match part of the data and fail for other parts, 
this is indicated by a star ({\Large $\star$}). In each case two symbols 
are shown. The first one corresponds to the in-medium scenario, i.e. 
the calculations using in-medium potentials, while the second one 
corresponds to calculations without kaon potentials. 
Hence a combination ({\Large $\bullet$}{\Large $\circ$}) means that the 
corresponding observable allows to clearly distinguish between the two 
scenarios and data support the existence of in-medium potentials, 
({\Large $\circ$}{\Large $\bullet$}) would mean the opposite. 

It is clear that such a classification is rough since the data situation 
strongly differs from observable to observable, the transport models 
differ at least partially in their input and the level of sophistication 
and do not in each case compare to the complete set of available data 
in one observable class. Nevertheless, 
there exists the necessity to bring some systematics into the theoretical 
predictions. 
\begin{table}
\begin{center}
\begin{tabular}{|c|l|l|l|l|l|}
\hline
model & QMD$^1$  & IQMD$^2$   & RBUU$^3$ & RBUU$^4$  & BUU$^5$ \\ 
\hline\hline
$K^+$ multiplicity &{\Large $\bullet$}{\Large   $\circ$} \protect\cite{fuchs01,zheng03}$\dagger$ & 
{\Large $\bullet$}{\Large $\circ$} \protect\cite{hartnack01} & 
 {\Large $\bullet$}{\Large $\circ$} \protect\cite{mishra04} & 
{\Large $\bullet$}{\Large $\circ$} \protect\cite{li97,li97b,chen04} & 
{\Large $\bullet$}{\Large $\circ$} \protect\cite{larionov04}\\
\hline
$K^+$ flow &{\Large $\bullet$}{\Large $\star$} \protect\cite{wang97,zheng03}$\dagger$ & 
{\Large $\circ$}{\Large $\star$} \protect\cite{nantes99} & 
 {\Large $\bullet$}{\Large $\circ$} \protect\cite{cassing99,mishra04} & 
{\Large $\bullet$}{\Large $\circ$} \protect\cite{ko95,li98c} & 
{\Large $\bullet$}{\Large $\circ$}\protect\cite{larionov04} \\
\hline
$K^+$ squeeze &{\Large $\bullet$}{\Large $\circ$} \protect\cite{wang99}$\dagger$ & 
{\Large $\bullet$}{\Large $\bullet$} \protect\cite{kaos04} & 
{\Large $\bullet$}{\Large $\circ$}  \protect\cite{mishra04} & 
{\Large $\bullet$}{\Large $\circ$} \protect\cite{shin98} & 
{\Large $\bullet$}{\Large $\circ$}\protect\cite{larionov04} \\
\hline\hline
$K^-/K^+$ ratio & --- & 
 ---- & 
 {\Large $\bullet$}{\Large $\circ$} \protect\cite{cassing99} & 
{\Large $\bullet$}{\Large $\circ$} \protect\cite{li97,li98d} & 
{\Large $\circ$}{\Large $\bullet$} \protect\cite{schaffner00}\\
\hline
$K^-$ multiplicity & --- & 
 ---- & 
 {\Large $\star$}{\Large $\star$} \protect\cite{cassing97a,cassing03,mishra04} & 
{\Large $\bullet$}{\Large $\circ$} \protect\cite{li97,chen04} & 
----\\
\hline
$K^-$ squeeze &{\Large $\circ$}{\Large $\circ$}\protect\cite{wang99}  & 
{\Large $\bullet$}{\Large $\circ$} \protect\cite{kaos04} & 
 {\Large $\circ$}{\Large $\circ$} \protect\cite{mishra04},  
{\Large $\bullet$}{\Large $\circ$} \protect\cite{cassing03}& 
 ---- &  ----\\
\hline
\end{tabular}
\end{center}
\caption{\label{tab_model1}
Comparison of various transport model calculations with existing data: the calculations are 
from $^1$T\"ubingen ($\dagger$ besides the given references also results from the present work are 
included), $^2$Nantes, $^3$Giessen (RBUU/HSD), 
$^4$Texas/Stony Brook, $^5$Giessen(Berkeley). Symbols denote: 
{\Large $\bullet$}$\equiv$ good description of available data within error bars; 
{\Large $\circ$}$\equiv$ clear failure to describe available data within error bars; 
{\Large $\star$}$\equiv$ situation unclear. The first symbol corresponds to calculations 
based on kaon in-medium potentials while the second one corresponds to the case 
w/o in-medium potential. 
}
\end{table}

As already discussed in Chapters 3 and 4 the various transport approaches  
differ partialy in the elementary input and in the technical realizations. 
Table \ref{tab_model2} summarizes 
the most relevant differences. These lie in usage of different paramterizations for 
the strangeness production cross sections $NN\mapsto NYK^+$\footnote{In 
QMD different parameterizations are applied for the $\Lambda$
and $\Sigma$ channel which is indicated in Tab. \protect\ref{tab_model2}.}, 
$N\Delta \mapsto NYK^+$ and the medium dependence of the strangeness exchange 
cross sections $N\pi \leftrightarrow YK^-$. 
The $K^+$ production cross sections are either based on one-boson-exchange
(OBE) or resonance model (R) calculations where parameters are fixed by the 
measured $pp\mapsto p\Lambda K^+$ reaction. For the pion induced $K^+$
production all approaches apply the cross section of \cite{tuebingen}. 
(P) denotes parameterizations of 
experimental cross sections which are in particular used for the well
constrained strangeness exchange reactions. However, here the 
intrinsic medium dependence of these cross sections is crucial and has been explored 
in chiral coupled channel (CC) \cite{schaffner00} and 
coupled channel G-matrix (CCG) \cite{tolos02} calculations. The medium 
dependence due to shifts of the threshold by the attractive/repulsive 
$K^\pm$ mean field are usually taken into account (mass shifts). Concerning the dynamics 
one has to distinguish between a non-relativisitc treatment of the mean 
field (nonrel.) and full covariant dynamics (covariant) which includes 
the vector Lorentz force according to 
(\ref{lorentz}) and the off-shell dynamics of \cite{cassing00,cassing03}.  
\begin{table}
\begin{center}
\begin{tabular}{|l|l|l|l|c|c|}
\hline
Model & $NN\mapsto XK^+$  &  $N\Delta\mapsto XK^+$   &  $\pi B\mapsto
YK^-$ & cross section & dynamics \\ 
\hline\hline
QMD: & & & & & \\
\protect\cite{wang99} &  
\protect\cite{sibirtsev95} $\Lambda$ (OBE) &  \protect\cite{tsushima97} (R) &
\protect\cite{cassing99} (P) & free & nonrel. \\
 & \protect\cite{tsushima97} $\Sigma$ (R)& & & & \\
\protect\cite{fuchs01,zheng03}&  
\protect\cite{sibirtsev95} $\Lambda$  (OBE) &  \protect\cite{tsushima97} (R) &
 -- & mass shift & covariant \\
present&  \protect\cite{tsushima97} $\Sigma$ (R)   & & & & \\
\hline
IQMD: & & & & & \\
\protect\cite{nantes99,hartnack01} &  
\protect\cite{nantes99} (P) &  \protect\cite{nantes99} (P) &
-- & mass shift & nonrel. \\
\protect\cite{kaos04} & \protect\cite{sibirtsev95} $\Lambda$  (OBE) &  \protect\cite{tsushima97} (R) &
 (P) & mass shift & nonrel. \\
    &  \protect\cite{tsushima97} $\Sigma$ (R)   & & & & \\

\hline
RBUU: & & & & & \\
\protect\cite{cassing97a,cassing99} &  
\protect\cite{cassing99} (OBE) &  \protect\cite{cassing99} (OBE) &
\protect\cite{cassing99} (P) & mass shift & nonrel. \\
\protect\cite{cassing03} &  
\protect\cite{cassing99} (OBE) &  \protect\cite{cassing99} (OBE) &
\protect\cite{tolos02} (CCG) & mass shift ($K^+$)  & nonrel. ($K^+$) \\
 & & & & off-shell ($K^-$) & off-shell ($K^-$) \\
\protect\cite{mishra04} &
\protect\cite{cassing99} (OBE) &  \protect\cite{tsushima97} (R) &
\protect\cite{tolos02} (CCG) & mass shift & nonrel. \\
\hline
RBUU: & & & & & \\
\protect\cite{li97b,li97,shin98} &  
\protect\cite{li97} (OBE) &  \protect\cite{li97} (OBE) &
\protect\cite{li97} (P) & mass shift & nonrel. \\
\protect\cite{li98c,li98d,chen04} & & & & & \\
\hline
BUU: & & & & & \\
\protect\cite{schaffner00} &
\protect\cite{schaffner00} (OBE) &  -- &
\protect\cite{schaffner00} (CC) & mass shift ($K^+$) & nonrel. \\
 & & & & in-med. ($K^-$) &  \\
\protect\cite{larionov04} &
\protect\cite{tsushima97} (R) &  \protect\cite{tsushima97} (R) &
 -- & mass shift & covariant \\
\hline
\end{tabular}
\end{center}
\caption{\label{tab_model2}
Elementary input, realization of the medium dependence of the cross sections and 
mean field dynamics in the transport 
models summarized in Table \protect\ref{tab_model1}.
}
\end{table}

By far the best measured quantity are $K^+$ multiplicities. High precision 
data exist for a broad range of energies and different mass systems. For the 
dynamical observables $K^+$ flow and squeeze, data are in the meantime also 
precise enough to constrain the models. For $K^+$ mesons one can summarize 
the situation as follows: there exist no data which contradict 
the in-medium scenario. In contrast, most observables can {\it only} be described 
within the   in-medium scenario. For $K^-$ the data situation as well as 
the theoretical situation is much less satisfying. No consistent picture has 
yet emerged which allows to discriminate between the two scenarios. 
\subsubsection{$p+A$ reactions}
The picture was recently
complemented by measurements of the $K^{+}$ production in proton-nucleus 
reactions \cite{anke,anke04}. Although such reactions test only subnormal 
nuclear densities they are much easier to handle than the complicated 
dynamical evolution of heavy ion reactions. 
\begin{figure}[h]
\unitlength1cm
\begin{picture}(6.5,11.0)
\put(4.5,0){\makebox{\epsfig{file=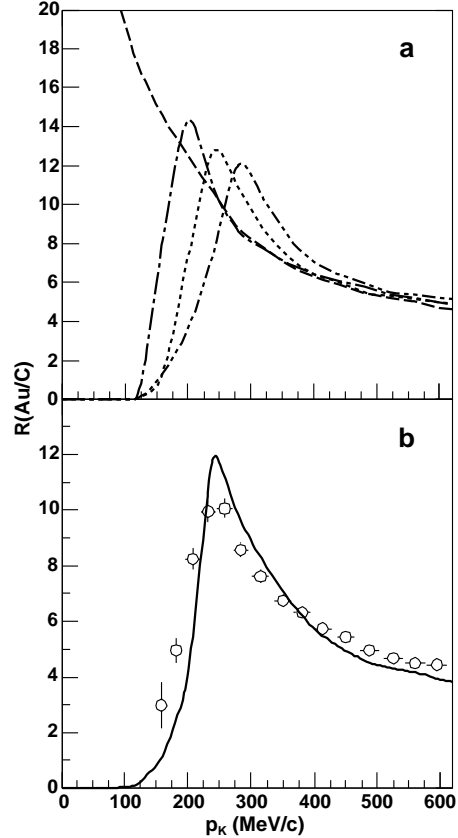,width=6.0cm}}}
\end{picture}
\caption{Ratios of $K^+$ production cross sections for p+Au/p+C 
at $T_p=2.3\,\mathrm{GeV}$ as a function of the kaon momentum. 
a) Transport calculations including
only the Coulomb potential (dash-dotted) and in addition a 
kaon potential of different strength, i.e.  20~MeV (dotted) and 
40 MeV (dashed-double-dotted) at $\rho_0$. The broken line
corresponds to simulations without Coulomb and nuclear kaon potentials.
In all cases considered here, $K^+$ rescattering in the nucleus 
has been taken into account. b) The open circles are the experimental data.
The solid line shows the result of RBUU transport calculations starting from 
the dotted line in the top figure with a baryon potential added. The 
figure is taken from \protect\cite{anke}.
}
\label{pA_ANKE_fig}
\end{figure}
A particularly sensitive 
observable was based on the measurement of the kaon momenta in p+Au 
and p+C reactions. A repulsive potential leads to a momentum shift which, 
in the absence of further final state interactions, results in minimal 
possible momenta 
\beq
p_{\rm min} = \sqrt{ 2\mk (V_C(r)+V_0(r))}~~.
\eeq
This feature is, e.g. well known from the suppression of $\beta^+$ emission 
in heavy nuclei at low positron momenta due to the Coulomb shift.  
Naturally, such a low momentum shift is more pronounced in a heavy 
nucleus. Building now the ratio between the heavy and the light system one 
is sensitive to density effects and thus to the size of the potential. 

This is demonstrated in Fig. \ref{pA_ANKE_fig} where the ratio of the 
$K^+$ cross section in p+Au/p+C reactions is shown 
as a function of the kaon momentum. Taking the Coulomb potential $V_C$ 
 into account one can determine the strength of $V_0$ from the peak in the 
data. This has been done in   \cite{anke,rudy02} using transport calculations and 
a repulsive kaon potential of  $V_0\sim 20\pm 5$ MeV at $\rho_0$ has been 
extracted. Hence the result is consistent with heavy ion reactions and 
the magnitude of $V_0$ as predicted by 
effective chiral Lagrangiens. Such a potential was also found to be 
consistent with the measured $K^+$ spectra in p+A reactions at 
subthreshold energies \cite{rudy02}. 

\setcounter{footnote}{0}
\section{Probing the nuclear equation of state}
Heavy ion reactions provide the only possibility to reach nuclear
 matter densities beyond saturation density $\rho_0 \simeq 0.16~{\rm
 fm}^{-3}$. Transport calculations indicate that in the intermediate energy range 
$E_{\rm lab}\simeq 1$ AGeV nuclear densities between $2\div 3 \rho_0$ 
are accessible while the highest baryon densities ($\sim 8 \rho_0$)  
will probably be reached in the energy range of 
the future GSI facility FAIR \cite{sis200} between $20\div 30$ AGeV. At even higher 
incident energies transparency sets in and the matter becomes less baryon 
rich due to the dominance of meson production. Since the knowledge of the 
nuclear equation-of-state (EOS) at supra-normal densities is essential for 
our understanding of the nuclear forces as well as for astrophysical 
purposes, the determination of the EOS was already one of the primary 
goals when first relativistic heavy ion beams started to operate in 
the beginning of the 80ties \cite{bevelac84}. In the following we will 
briefly report the knowledge on the nuclear EOS from a theoretical point of 
view, then turn to the compression phase in heavy ion reactions, 
give a short review on possible observables and finally discuss the 
recent progress achieved by the kaon measurements. 
\subsection{Modeling the nuclear EOS}
\subsubsection{Predictions for the nuclear EOS}
Models which make predictions on the nuclear EOS can roughly be divided 
into three classes:
\begin{enumerate}
\item {\bf Phenomenological density functionals}: These are models based on 
effective density dependent interactions such as 
Gogny or Skyrme forces \cite{reinhard04} or 
relativistic mean field (RMF) models \cite{rmf}. 
The number of parameters which are fine tuned to the nuclear chart is 
usually larger than six and less than 15. This type of models 
allows the most precise description of finite nuclear properties.
\item {\bf Effective field theory approaches}: Models where 
the effective interaction is determined within the spirit of 
effective field theory (EFT) become recently more and more popular. 
Such approaches lead to a more systematic expansion of the 
EOS in powers of density, respectively the Fermi momentum $k_F$. They 
can be based on density functional theory \cite{eft1,eft2} or e.g. on 
chiral perturbation theory \cite{lutz00,finelli}. The advantage of EFT 
is the small number of free parameters and a correspondingly higher 
predictive power. However, when high precision fits to finite nuclei 
are intended this is presently only possible by the price of 
fine tuning through additional parameters. Then functionals based on 
EFT have approximately the same number of parameters as  
phenomenological density functionals.  
\item {\bf Ab initio approaches}: Based on high precision 
free space nucleon-nucleon interactions, the nuclear many-body 
problem is treated microscopically. Predictions for the nuclear EOS 
are parameter free. Examples are variational 
calculations \cite{wiringa79,akmal98}, 
relativistic \cite{bm90,boelting99,honnef,dalen04} 
or non-relativistic Brueckner calculations and Greens functions 
Monte-Carlo approaches \cite{muether00}. 
\end{enumerate}
Phenomenological models as well as EFT contain parameters which have 
to be fixed by nuclear properties around or below saturation density  
 which makes the extrapolation to supra-normal densities somewhat 
questionable. However, in the EFT case such an extrapolation is safer 
due to a systematic density 
expansion. One has, nevertheless, to keep in mind that EFT approaches are 
in general based on low density expansions. Many-body calculations, 
on the other hand, 
have to rely on the summation of relevant diagram classes and are still 
too involved for systematic applications to finite nuclei. In the 
following we will restrict the discussion mainly to the prediction 
from many-body calculations.

In the relativistic Brueckner approach the nucleon 
inside the nuclear medium is viewed as a dressed particle in consequence
of its two-body interaction with the surrounding nucleons. 
The in-medium interaction of the nucleons is treated in the ladder
approximation of the relativistic Bethe-Salpeter (BS) equation
\beq
T= V + i \int  VQGGT
\quad ,
\label{BSeq}
\eeq
where $T$ denotes the T-matrix, while $V$ is the bare nucleon-nucleon interaction. 
The intermediate off-shell nucleons in the 
scattering equation are described by a two-particle propagator $iGG$.
The Pauli operator $Q$ accounts for the 
influence of the medium by the Pauli-principle and projects the 
intermediate scattering states out of the Fermi sea. 
The Green's function $G$ fulfills the Dyson equation
\beq
G=G_0+G_0\Sigma G 
\quad .
\label{Dysoneq}
\eeq 
$G_0$ denotes the free nucleon propagator while the influence of the 
surrounding nucleons is expressed by the nucleon self-energy $\Sigma$. 
In Brueckner theory this self-energy is determined by summing up the 
interaction with all the nucleons inside the Fermi sea in Hartree-Fock 
approximation
\beq
\Sigma = -i \int\limits_{F} (Tr[G T] - GT )
\quad .
\label{HFselfeq1}
\eeq
The coupled set of equations (\ref{BSeq})-(\ref{HFselfeq1}) represents
a self-consistency problem. 
\begin{figure}[h]
\unitlength1cm
\begin{picture}(10.,8.5)
\put(3.5,0){\makebox{\epsfig{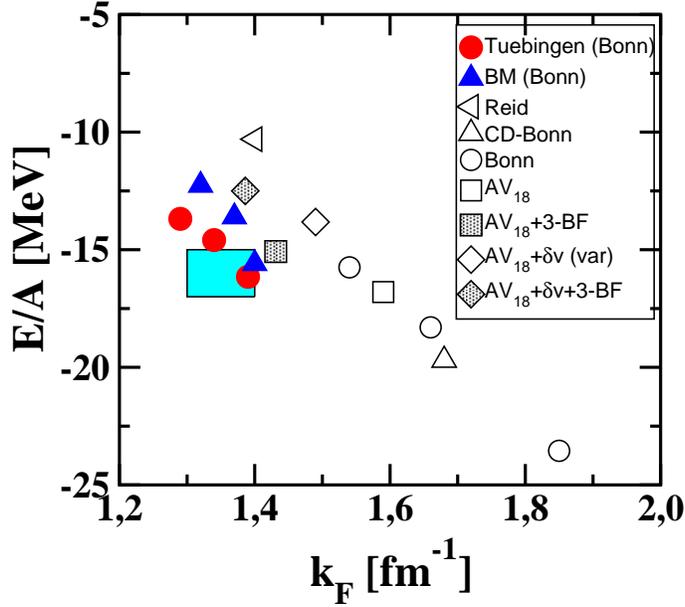}}}
\end{picture}
\caption{Nuclear matter saturation points from relativistic (full symbols) 
and non-relativistic (open symbols) Brueckner-Hartree-Fock calculations 
based on different nucleon-nucleon forces. The diamonds show results from 
variational calculations. Shaded symbols denote calculations which 
include 3-body forces. The shaded area is the empirical region of 
saturation.
}
\label{coester_fig}
\end{figure}

Fig. \ref{coester_fig} compares the saturation points of nuclear matter 
obtained by relativistic Dirac-Brueckner-Hartree-Fock (DBHF) calculations 
using the  Bonn potentials \cite{bonn} as bare $NN$ interactions  
to non-relativistic Brueckner-Hartree-Fock calculations for various 
 $NN$ interactions. The DBHF results are taken from Ref. \cite{bm90} (BM) 
and more recent calculations based on improved techniques are 
from \cite{boelting99} (T\"ubingen). In these calculations the 
Bonn A interactions matches with the empirical 
region of saturation, however, still at a slightly too high 
density. For various $NN$ potentials the  
non-relativistic results lie on the so called {\it Coester line} which 
misses the empirical region of saturation. 
Only by the inclusion of 3-body-forces (shown by the dashed 
square in Fig. \ref{coester_fig}) the situation can be improved 
\cite{zuo02,zuo04}. The contributions from 3-body-forces (3-BFs) are in total  
repulsive which makes the EOS harder and non-relativistic 
calculations come close to their relativistic counterparts 
when 3-BFs are included. The same effect is observed in variational 
calculations \cite{akmal98}. The variational approach shown 
contains relativistic boost corrections to the potential 
which lead to additional repulsion \cite{akmal98}. Both, the BHF 
calculations from \cite{zuo02} and the variational calculations from 
 \cite{akmal98} are based on the latest ${\rm AV}_{18}$ version of the Argonne 
potential. In both cases phenomenological 3-body-forces are used,  
the Tucson-Melbourne 3-BF in  \cite{zuo02} and 
the Urbana IX 3-BF (using boost corrections 
the repulsive contributions of the UIX interaction are 
reduced by about 40\% compared to the original ones) in \cite{akmal98}. 
Except of ChPT \cite{klock94} there exists no systematic generation 
of 3-BF contributions 
\footnote{Next to leading order all 3-BFs cancel 
while non-vanishing contributions appear at NNLO \protect\cite{klock94}.}.  
On the other hand, contributions from  3-body-forces 
are to large extent canceled by box diagrams containing resonance 
excitations and/or are partially effectively included in the relativistic 
approach (see e.g. the discussion in \cite{honnef,machleidt01}). This fact 
should make an application of DBHF at supra-normal densities more reliable. 
\begin{figure}[h]
\unitlength1cm
\begin{picture}(10.,8.0)
\put(3.5,0){\makebox{\epsfig{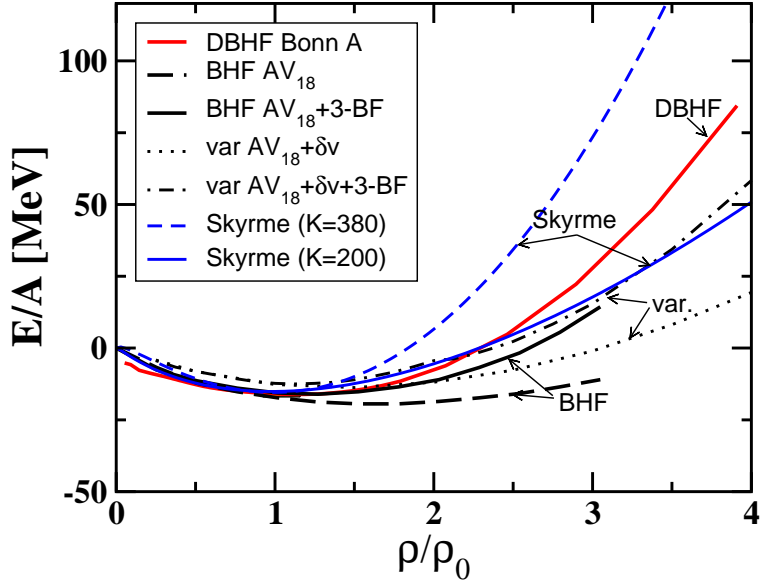}}}
\end{picture}
\caption{The nuclear matter EOS from soft and hard Skyrme forces 
are compared to the predictions from microscopic ab initio 
calculations, i.e. relativistic DBHF \protect\cite{boelting99}, 
non-relativistic BHF \protect\cite{zuo02} and variational 
\protect\cite{akmal98} calculations. 
}
\label{dbeos_fig}
\end{figure}
Fig. \ref{dbeos_fig} compares the equations of state from the 
different approaches, i.e. relativistic DBHF from Ref. \cite{boelting99}  
based on the Bonn A interaction\footnote{The high density behavior of the 
EOS obtained with different interaction, e.g. Bonn B or C is 
very similar. \cite{boelting99}}  \cite{bonn}, 
non-relativistic BHF  \cite{zuo02} 
and variational calculations \cite{akmal98}. The latter ones are 
 based on the Argonne ${\rm AV}_{18}$ potential and include 
3-body forces. All the approaches use modern high precision $NN$ 
interactions and represent state of the art calculations. The  
corresponding EOSs can also be compared to phenomenological 
parameterizations. 

In Fig. \ref{dbeos_fig} two Skyrme EOSs are shown  
which correspond to the limiting cases of a soft (K=200 MeV) and 
a hard (K=380 MeV) EOS. One can conclude from Fig. \ref{dbeos_fig} 
that {\it ab initio} calculations predict throughout a soft EOS in the density range 
relevant for heavy ion reactions at SIS energies, i.e. up to 
about three times $\rho_0$. There seems to be no way to obtain an 
EOS as stiff as the hard Skyrme force shown in Fig. \ref{dbeos_fig}. This 
observation stands somehow in contrast to the observations made from fits to
finite nuclei. When density functionals are fine tuned to the 
nuclear chart, e.g. in RMF theory, the corresponding EOS turns out 
to be relatively stiff \cite{rmf}. The same observation can be made 
within EFT. E.g. the EOS obtained from chiral pion-nucleon dynamics 
by Finelli et al. \cite{finelli} is rather soft but when phenomenological 
correction terms are added in order to improve the description of the 
finite nuclei this results in much stiffer EOS. However, one has to keep 
in mind that finite nuclei constrain the interaction at saturation density 
and below. The predictive power of such density functionals  at supra-normal 
densities is therefore restricted.  

\subsubsection{Skyrme forces in QMD}
Like for QMD calculations shown in the foregoing chapter the 
following investigations are based on the soft and hard 
Skyrme parameterizations. These forces are easy to handle, cover 
the range of uncertainty concerning the EOS for isospin symmetric 
nuclear matter   and are therefore widely used in transport 
calculations for heavy ion collisions.

The QMD N-particle Hamiltonian is given by \cite{Ai91,bass98,aist86}
\beq
H = \sum_{i} \sqrt{{\bf k}_{i}^2 + m_{i}^2} + 
\frac{1}{2} \sum_{i,j\atop (j\neq i)} \left( V_{ij}^{\rm Sk} 
+ V_{ij}^{\rm Yuk} 
+ V_{ij}^{\rm Coul}\right)
\quad .
\label{hamB}
\eeq
The Hamiltonian (\ref{hamB}) contains 2-body 
interactions which are determined as classical expectation values
from local Skyrme forces $V_{ij}^{\rm Sk} $ supplemented by a phenomenological
momentum dependence, an effective 
Coulomb interaction $V_{ij}^{\rm Coul}$ and a  Yukawa-type potential 
$V_{ij}^{\rm Yuk}$. The Yukawa potential mainly serves to improve 
the surface properties and the stability of the initialized nuclei 
when used in heavy ion collisions. The individual nucleons are 
described by Gaussian wave packets with fixed width $2\sqrt{L}$. This 
leads to a one-particle Wigner density
\beq
f_{i} ({\bf q},{\bf k},t) =   \frac{1}{\pi^3} 
e^{-( {\bf q} -{\bf q}_i (t))^2 2/L} ~e^{-( {\bf k} -{\bf k}_i (t))^2 L/2}~~.
\label{wigner}
\eeq
 The Skyrme interaction contains an attractive local two-body part,
 an effective density dependent repulsive two-body part and an 
nonlocal momentum dependent  two-body part. The elementary two-body 
potentials entering into (\ref{hamB}) read then
\beqa
V^{\rm Sk}  &=& t_1 \delta^3 ({\bf q} - {\bf q}^\prime) + 
 t_2 \delta^3 ({\bf q} - {\bf q}^\prime) \rho^{\gamma -1}({\bf q}) 
+ t_3 {\rm ln}^2\left(\epsilon |{\bf k} -{\bf k}^\prime |^2 +1 \right)  
\delta^3 ({\bf q} - {\bf q}^\prime)
\label{skyrme1} 
\\
V^{\rm Yuk}  &=& t_4 
\frac{e^{-|{\bf q} - {\bf q}^\prime|/\mu}}{|{\bf q} - {\bf q}^\prime|/\mu}~~,~~
 V^{\rm coul} = {\left ( \frac{Z}{A}\right )}^2 
\frac{e^2}{|{\bf q} - {\bf q}^\prime |}~~.
\eeqa
In the case $\gamma=2$ the density dependent interaction 
can be derived from a local three-body interaction. The parameterizations of 
\cite{Ai91,bass98,aist86} treat  $\gamma$ as a phenomenological 
parameter. The folding over the Wigner distributions (\ref{wigner}) 
yields the expectation values 
\beqa
V_{ij}^{\rm Sk}  &=& \int d^3q~d^3q^\prime~ d^3k~d^3k^\prime~  
f_i({\bf q},{\bf k},t) ~V^{\rm Sk}({\bf q},{\bf k};{\bf q}^\prime,{\bf k}^\prime)~ 
f_j({\bf q^\prime},{\bf k^\prime},t) 
\nonumber \\ 
&=& \alpha\left(\frac{\rho_{ij}}{\rho_0}\right)
+\beta\left(\frac{\rho_{ij}}{\rho_0}\right)^{\gamma}
+\delta {\rm ln}^2\left(\epsilon |{\bf k}_i -{\bf k}_j |^2 +1 \right)
\frac{\rho_{ij}}{\rho_0},
\label{skyrme2} 
\eeqa
where $\rho_{ij}$ is an  interaction density 
\beq
\rho_{ij} = \frac{1}{(4\pi L)^{\frac{3}{2}}}
e^{-( {\bf q}_{i}- {\bf q}_{j})^2 /4L},
\label{dens1}
\eeq
which arises due to the folding of the two Gaussian wave packets 
with fixed width $2\sqrt{L}$ in coordinate space. In analogous way 
the expectation values of the Yukawa and Coulomb potentials are obtained. 
The parameters $\alpha, \beta, \gamma, 
\delta, \epsilon$ in Eq. (\ref{skyrme2})   are fitted to the saturation point 
($\rho_0 = 0.16~{\rm fm}^{-3}, ~ E_B=-16$ MeV) and the 
momentum dependence of the real part of the 
nucleon-nucleus optical potential. The linear density dependence 
of $V^{\rm Sk} $ is obtained from the point-like 2-body 
interaction in Eq. (\ref{skyrme1}) while 
the nonlinear density dependence is motivated by point-like 3-body 
interactions\footnote{A parameter $\gamma \neq 2$ is purely phenomenological.}
With this Hamiltonian (\ref{hamB}) the EOS of isospin saturated nuclear 
matter, i.e. the binding energy per particle, is of the simple form\cite{Ai91}
\footnote{The small Coulomb and Yukawa terms as well as relativistic corrections 
to the kinetic energy corrections have been suppressed.}
\beqa
E_{\rm bind}= \frac{E}{A} = \frac{3 k_{F}^2}{10 M} 
+ \frac{\alpha}{2}\left(\frac{\rho}{\rho_0}\right) 
+ \frac{\beta}{1+\gamma} \left(\frac{\rho}{\rho_0}\right)^\gamma 
+ \frac{\delta}{2}  {\rm ln}^2\left(\epsilon \left(\frac{\rho}{\rho_0}\right)^2 +1 \right)
\frac{\rho}{\rho_0}~~.
\label{skeos}
\eeqa
In contrast to the Skyrme functional (\ref{skeos}) 
where the high density behavior is fixed by the compression modulus, 
in microscopic approaches the compression modulus is 
only loosely connected to the curvature at saturation density. 
Below $3\rho_0$, e.g. the DBHF EOS with K=230 MeV is close to the 
soft Skyrme EOS but becomes significantly stiffer at higher 
densities. 
\subsection{Particle production and the compression phase in HICs}
\subsubsection{Pions}
With the start of the first relativistic heavy ion programs the 
hope was that particle production would provide a direct experimental 
access to nuclear EOS \cite{stock86}. Without additional compression two times 
saturation density  should be reached in the participant zone of 
the reactions where the difference between the soft and hard Skyrme EOS is about 
13 MeV in binding energy. If the matter could be compressed up to 3$\rho_0$ 
the difference is already $\sim 55$ MeV. It was expected that the  
compressional energy should be released into the creation of 
new particles, primarily pions, when the matter expands \cite{stock86}. 
However, 
pions have large absorption cross sections and they turned out not to 
be suitable messengers of the compression phase. They undergo several 
absorption cycles through nucleon  resonances \cite{bass95,Teis,uma97} 
and freeze out at final stages of the reaction 
and at low densities. Hence pions loose most of their knowledge on the 
compression phase and are not very sensitive probes for 
stiffness of the EOS. This fact is illustrated in  Fig. \ref{pi+spec_fig} which 
shows the $\pi^+$ excitation function in minimal bias Au+Au and C+C 
reactions and compares to data from KaoS \cite{kaos99,sturm01}
\footnote{Details on the treatment of pions in the present QMD 
calculations can be found in \protect\cite{uma97,shekhter03}}.  
For C+C there exists practically no dependence of the pion 
yield on EOS, in Au+Au the dependence is moderate, i.e. of the order of 
$\sim 15-20\%$. 
\begin{figure}[h]
\unitlength1cm
\begin{picture}(9.,9.5)
\put(3.5,0){\makebox{\epsfig{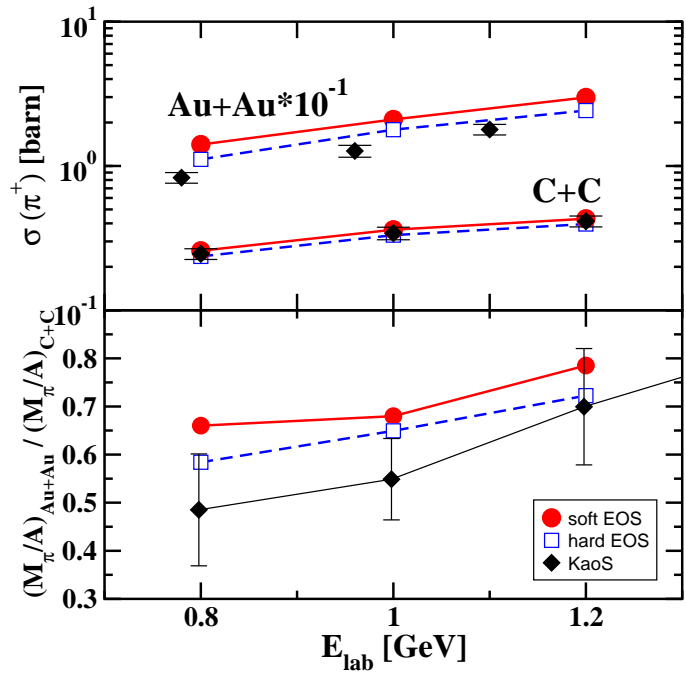}}}
\end{picture}
\caption{Excitation function of the $\pi^+$ production cross section 
in Au+Au (scaled by $10^{-1}$) and C+C reactions (top), and ratio 
of the total pion multiplicities in Au+Au over C+C. 
The calculations are performed using 
a hard/soft nuclear EOS and are compared to data from KaoS 
\protect\cite{kaos99,sturm01}. 
}
\label{pi+spec_fig}
\end{figure}

A second and important observation, in particular with 
respect to the $K^+$ production discussed below, is the fact that 
the enhancement of the pion multiplicities in Au+Au 
when using a soft compared to a hard EOS is almost completely independent 
of the beam energy. This feature becomes even more transparent from the 
lower part of Fig. \ref{pi+spec_fig} which shows the ratio of the total 
pion multiplicities in  $Au+Au$ over $C+C$ reactions scaled by the 
corresponding mass numbers. For this observable the data from 
KaoS \cite{kaos99,sturm01} are qualitatively well reproduced. The usage of 
different nuclear forces leads to a small shift of the theoretical curves 
but does not change their slope.  The fraction is generally 
below unity indicating the larger absorption rate in $Au+Au$ 
compared to $C+C$. With rising energy the pion suppression in 
the heavy compared to the light system becomes smaller as also seen in 
the data. The dependence of this observable on the nuclear EOS is 
rather moderate and would not allow to draw some definite conclusions from 
the model calculations. 
\begin{figure}[h]
\unitlength1cm
\begin{picture}(9.,9.5)
\put(3.5,0){\makebox{\epsfig{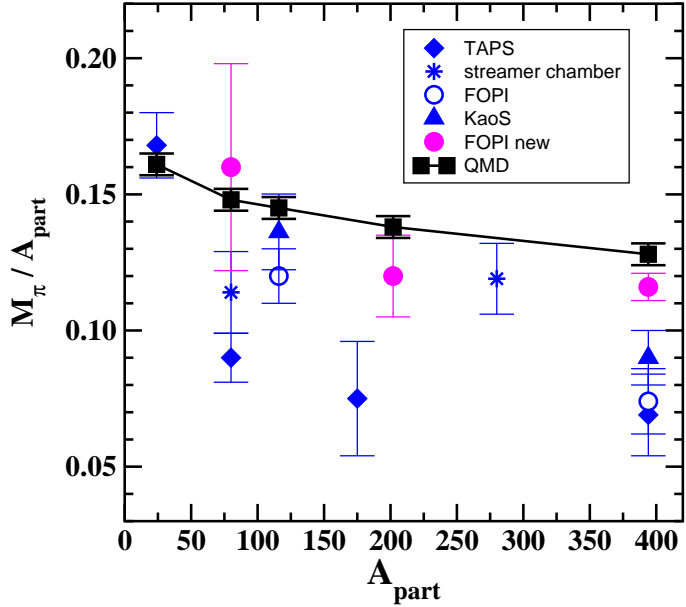}}}
\end{picture}
\caption{Pion multiplicities as a function of $A_{\rm part}$ for various 
mass systems at 1 AGeV incident energy. QMD calculations for 
central reactions (b=0 fm) are 
compared to a compilation of experimental data 
taken form \protect\cite{senger99}. Recent preliminary  
results from FOPI \protect\cite{fopinew} are denoted as FOPI new.
}
\label{pi_apart_fig}
\end{figure}

The other observation  from Fig. \ref{pi+spec_fig}, namely that 
experimental pion multiplicities are in the SIS energy range systematically 
overestimated in large systems like Au+Au is a common (and well known) 
feature of transport models, 
independent if they are of BUU or QMD type
\cite{bass95,Teis,fuchs97,uma97,larionov01}. It demonstrates that the 
pion production and/or absorption mechanisms are theoretically 
not yet fully understood. The pion as a Goldstone boson is the lightest meson 
with a large Compton wave 
length and quantum aspects beyond the semi-classical treatment 
in transport models could be of particular relevance. However, several 
attempts to go beyond the standard treatments and to account for particular 
quantum effects such as the coupling to $\Delta N^{-1}$ excitations 
\cite{helg95,helg98,fuchs97} 
or non-localities, i.e. memory effects and time delays in 
resonance decays \cite{dani96,morawetz} did not 
really improve on this discrepancy and solve the problem.  
Since pions are a source for kaon production through pion induced 
reactions one could consider this fact as a severe uncertainty. 
Fortunately, the situation is less worse since the overestimation of 
the experimental pion multiplicities is restricted to low momentum pions which 
dominate the yields. These pions are, however, not energetic enough 
to contribute to subthreshold 
kaon production and the relevant high momentum tails of the pion spectra are 
generally well described by the transport models \cite{bass95,Teis,uma97}. 

However, for a fair comparison of experiment and theory it should 
be mentioned that also the experimental situation, in particular  
the $A_{\rm part}$ dependence of the multiplicities,  
is not yet completely settled for the SIS/BEVALAC 
energy range. Fig.\ref{pi_apart_fig} 
shows a compilation of published pion multiplicities 
$(\pi^- +\pi^0 + \pi^+)$ for different mass systems as a function of 
$A_{\rm part}$ \cite{senger99} at 1 AGeV 
incident energy. The data set consists of measurements with the 
streamer chamber at LBL 
\cite{harris87} and with the TAPS \cite{berg94,schwalb94,averbeck97}, 
KaoS \cite{wagner94,kaos99} and FOPI \cite{pelte97} 
spectrometers at the GSI. In none of the experiments all pion charges 
have been measured simultaneously and thus each data set contains 
extrapolations in isospin space which are based on the isotopic  
relations for the $\Delta$ production and decay \cite{pelte97}. 
The largest experimental uncertainty lies probably in the determination 
of $A_{\rm part}$. On the one side this quantity is biased by the impact 
parameter resolution but it depends on model assumptions as well.
\footnote{For the determination of the overlapping volume of the 
two interpenetrating nuclei as function of the impact parameter.} The 
corresponding QMD calculations are performed for central reactions 
(b=0 fm) with $A_{\rm part} = A+A$. The experimental yields are well 
described in light and intermediate mass systems but are 
significantly overestimated in heavy systems with $A_{\rm part} \ge 120$. 
The La+La data point \cite{harris87}, which falls off the general systematics, is 
compatible with the transport result within error bars.   
Interesting is, however, that the FOPI Collaboration re-measured 
$\pi^\pm$ yields and a new analysis, denoted in Fig.\ref{pi_apart_fig} as 
FOPI new, is compatible with the streamer chamber systematics and 
also with the transport model predictions. 
\subsubsection{Kaons - historical overview}
\begin{figure}[h]
\unitlength1cm
\begin{picture}(12.,10.5)
\put(1.5,0){\makebox{\epsfig{file=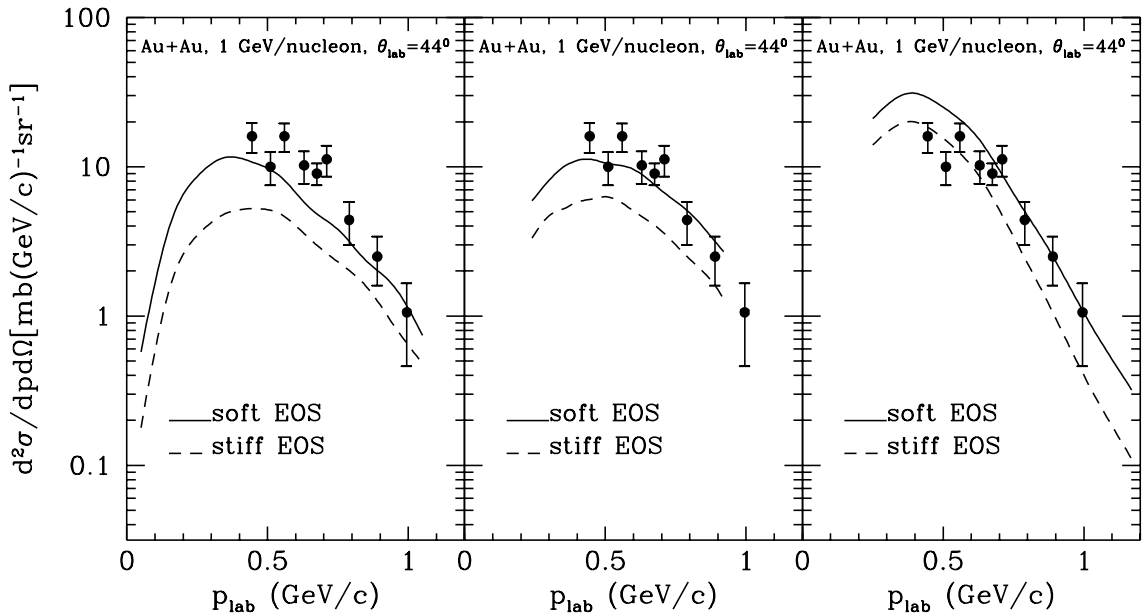,width=12.0cm}}}
\end{picture}
\vspace*{-1.5cm}
\caption{QMD and RBUU transport model calculations from 
Nantes, Texas and Giessen are compared to the 
first KaoS data \protect\cite{kaos94} for $Au+Au$. 
The figure is taken from \protect\cite{ko96}.
}
\label{eos1_fig}
\end{figure}

After pions turned out to fail as suitable messengers, 
$K^+$ mesons were suggested as promising tools to probe the nuclear 
EOS. This idea was first put forward by Aichelin and Ko 
almost 20 years ago \cite{AiKo85}. At subthreshold energies 
$K^+$ mesons are produced in the high density phase and due to 
the absence of absorption reactions they have a long mean free path 
and leave the matter undistorted by strong final state interactions. 
Moreover, at subthreshold energies nucleons have to accumulate energy by 
multiple scattering processes in order to overcome the threshold 
for kaon production and therefore these processes should be particularly 
sensitive to collective effects. 

Already in the first theoretical 
investigations by transport models it was noticed that the $K^+$ 
yield reacts rather sensitive on the EOS 
\cite{lang92,huang93,hartnack94,li95b}. 
Both, in non-relativistic QMD calculations 
based on soft/hard Skyrme forces \cite{huang93,hartnack94} 
and in RBUU \cite{lang92,li95b,giessen94} with soft/hard versions 
of the (non-linear) 
$\sigma\omega$--model for the nuclear mean field it turned out 
that the $K^+$ yield is about a factor 2--3 larger when a soft EOS 
is applied compared to a hard EOS. At that time the available 
data favored a soft equation of state. This fact is 
illustrated in Fig. \ref{eos1_fig} which compares QMD calculations from 
the Nantes \cite{hartnack94} and RBUU  calculations from the Texas/Stony Brook 
\cite{li95b} and Giessen \cite{giessen94} groups to the first KaoS data for 
$Au+Au$ \cite{kaos94}. However, at that stage the 
theoretical calculations were still burdened with large uncertainties. 
First of all, it was noticed \cite{huang93,hartnack94} that 
the influence of the repulsive momentum dependent 
part of the nuclear interaction, Eq. (\ref{skyrme1}), leads to 
a strong suppression of the kaon abundances which made a quantitative 
description of the available data more difficult. Moreover, at that 
time the pion induced reaction channels $\pi B\longrightarrow YK^+ $ 
have not yet been taken into account explicitely. In \cite{lang92} the frozen 
$\Delta$ approximation has been used which includes pionic degrees of freedom 
implicitly. The importance of this 
channel was first pointed out by Fuchs {\it et al.}  \cite{fuchs97b}. 
These additional channels which contribute up to $30\div 50\%$ to the 
total yield allowed to explain the measured yields with 
realistic momentum dependent interactions \cite{fuchs97b,brat97}. 
However, the dependence of the total $K^+$ yield on the nuclear EOS 
turned now out to be much smaller than originally expected, 
i.e. in the order of 15--20\%.  

A further breakthrough was achieved when the  COSY-11 Collaboration 
measured the $pp \longrightarrow p K^+ \Lambda$ reactions at 
threshold \cite{cosy11}. The strangeness production cross 
sections $NN \longrightarrow N K^+ Y$ \cite{sibirtsev95,tsushima99} 
which are nowadays in use are based on these data and are in 
particular close to threshold three orders of magnitude smaller 
than the parameterizations of Randrup and Ko \cite{ran80} which 
were used in the early QMD and RBUU/Texas calculations shown in Fig. \ref{eos1_fig}. 

\subsection{The ratio Au+Au/C+C}
\begin{figure}[h]
\unitlength1cm
\begin{picture}(14.,10.0)
\put(1.5,0){\makebox{\epsfig{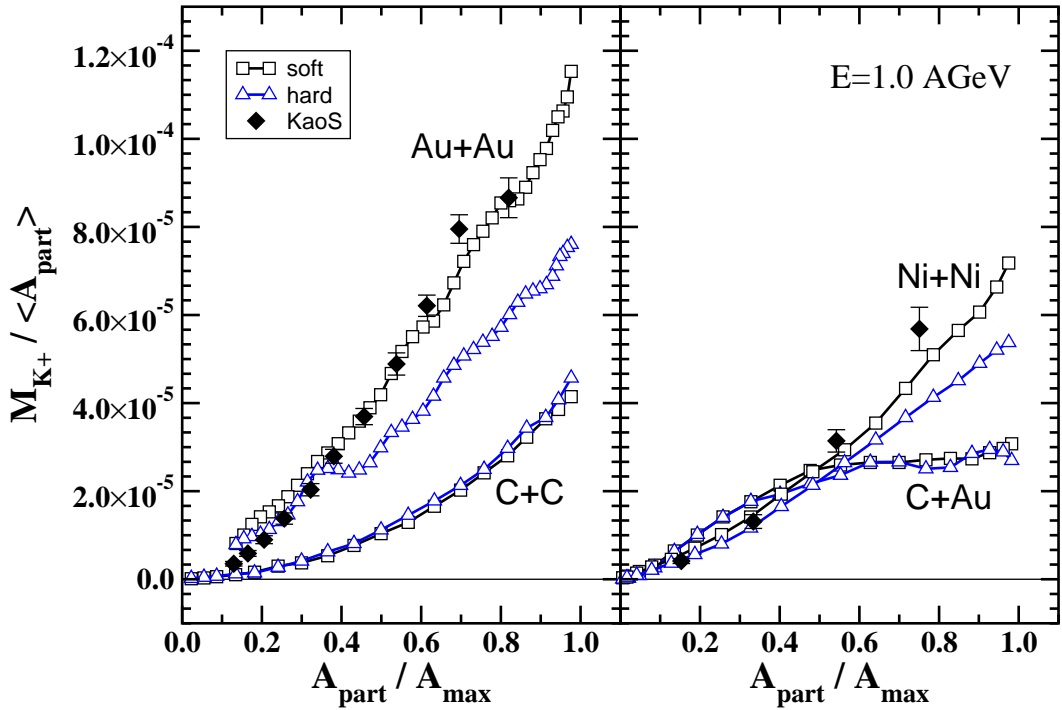}}}
\end{picture}
\caption{EOS dependence of the 
$K^+$ multiplicities as a function of $A_{\rm part}$ in Au+Au, 
Ni+Ni, C+Au and C+C reactions at 1 AGeV. For Au+Au and Ni+Ni data from KaoS 
\protect\cite{senger99,barth} are shown as well. 
}
\label{fig_aparteos}
\end{figure}
Within the last decade the KaoS Collaboration has performed systematic 
measurements of the $ K^+$ production far below threshold 
\cite{kaos94,barth,kaos99,laue00,sturm01}. Based on the new data situation, 
in Ref. \cite{fuchs01} the question 
if valuable information on the nuclear EOS can be extracted 
has been revisited and it has been 
shown that subthreshold  $K^+$ production provides indeed a suitable and 
reliable tool for this purpose. These results have been confirmed by 
the Nantes group later on \cite{hartnack01}. In subsequent publications 
these findings were worked out in more detail \cite{fuchs01b,fuchs04}. 
Here we summarize and complete these investigations.  If not denoted differently, 
throughout this section all model calculations contain a repulsive 
in-medium $K^+$ potential as discussed in Chap. 5. 

In Chap. 3 we discussed already the $A_{\rm part}$ dependence of the 
$K^+$ yield on a qualitative bases. The calculations shown in 
Fig.  \ref{fig_apart_2} were based on a soft EOS. Fig.  \ref{fig_aparteos} 
demonstrates the interplay between $A_{\rm part}$, system size and 
the nuclear EOS. It shows the $K^+$ multiplicities as a 
function $A_{\rm part}$ in Au+Au, Ni+Ni, C+Au and C+C reactions at 1 AGeV. 
The multiplicities are normalized to the mean  $A_{\rm part}$: 
$\langle A_{\rm part} \rangle = A_{\rm max}/2$ with $A_{\rm max}=A+A$ 
for symmetric systems and $A_{\rm max}=56$ for C+Au. 
A significant dependence of the kaon multiplicities on 
the nuclear EOS requires a large amount of collectivity which is easiest 
reached in central reactions of heavy mass systems. Consequently, the 
EOS dependence is most pronounced in central Au+Au reactions. Also 
in Ni+Ni effects are still sizable while the small C+C 
system is completely insensitive on the nuclear EOS even in most central 
reactions. The available data for Au+Au and Ni+Ni  
\protect\cite{senger99,barth} support the soft EOS. 
\begin{figure}[h]
\unitlength1cm
\begin{picture}(9.,6.5)
\put(3.5,0){\makebox{\epsfig{file=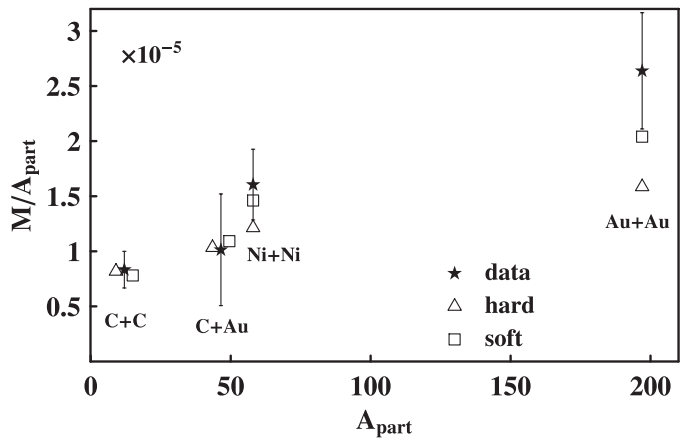,width=10.0cm}}}
\end{picture}
\caption{$K^+$ multiplicities in inclusive C+C,~Ni+Ni,~Au+Au and 
C+Au reactions at 1 AGeV. QMD calculations using a hard/soft nuclear EOS are 
compared to KaoS data \protect\cite{kaos05}. The figure is taken 
from   \protect\cite{kaos05}.
}
\label{fig_aparteos2}
\end{figure}
Interesting is in this context the asymmetric C+Au system: Though in 
central C+Au reactions the number of participants is comparable to 
Ni+Ni the $K^+$ yield does not depend on the EOS. This indicates again 
that a sensitivity on the  EOS is not only a question of  $A_{\rm part}$ but 
of the compression which can be reached by the colliding system. Remarkable 
is the saturation of the kaon yield as a function of  $A_{\rm part}$ predicted 
by the transport calculations. It stands in clear contradiction to the 
$A_{\rm  part}$ dependence predicted by the thermal model
\cite{cleymans99,cleymans00} and a measurement of the quantity would 
allow to distinguish between these two approaches. In Fig. 
\ref{fig_aparteos2} the QMD calculations for inclusive reactions 
are compared to KaoS data from \cite{kaos05}. The data 
are generally described within error bars by the soft EOS. 
Compared to Fig.  \ref{fig_aparteos} in inclusive reactions the EOS dependence 
survives for large mass systems although 
the large difference between the soft and hard -- seen 
in most central Au+Au reactions -- is washed out 
to some extent. In minimal bias reactions the bulk of kaons 
originates from semi-central reactions $b\sim 5$ fm, corresponding 
to $A_{\rm part}/A_{\rm max}\sim 0.7$. The fact that the $K^+$ 
multiplicities in C+Au are significantly smaller than in  Ni+Ni reactions 
although both systems correspond to comparable mean  $A_{\rm part}$  is 
a strong experimental evidence for a strong sensitivity of the kaon 
production on the compression. As also discussed in \cite{kaos05} 
even in central reactions the carbon projectile is simply to small 
in order to achieve a high compression of the  gold target. The 
situation is different in central Ni+Ni reactions. Hence the 
mass dependence  of the kaon multiplicities seems not primordially 
to be an $A_{\rm part}$ effect but a compression effect. This is also 
clearly reflected by the transport calculations.
\begin{figure}[h]
\unitlength1cm
\begin{picture}(9.,9.5)
\put(3.5,0){\makebox{\epsfig{file=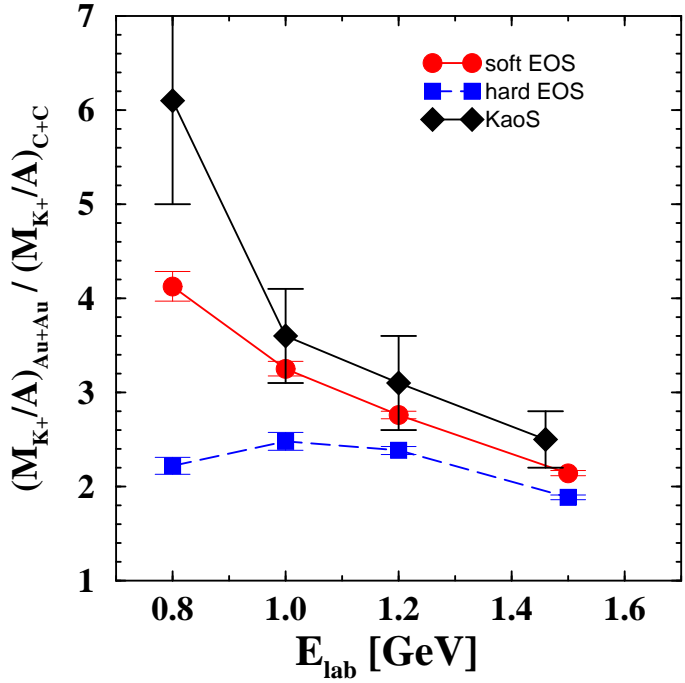,width=9.0cm}}}
\end{picture}
\caption{Excitation function of the ratio $R$ of $K^+$ 
multiplicities obtained in inclusive Au+Au over C+C 
reactions. The calculations are performed with a hard/soft nuclear EOS and 
compared to the data from the KaoS Collaboration \protect\cite{sturm01}. 
}
\label{fig_ratio_1}
\end{figure}
The next and natural step is to consider the energy dependence of the EOS 
effect. It is expected to be most pronounced most far below threshold because 
there the highest degree of collectivity, reflected in multi-step collisions,
is necessary to overcome the production thresholds (see also discussion in 
chapter 3.2).  The calculations for the excitation function shown in 
Fig. \ref{fig_ex_2} were obtained for a soft EOS and are 
performed under minimal bias conditions with $b_{{\rm max}}=11$ fm 
for $Au+Au$ and $b_{{\rm max}}=5$ fm for $C+C$ and normalized to the 
experimental reaction cross sections \cite{sturm01,kaos99}.
For both systems the agreement with the 
KaoS data is quite good. 

The effects become even more evident when the ratio $R$ of the 
kaon multiplicities obtained in Au+Au over C+C 
reactions (normalized to the corresponding mass numbers) is built 
\cite{fuchs01,sturm01}. Such a ratio has moreover the advantage that 
possible uncertainties which 
might still exist in the theoretical calculations should cancel out 
to large extent. 

This ratio is shown in Fig. \ref{fig_ratio_1}. Both, soft and hard EOS, 
show an increase of $R$ with decreasing energy 
down to 1.0 AGeV. However, this increase is much less 
pronounced when the stiff EOS is employed. 
In the latter  case $R$ even decreases at 0.8 AGeV 
whereas the soft EOS leads to an unrelieved increase of $R$. 
At 1.5 AGeV which is already very close to threshold 
the differences between the two models become small. 
The strong increase of $R$ can be directly related to 
higher compressible nuclear matter. The comparison to the experimental 
data from KaoS \cite{sturm01}, where the increase of $R$ is even more 
pronounced, strongly favors a soft equation of state.

\subsubsection{Phase space dependence}
\begin{figure}[h]
\unitlength1cm
\begin{picture}(13.,9.0)
\put(1.5,0){\makebox{\epsfig{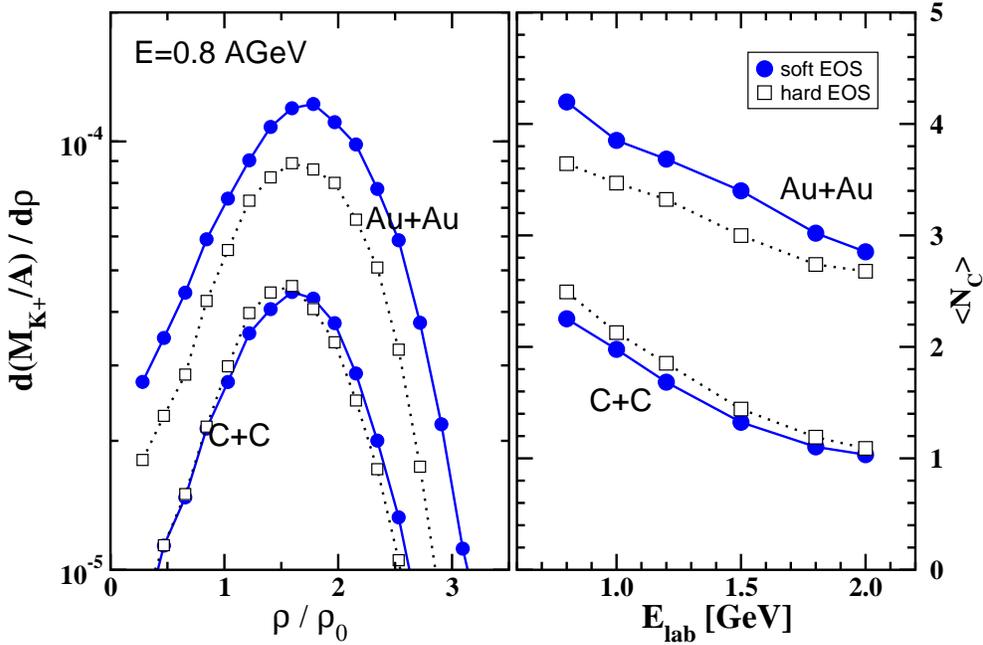}}}
\end{picture}
\caption{Phase space dependence of the $K^+$ production. Left: 
nuclear density at the production of $K^+$ mesons. 
Right: number of collisions 
which the particles encountered prior to $K^+$ production. In both 
cases the EOS dependence in central Au+Au and C+C reactions
is studied. 
}
\label{EOS3_fig}
\end{figure}
To obtain a quantitative picture of the explored density 
effects in Fig. \ref{EOS3_fig} the baryon densities are shown at 
which the kaons are created. The energy is chosen most below 
threshold, i.e. at 0.8 AGeV and only central collisions 
are considered where the effects are maximal. $dM_{K^+}/d\rho$ 
is defined as in Eq. (\ref{densdist}). For 
the comparison of the two systems the curves are normalized to the 
corresponding mass numbers. 

Fig. \ref{EOS3_fig} illustrates several features: 
Only in the case of a soft EOS the mean densities at which kaons 
are created differ significantly for the two different reaction 
systems, i.e. $<\rho /\rho_{0} >$=1.46/1.40 for C+C  
and 1.47/1.57 for Au+Au using  
the hard/soft EOS. Generally, in C+C reactions densities above 
$2\rho_{0}$ are rarely reached whereas in Au+Au the kaons are 
created at densities up to three times saturation density. 
Furthermore, for C+C the density distributions are weakly 
dependent on the nuclear EOS. The situation changes 
completely in Au+Au. Here the density profile 
shows a pronounced EOS dependence \cite{li95b}. 
Moreover, the excess of kaons obtained with the soft EOS 
originates almost exclusively from 
high density matter. A second quantitative measure 
for the collectivity is the average number of collisions for those 
hadrons which were involved in the $K^+$ production 
displayed in Fig. \ref{EOS3_fig}. 
Again only central collisions are considered where the 
effects are maximal. $<N_C >$ is defined as in 
Fig. (\ref{fig_ncoll_1}).  In average the particles 
undergo about twice as much relevant collisions in the heavy 
compared to the light system. Furthermore, the collectivity, i.e. 
the accumulation of energy by multiple scattering, increases 
with decreasing energy. Thus one can conclude that the 
increase of $R$ is not due to a trivial phase space effect, namely 
the fact that far below threshold the C+C system is  
simply too small to provide enough collectivity for the 
kaon production. If such a scenario - which in principle 
also explain the rise of $R$ seen in the KaoS data - 
would be true, $<N_C >$ would have to saturate for C+C 
at low energies. This demonstrates that $K^+$ production 
far below threshold always requires a certain amount of collectivity 
which can be provided also in a very small colliding system, though 
such processes are rare. There is, however, no sharp limit 
were such collision histories become impossible. 
Thus trivial phase space effects can be excluded 
for an explanation of the increase of $R$. In \cite{sturm01} 
a similar argument was based on the measurement 
of high energy pions which can test the phase space available for 
particle production.

\subsubsection{Stability of the EOS dependence}
\begin{figure}[h]
\unitlength1cm
\begin{picture}(9.,9.0)
\put(3.5,0){\makebox{\epsfig{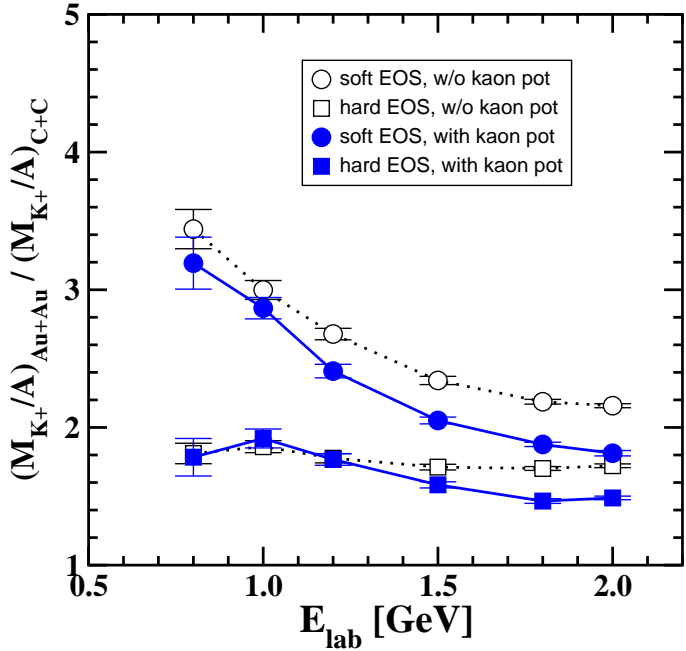}}}
\end{picture}
\caption{Excitation function of the ratio of $K^+$ 
multiplicities obtained in central  (b=0 fm) Au+Au over C+C 
reactions. The calculations are performed with/without 
in-medium kaon potential and using a hard/soft nuclear EOS. 
The figure is taken from \protect\cite{fuchs01}.
}
\label{fig_ex_4}
\end{figure}
Now the question arises, how firm the conclusions on the
nuclear EOS are. The influence of the repulsive in-medium potential 
has been discussed in \cite{fuchs01} and can be seen from 
Fig. \ref{fig_ex_4} which shows the ratio $R$ in central 
reactions from simulations with and without kaon potential. It is 
remarkable and at a first glance surprising that $R$ shows for 
both cases qualitatively the same behavior. The 
in-medium kaon potential acts opposite to the EOS 
effect: a higher compression 
increases the kaon yield but also the value of the in-medium 
kaon mass which, on the other hand, tends to lower the yield again. 
However, the increase of the in-medium mass goes linear with 
density whereas the collision rate per volume increases approximately with 
$\rho^2$. E.g. in central Au+Au reactions at 0.8 A GeV the 
average density $<\rho >$ at kaon production 
is enhanced from 1.47 to 1.57 $\rho_0$ switching from the 
hard to the soft EOS. This leads to an average shift of the in-medium 
mass (\ref{effmass}) compared to the vacuum value of 55/61 MeV using 
the hard/soft EOS, i.e. a relative shift of 6 MeV between 
soft and hard. However, collective effects 
like the accumulation of energy by multiple scattering 
show a higher sensitivity on the compression resulting in an 
enhancement of the available energy $<\sqrt{s}>=90$ MeV applying 
the soft EOS. 
\begin{figure}[h]
\unitlength1cm
\begin{picture}(14.,9)
\put(1.5,0){\makebox{\epsfig{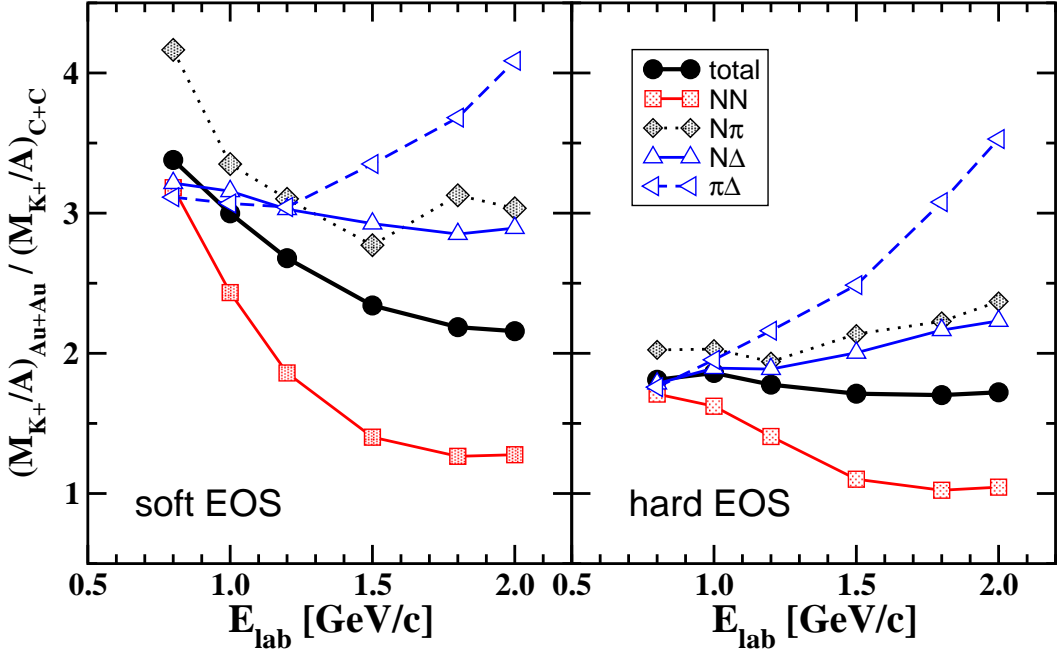}}}
\end{picture}
\caption{Dependence of the excitation function of $R$ on the 
various $K^+$ production channels. Central (b=0 fm) Au+Au and C+C 
reactions are considered. The results are taken from 
\protect\cite{fuchs01b}.
}
\label{fig_chan}
\end{figure}

The next question concerns the  knowledge 
of the elementary reaction cross sections. As discussed in 
Chapter 4 the $NN$ and $ \pi N$ cross sections are well under
control since these channels are constrained by data. The reactions which 
involve nucleon resonances in the 
initial states ($ i= N\Delta, \pi\Delta, \Delta\Delta$) are less secure 
due to the lack of data and one has to rely on model assumptions. 
The cross sections which have been used in the present transport 
calculations are based on the effective Lagrangian model of 
Refs. \cite{tsushima99,tuebingen,tsushima00}. The isospin dependence 
of the cross sections was determined 
by isotopic relations assuming isospin independent matrix elements.   

In summary, some uncertainty in the transport calculations is still 
existing due to the fact that elementary production channels involving 
$\Delta$ resonances are not constrained by data \footnote{Elementary reactions 
with a $N^*(1440)$ resonance in the initial states are included in the 
present calculations. For these reactions the same 
cross sections as for nucleons are used. Higher lying nucleon 
resonances can be neglected at subthreshold energies.}. The ratio built 
from different mass systems should, however, be robust against such 
uncertainties:
\begin{itemize}
\item Changes of the production cross sections shift absolute yields   
but considering the ratio possible errors 
drop out in leading order.
\item Conclusions are based on the slope of this ratio as a function
of energy. It is extremely unlikely that an incomplete knowledge 
of  cross sections, i.e. an unknown isospin dependence, 
can create the observed energy dependence. 
The systematics of spurious contributions should rather 
be flat as a function of energy. 
\end{itemize}
To illustrate these arguments the influence of the different elementary 
channels is shown in Fig. \ref{fig_chan}. 
There the ratios $R_i$ are built separately for the production 
channels with initial states $ i= NN, \pi N, N\Delta, \pi\Delta, 
\Delta\Delta$. The shape of $R$ is not strongly influenced 
by the $N\Delta~, \pi\Delta$ channels which are the most insecure ones. 
The excitation function for 
the $N\Delta$ contribution varies only little as a function of energy 
and is similar using the different EOSs. The contribution of the 
$\pi\Delta$ channel is decreasing for 
both, a hard and a soft EOS. The shape of $R$ is to most extent 
determined by the $NN$ and $ \pi N$ contributions which are well under 
control. 
\begin{figure}[h]
\unitlength1cm
\begin{picture}(9.5,9.0)
\put(3.5,0){\makebox{\epsfig{file=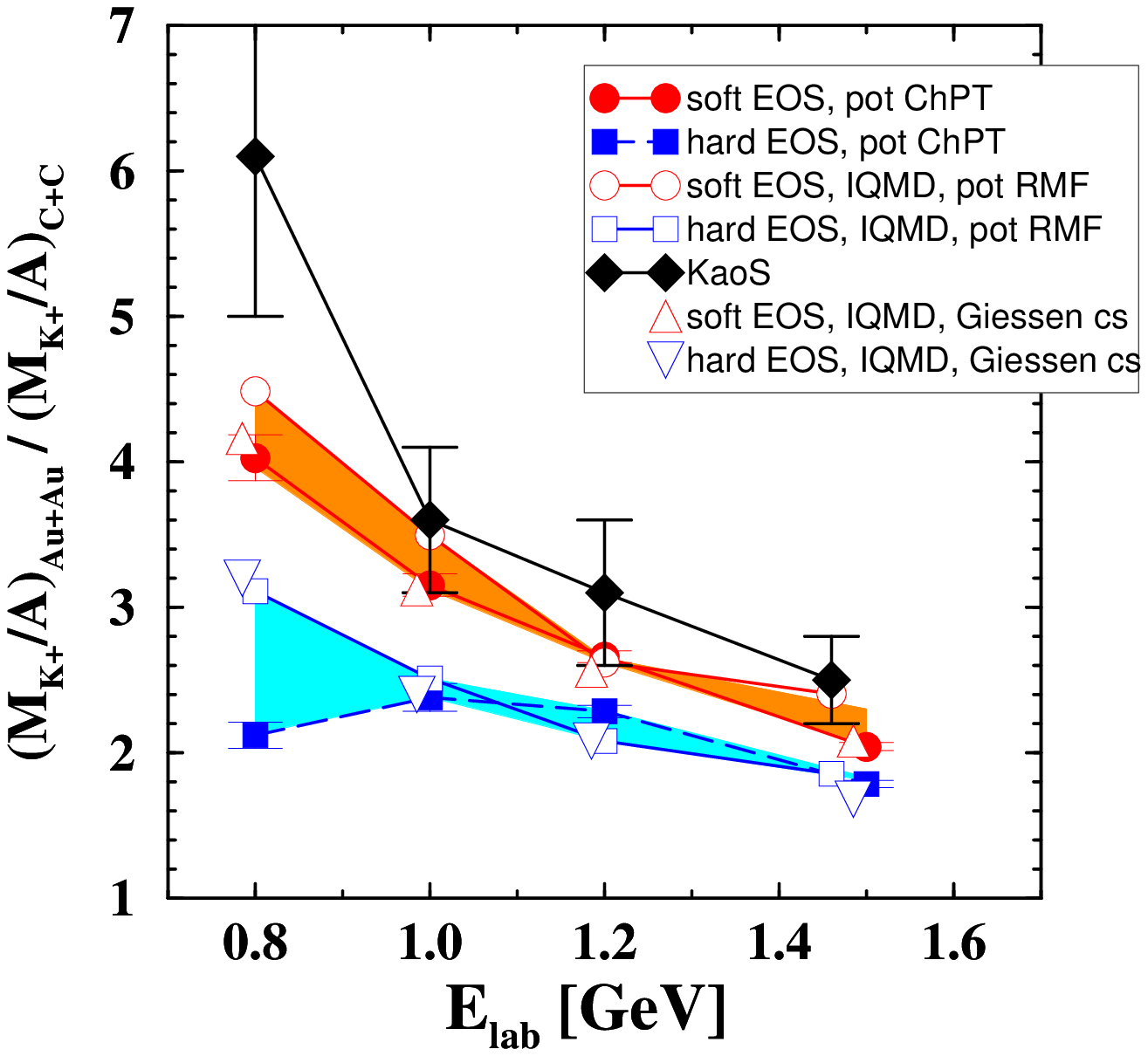,width=9.5cm}}}
\end{picture}
\caption{Excitation function of the ratio $R$ of $K^+$ 
multiplicities obtained in inclusive Au+Au over C+C 
reactions. Our results are compared to IQMD 
calculations \protect\cite{hartnack01}. The shaded area indicates
thereby the range of uncertainty in the theoretical models. 
In addition IQMD results based on an alternative set of 
elementary $K^+$ production cross sections are shown.
}
\label{FigIQMD}
\end{figure}

These findings are generally confirmed by 
independent transport calculations of 
the Nantes group using the IQMD transport model \cite{hartnack01} shown 
in Fig. \ref{FigIQMD} together with the present QMD results. 
The IQMD calculations include an in-medium kaon potential derived in 
relativistic mean field theory (RMF) \cite{schaffner97} 
which is somewhat less repulsive than that one 
used in the present calculations. For the soft 
EOS the IQMD calculations coincide almost with the present 
results \cite{fuchs01}. The two 
sets of transport calculations show a good overall agreement and 
both rule out the hard EOS from the comparison with data. The 
shaded area in Fig. \ref{FigIQMD} can be taken as the existing range of 
uncertainty in the theoretical model description of the considered 
observable. 

Moreover, the IQMD calculations were also repeated with an alternative 
set of $N\Delta; \Delta\Delta \mapsto NYK^+$ cross sections taken from
\cite{cassing99} which are almost one order of magnitude smaller than those  
from Tsushima et al. \cite{tsushima99} (see Fig. \ref{fig_cross2}). 
The ratio $R$ is almost completely independent on this change in 
elementary cross sections and also total yields. An even more systematic 
study of possible uncertainties concerning this observable has been 
performed by Hartnack in \cite{hartnack05}. A possible medium dependence 
of the elementary cross sections, i.e. a density dependent reduction, 
as well as a general scaling has been investigated. Also in these 
extreme cases the EOS dependence of the ratio $R$ survived and 
conclusions stayed stable. This demonstrates once more 
the robustness of this observable. 

\subsection{Constraints from other sources}
\subsubsection{Nucleon flow}
\begin{figure}[h]
\unitlength1cm
\begin{picture}(10.,8.0)
\put(3.5,0){\makebox{\epsfig{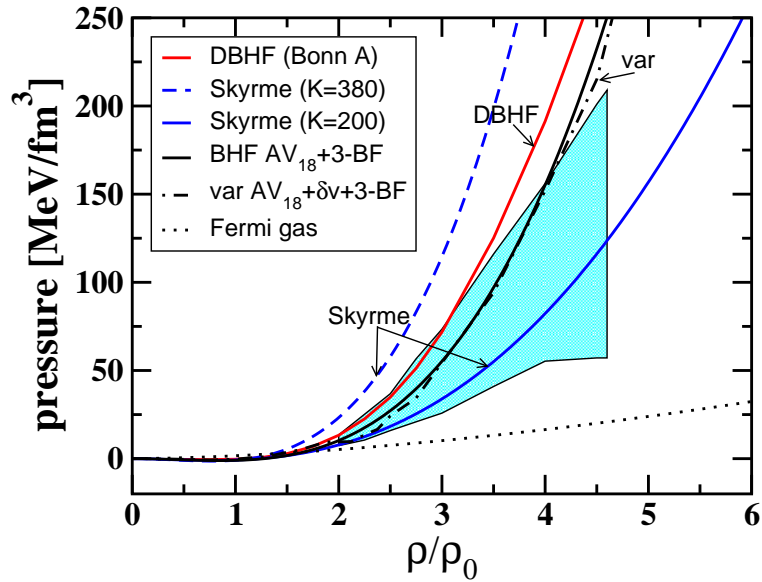}}}
\end{picture}
\caption{Constraints on the nuclear EOS from heavy ion flow data. 
The shaded area shows the pressure-density which is compatible with 
heavy ion flow data according the analysis on \protect\cite{dani02}. 
The equations-of-state from the models 
shown in Fig. \protect\ref{dbeos_fig} are displayed.
}
\label{eos_pressure_fig}
\end{figure}

Concerning the nuclear equation of state one has to confront the 
information from subthreshold $K^+$ production with the knowledge 
obtained from other sources: 
At intermediate energies heavy ion reactions test the density 
range between one and two, maximally 
three times nuclear density. The information from kaon 
production implies  
that in this density range the EOS shows a soft behavior. 
One has of course to be aware that the adopted Skyrme forces are 
simplified interactions which are easy to handle but must not be 
very realistic.

Another observable which helps to constrain the nuclear mean field 
and the underlying EOS at supra-normal densities 
is the collective nucleon flow \cite{norbert99}. 
The transverse flow $v_1$ has been found to be sensitive to the EOS
and, in particular in peripheral reactions, to the momentum dependence 
of the mean field \cite{dani00,gaitanos01,larionov00}. The elliptic flow $v_2$, 
in addition, is very sensitive to the maximal compression reached in
the early phase of a heavy ion reaction. The cross over from
preferential in-plane flow $v_2 < 0$ to preferential 
out-off-plane flow  $v_2 > 0$ around 4-6 AGeV has also led to
speculations about a phase transition in this energy region which 
goes along with a softening of the EOS \cite{eflow}. However, the 
corresponding AGS data can also be explained conventionally \cite{sahu}.

The present situation can be summarized as follows: Flow data at SIS energies 
are consistent with a soft 
EOS \cite{hombach99,dani00}. The full flow excitation 
function, ranging from low SIS ($E_{\rm lab} \simeq 0.2\div 2$ AGeV) 
up to top AGS energies ($E_{\rm lab} \simeq 2\div 11$ AGeV), 
has been studied in \cite{dani02,sahu}. The conclusion from Ref. \cite{dani02}
was that, both, super-soft equations of state (K=167 MeV) 
as well as hard EOSs (K$>$300 MeV) are ruled out by  data. Fig. 
\ref{eos_pressure_fig} displays the pressure-density area which, 
according to the analysis of \cite{dani02}, is consistent with 
heavy ion flow data. The soft Skyrme EOS is in agreement with 
flow data. The boundaries of Fig. \ref{eos_pressure_fig} 
are the result of a compilation from the analysis of sideward and 
elliptic anisotropies. In the models used by Danielewicz et al. 
\cite{dani00,dani02} sideward flow favors indeed a rather soft 
EOS with K=210 MeV 
while the development of the elliptic flow requires slightly 
higher pressures. The BHF and variational calculations 
including 3-body-forces\protect\footnote{For the 
BHF + 3-BF calculation the pressure shown in Fig. 
\ref{eos_pressure_fig} has been determined from 
the parameterization given in \protect\cite{zuo04} which is 
based on the Urbana IX 3-BF different to that used in \protect\cite{zuo02}.} 
fit well into the constrained area up to 4$\rho_0$. At higher 
densities the microscopic EOSs, also DBHF, tend to be too 
repulsive.

However, conclusions from flow data are 
generally complicated by the interplay of the compressional part 
of the nuclear EOS and the momentum dependence of the nuclear 
forces. A detailed comparison to  $v_1$ and $v_2$ data below 1 AGeV from 
FOPI \cite{andronic99} and KaoS \cite{brill96} favors 
again a relatively soft EOS with a momentum dependence
 close to that obtained from microscopic DBHF calculations 
\cite{dani00,gaitanos01,gaitanos04}. In Fig. 
\ref{eos_pressure_fig} the microscopic DBHF EOS (K=230 MeV) 
lies at the upper edge of the boundary, but is still consistent  
in the density range tested at SIS energies, 
i.e. up to maximally 3 $\rho_0$. This fact is further 
consistent with the findings of Gaitanos et al. 
\cite{gaitanos01,gaitanos04} where a good 
description of $v_1$ and $v_2$ data at energies between 
0.2 and 0.8 AGeV has been found in RBUU calculations based 
on DBHF mean fields. However, as pointed out in Refs. 
\cite{gaitanos01,fuchs03,gaitanos04,lca} is thereby essential 
to account for non-equilibrium effects and the momentum 
dependence of the forces which softens the EOS compared to 
the equilibrium case shown in  Fig. \ref{eos_pressure_fig}.   
 
In summary, $K^+$ production and nucleon flow 
provide a consistent picture so far that stiff equations-of-state 
are ruled out. Details concerning the interplay between density and 
momentum dependence have still to be settled and require future efforts. 
\subsubsection{Neutron stars and symmetry energy}
Models for neutron stars are constrained by the lower limit of 
the maximal neutron star mass. This means that any equation-of-state 
must be stiff enough to produce a neutron star of mass greater than 
1.44 solar masses, the largest neutron star in the PSR 1913+16 system. 
The upper limit for a nuclear EOS is thereby obtained by the 
conventional neutron star consisting of neutrons and protons. 
The occurrence of additional degrees of freedom such as 
pion, $K^-$ or H-dibaryon condensation or the excitation of 
hyperons in hybrid stars softens the EOS and reduces the maximal 
neutron star mass. The same is true when quark cores or cores 
of strange matter are considered. 
Conventional neutron stars put only weak constraints on the 
nuclear EOS. Super-soft EOSs with a compressibility less than 
K$\lesssim$ 120 MeV can be ruled out \cite{prakash94}. The microscopic models 
discussed above yield maximal neutron star mass above two solar 
mass which are all very close: 
DBHF (Bonn A) \cite{dalen04} gives $M_{\rm max} = 2.26 M_\odot$, 
BHF + 3-BF $M_{\rm max} =2.3 M_\odot$ \cite{zuo04} and the variational 
calculations with  3-BFs + boost correction gives 
$M_{\rm max}= 2.21 M_\odot$ \cite{akmal98}. 
A soft EOS comparable to the soft Skyrme force, 
i.e. a chiral relativistic mean field model with K=194 MeV used 
in \cite{li97b} yields $M_{\rm max} = 2.0 M_\odot$. 

Since strangeness is not conserved in weak interactions 
$K^-$ condensation can occur in neutron stars 
at densities above $3\div 5\rho_0$ 
\cite{brown94,prakash94,li97b,weber05,kolom03}, depending on the strength 
of the $K^-N$ interaction and the nuclear EOS. The $K^-$ condensate 
introduces additional negative charge which enhances the proton 
fraction in the star and makes the EOS softer. Its influence depends 
therefore strongly on the high density behavior of the symmetry energy. 
It has been found that the maximal mass is reduced by about $20\div 25\%$ 
by a $K^-$ condensate \cite{prakash94,li97b} which would then rule 
out an EOS softer than K$\lesssim$ 180 MeV. Neutron stars constrain 
the isospin symmetric  EOS from below.
\begin{figure}[h]
\unitlength1cm
\begin{picture}(14.,10.0)
\put(1.0,0){\makebox{\epsfig{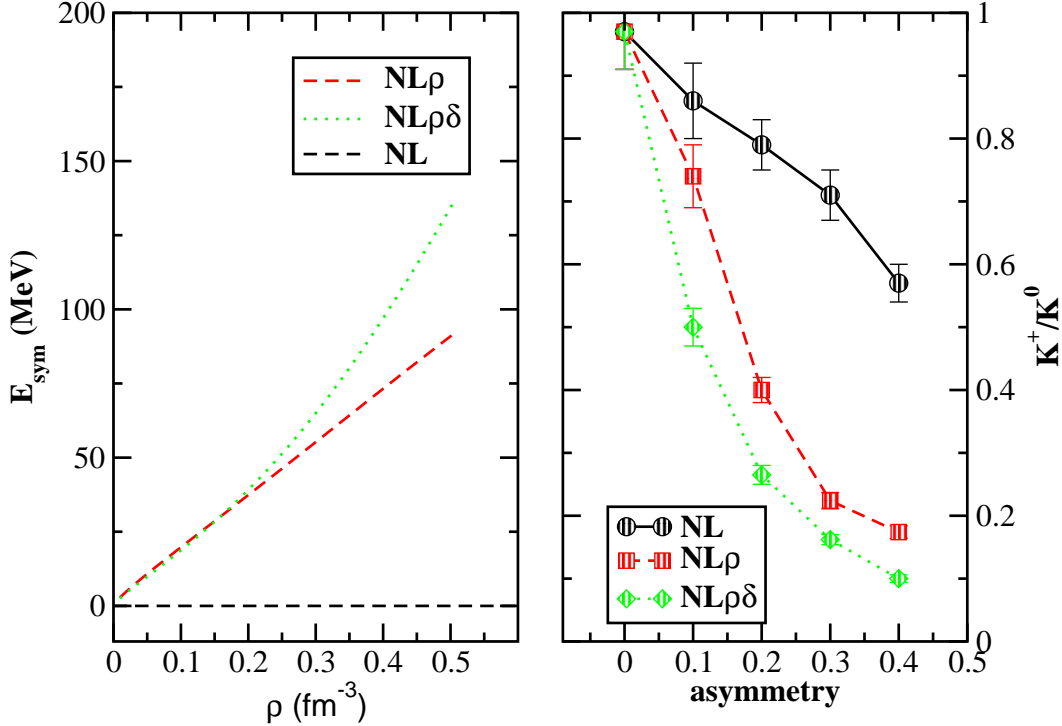}}}
\end{picture}
\caption{Left: symmetry energy of the various models as a function 
of density. Right: $K^+/K^0$ ratio in nuclear matter at  
density 2.5 $\rho_0$ and temperature $T=60$ MeV 
as a function of the proton-neutron asymmetry $\beta$ obtained 
from RBUU calculations with periodic boundary conditions. The results 
are taken from \protect\cite{ferini05}. 
}
\label{asym_fig}
\end{figure}


Of particular interest is in this context the symmetry energy 
which characterizes the isospin dependence of the nuclear forces 
and determines the  proton fraction inside a neutron star. 
In isospin asymmetric matter the binding energy is a functional 
of the proton and neutron densities.  Asymmetric matter is 
characterized by the  asymmetry parameter $\beta=Y_n-Y_p$ which 
is the difference of the neutron and proton fraction 
$Y_i=\rho_i/\rho~,i=n,p$.  The isospin dependence of the energy  
functional can be expanded in terms of $\beta$ 
\begin{eqnarray}
E(\rho,\beta) = E(\rho) + E_{\rm sym}(\rho) \beta^2 + {\cal O}(\beta^4) + 
\cdots ~~~. 
\label{esym}
\end{eqnarray}
The high density behavior of the symmetry energy is at present 
largely unconstrained and kaons could again turn out to be a suitable 
tool to derive constraints from heavy ion collisions. In this case  
the $K^+/K^0$ ratio has been suggested as a promising observable by 
the Catania group \cite{ferini05}. 
The isospin dependence of the elementary production 
cross sections reflects the asymmetry of the emitting 
source which is connected 
to the stiffness of symmetry energy and thus translated into  
the corresponding $K^+$ and $K^0$ yields. RBUU calculations with 
periodic boundary conditions for equilibrated nuclear matter at 2.5 times 
saturation density and a temperature of $T=60$ MeV, 
shwon in Fig. \ref{asym_fig}, indicate a strong sensitivity 
on $ E_{\rm sym}$ and the isospin dependence of the nuclear forces. 
The calculations are based on non-linear Walecka mean field including  
either the vector isovetor $\rho$-meson ($NL\rho$) or in addition the 
scalar isovector $\delta$-meson  ($NL\rho\delta$). The NL model contains 
only isoscalar $\sigma$ and $\omega$ mesons and has therefore 
$ E_{\rm sym} =0 $.   
The measurement of $K^{+,0}$ mesons in symmetric and highly isospin 
asymmetric colliding systems might therefore provide experimental 
access to the isospin dependent part of the nuclear EOS.   

\section{Summary and outlook}
Kaon and antikaon production in nucleus-nucleus collisions around 
threshold energies opens the possibility to attack a variety of 
physics questions which have important implications in nuclear 
physics, QCD and astrophysics. The present article tried to summarize 
the status of the field and to point out which problems could be 
settled and which questions are still open.

Theory predicts strong modifications of the kaon and antikaon properties 
in a dense hadronic environment. Mean field models as well as chiral 
perturbation theory predict a repulsive $K^+$ potential of about 
$V_{K^+} \simeq +(20\div 30)$ MeV at nuclear saturation 
density. Such a value is in agreement with empirical kaon-nucleon 
scattering. The $K^-$-nucleon interaction, in contrast, is resonant around 
threshold and requires non-perturbative approaches. The strength of the 
$K^-$ potential is still an open question. The depth of the attractive 
antikaon-nucleon potential  ranges from 
$V_{K^-} \simeq - (50\div 100)$ MeV, obtained within chiral coupled 
channel dynamics, to $V_{K^-} \simeq - (100\div 200)$ MeV predicted by 
mean field approaches and the analysis of kaonic atoms. In contrast to 
the $K^+$ mesons the  $K^-$ mesons develop complicated 
spectral properties in the medium. 

In heavy ion collisions at intermediate energies, i.e. at energies around 
the threshold region, strangeness is generally produced in the early and 
high density phase of the reaction. However, the freeze-out conditions 
for kaons and antikaons are completely different. Due to strangeness 
conservation $K^+$  mesons cannot 
be reabsorbed by the surrounding nucleons and their chemical 
freeze-out takes place early. Final state interactions, i.e. elastic 
scattering or charge exchange reactions and the influence of the optical 
kaon-nucleon potential change their dynamical pattern but not the abundances. 
This makes $K^+$ mesons to a suitable 'penetrating' probe to study the 
dense fireball created in a heavy ion reaction. Antikaons, in contrast, are 
strongly coupled to the environment through strangeness exchange reactions. 
This leads to a late freeze-out and a loss of memory on the early 
reaction stages. 

The link of the underlying physics to the heavy ion experiments must be provided by 
dynamical transport models. The precision of such models depends thereby 
crucially on the elementary input, i.e. the knowledge of the elementary reaction 
cross sections. This input is much better constrained by data 
for the $K^+$ mesons than for $K^-$. In addition, the quasiparticle picture 
which underlies all types of semiclassical transport approaches, 
is much better justified for the kaons than for the antikaons. 
Hence state of the art transport calculations have reached a reasonable 
degree of consistency concerning $K^+$ production and dynamics. The 
comparison to experiment concerning both, total yields as well as 
dynamical observables such as in-plane and out-off-plane flow pattern, 
supports the existence of a slightly repulsive in-medium potential as predicted 
by chiral dynamics. This picture is complemented by data from 
proton-nucleus reactions. 

Concerning the antikaons the 
situation is less satisfactory, both from the experimental 
as well as from the theoretical side. To settle the question of 
a strongly attractive $K^-$-nucleon potential is one of the major 
challenges of this field in future. If the $K^-$-nucleon potential is 
strong enough, this can lead to deeply 
bound $K^-$ states in nuclei  \cite{hirenzaki05} and even to light 
kaonic nuclear clusters \cite{yamazaki05}. Such 
states would be much stronger bound than the 
$\pi^-$ states observed in pionic atoms \cite{kienle04}. 
Some calculations predict 
even a collapse of the nuclear wave functions to densities significantly 
above saturation density in $K^-$-nuclear clusters which would allow to access the 
nuclear forces at very short distances. The existence of 
such molecule states would thus open a completely new field in hadron physics 
with strong implications on nuclear structure. Heavy ion reactions as well 
as proton-nucleus reactions are the tools to clarify the preconditions 
for this hypothesis.

To draw firm conclusions on the 
in-medium antikaon properties will, however, require significant efforts to control 
their off-shell dynamics. The same holds for other hadrons, e.g. 
vector mesons, where dramatic changes of their spectral properties 
are expected. First attempts towards a quantum transport have been made but an 
exact treatment of the quantum evolution equations requires a better 
knowledge of off-shell transition elements and in-medium spectral functions 
of all the involved hadron species. To develop a consistent 
quantum transport is  the major challenge of future theoretical 
heavy ion physics. The same holds for proton-nucleus reactions. Also 
there the quantitative understanding of high precision meson production data, 
e.g. from GSI, COSY or KEK, requires to control the off-shell dynamics 
of such processes.

Finally the kaons turned out to provide a suitable tool to attack another 
longstanding question, namely the stiffness of the nuclear 
equation-of-state. The high density behavior of the EOS has severe 
astrophysical consequences since it determines e.g. the maximal mass 
and the radii of neutron stars. Although heavy ion reactions test 
mainly isospin symmetric matter they put constraints on theoretical 
models which are also applied to neutron stars. The systematic measurement  
of the $K^+$ excitation function in heavy and light systems down to energies 
far below threshold can here be considered as a breakthrough. Extreme 
subthreshold energies exclude distortions from surface effects and ensure 
that the $K^+$ mesons originate from supra-normal nuclear densities. To 
overcome the production thresholds 
requires a high degree of collectivity which, on 
the other hand, introduces the sensitivity on the compression achieved 
in the reaction. The comparison with data strongly supports an EOS which is 
'soft' in a density regime between 1-3 times saturation density. Such 
a behavior is consistent with the predictions from microscopic 
many-body calculations and the constraints obtained from nucleon flow 
data in heavy ion reactions. Based on similar arguments the $K^+/K^0$ ratio 
has recently been proposed as a tool to access the isospin dependence of the 
nuclear EOS and to constrain the symmetry energy. If this turns out to be 
true, subthreshold kaon production can probably 
be considered as the most successful 
observable to constrain nuclear forces at high densities.\\

{\bf Acknowledgments:}\\
The author would like to thank the following people 
for valuable discussions and/or for providing experimental data 
or results of theoretical calculations:\\
J. Aichelin, H.-W. Barz, E. Bratkovskaya, G. Burau, 
W. Cassing, L.-W. Chen, A. Faessler, T. Gaitanos, 
T. Gutsche, Ch. Hartnack, N. Herrmann, E. Kolomeitsev, C.M. Ko, A. Larionov, M. Lutz,  
V. Lyubowitzki, H. Oeschler, A. Ramos, P. Senger, C. Sturm, 
L. Tolos, K. Tsushima, Y.-M. Zheng



\end{document}